\documentclass[notitlepage, nofootinbib, amsmath,amssymb, aps, prd]{revtex4-1}

\usepackage{graphicx,grffile}
\usepackage{dcolumn}
\usepackage{bm,physics}
\usepackage{amsthm}
\usepackage{ulem}
\usepackage{hyperref}

\usepackage{braket}

\begin{document}
 
\title{Multipartite Optimized Correlation Measures and Holography}


\author{Oliver DeWolfe}
\email{oliver.dewolfe@colorado.edu}
\affiliation{Department of Physics and Center for Theory of Quantum Matter, University of Colorado, Boulder CO 80309, USA}
\author{Joshua Levin}
\email{joshua.t.levin@colorado.edu}

\affiliation{Department of Physics, University of Colorado, Boulder CO 80309, USA}
\affiliation{JILA, University of Colorado/NIST, Boulder, CO, 80309, USA}

\author{Graeme Smith}
\email{graeme.smith@colorado.edu}
\affiliation{Department of Physics and Center for Theory of Quantum Matter, University of Colorado, Boulder CO 80309, USA}
\affiliation{JILA, University of Colorado/NIST, Boulder, CO, 80309, USA}

\date{\today}

\begin{abstract}
We explore ways to quantify multipartite correlations, in quantum information and in holography. We focus on optimized correlation measures, linear combinations of entropies minimized over all possible purifications of a state that satisfy monotonicity conditions. These contain far more information about correlations than entanglement entropy alone. We present a procedure to derive such quantities, and construct a menagerie of symmetric optimized  correlation measures on three parties.  These include tripartite generalizations of the entanglement of purification, the squashed entanglement, and the recently introduced $Q$-correlation and $R$-correlation. Some correlation measures vanish only on product states, and thus quantify both classical and quantum correlations;  others vanish on any separable state, capturing quantum correlations alone. We then use a procedure  motivated by the surface-state correspondence to construct holographic duals for the correlation measures as linear combinations of bulk surfaces.  The geometry of the surfaces can preserve, partially break, or fully break the symmetry of the correlation measure. The optimal purification is encoded in the locations of certain points, whose locations are fixed by constraints on the areas of combinations of surfaces. This gives a new concrete connection between information theoretic quantities evaluated on a boundary state and detailed geometric properties of its dual. 
\end{abstract}

\maketitle


\section{Introduction}
Finding ways to quantify the correlations between multiple parties in a quantum system is of natural importance. A fundamental concept is the entanglement entropy, which for a bipartite pure state captures all the essential information. However, for systems with three or more parties and/or in mixed states, other measures may quantify other aspects of the correlations.  Linear combinations of entropies satisfy inequalities placing them in a cone (see \cite{Pip03} for an introduction) and produce useful measures such as mutual informations and conditional mutual informations; however, more refined quantities are also possible. One such class is the {\it optimized information measures}, constructed by adding additional degrees of freedom to a quantum state and optimizing a linear combination of entropies over all such extensions; in all examples discussed here, these extensions will make the original (generally mixed) state into a pure state, and then optimizing over all such purifications. The entanglement of purification $E_P$ \cite{EP02} is one such optimized information measure.

Meanwhile, the relationship between entanglement and geometry has become a subject of great interest. One of the best arenas for such investigations has been the AdS/CFT correspondence \cite{maldacena1999}, which proposes that a theory of quantum gravity in a $D$-dimensional asymptotically anti-de Sitter (AdS) spacetime is exactly equivalent to a non-gravitational conformal field theory (CFT) living on the spacetime boundary, a duality known as holography. Moreover, the geometry of the bulk spacetime encodes the entanglement structure of the boundary theory, exemplified by the Ryu-Takayanagi (RT) formula \cite{RT} and its covariant generalization due to Hubeny, Rangamani and Takayanagi \cite{Hubeny:2007xt}, which states that the entanglement entropy of a region $A$ of the boundary CFT is proportional to the minimal area of a bulk surface which shares its boundary with $A$ and is homologous to $A$. The RT formula allows one to naturally associate the value of any linear entropic formula with a geometric object in the bulk. It emerges that the allowed entanglement entropies of multiparty systems that admit a classical geometric dual are more restricted than those for generic quantum states; for example, the tripartite information $I_3$, which is a linear entropic formula, cannot be positive in a holographic state\footnote{For us ``holographic state" is shorthand for ``state that admits a holographic dual well-described by classical gravity". Other states may well have holographic duals, but involving strong quantum gravitational effects.}, an inequality dubbed the monogamy of mutual information (MMI) \cite{Hayden:2011ag}, while in a generic quantum state it may have any sign.  In \cite{HEC15}, an infinite family of holographic entropy inequalities for arbitrarily large number of parties was obtained, and building on this the full five-party holographic entropy cone was subsequently worked out in \cite{cuenca2019holographic}, showing the richness of the entanglement structure of holographic states; for further work see for example \cite{Marolf:2017shp, Hubeny:2018trv, Hubeny:2018ijt, He:2019ttu, He:2020xuo, Bao:2020zgx}.

It is then natural to ask how more general information measures are encoded in geometric data. Such inquiry has the potential to illuminate both the subjects of quantum information and quantum gravity, and to further reveal how spacetime and geometry are encoded within information-theoretic concepts. A step in that direction was taken in \cite{TU18} and \cite{Swing18}, where it was conjectured that a geometric object called the {\it entanglement wedge cross-section} (EWCS) is dual to the entanglement of purification $E_P$ for bipartite systems in AdS/CFT, and arguments for this based on the matching of several inequalities between $E_P$ and the EWCS were presented. The EWCS is the minimal surface stretching across the entanglement wedge, the bulk region bounded by the RT surfaces of the boundary state. Several other proposals in addition to $E_P$ have been made as the boundary dual of the EWCS; in \cite{dutta2019canonical} it was argued, using ideas from \cite{engelhardt2019coarse,engelhardt2018decoding}, that the EWCS is dual to a quantity called the {\it reflected entropy}.  In \cite{kudler2019entanglement}, the {\it log negativity} \cite{vidal2002computable} was proposed as the EWCS dual, while in \cite{tamaoka2019entanglement}, a quantity called the {\it odd entanglement entropy} was argued for.  Indeed, investigations of the EWCS and its proposed boundary duals have been the subject of a large body of work \cite{bao2018holographic,hirai2018towards, espindola2018entanglement,bao2019conditional, bao2019entanglement,cheng2020optimized,ghodrati2019connection, chu2020generalizations,velni2019some,velni2020evolution,agon2019geometric,caputa2019holographic, kudler2019holographic,jokela2019notes,kudler2020quantum, kusuki2019dynamics,kusuki2019derivation,jeong2019reflected, kusuki2020entanglement}.  Many of these proposed dualities coincide with $E_P$ in the limit of classical geometry. The $E_P$ proposal and the entanglement wedge cross-section were generalized to multipartite systems in \cite{umemoto2018entanglement}, which also discussed the holographic dual of another optimized information measure, the squashed entanglement $E_{\rm sq}$ \cite{Tucci, Christandl}, obtaining that holographically this coincides with half the mutual information, though in generic quantum states this need not be true.

The entanglement of purification and the squashed entanglement are examples of optimized information measures that are also {\it monotonic}, meaning they cannot increase under local processing (or equivalently, under partial trace) by any of the parties; removing degrees of freedom cannot increase the measure. Such a monotonic optimized information measure was called an {\it optimized correlation measure}, or simply a correlation measure, in \cite{LS20}, where the bipartite cases were studied systematically. There it was found that in addition to $E_P$ and $E_{\rm sq}$, there are two more optimized correlation measures on two parties that treat the parties symmetrically, there called the $Q$-correlation $E_Q$ and the $R$-correlation $E_R$. (The mutual information $I(A:B)$ is also a monotonic measure on two parties, and thus may be considered a bipartite correlation measure, though it does not involve an optimization.) It was shown in \cite{LS20} that $E_Q$ and $E_R$ satisfy the same set of inequalities as $E_P$, suggesting further arguments for which if any are dual to the EWCS would be needed.

A program to systematically construct holographic duals for optimized correlation measures was undertaken in \cite{LSD20}. To do so, one needs an argument for what the set of purifications of a given state looks like in the holographic context. The idea employed comes from the surface-state correspondence \cite{Miyaji:2015yva}, a conjectured generalization of AdS/CFT which associates a pure state in a Hilbert space to any closed convex spacelike surface in the bulk geometry, not just on the boundary. Viewing the quantum multipartite state as described geometrically by subregions of the boundary, the set of geometric purifications is then all the convex closures of these regions. Moreover, the surface-state correspondence associates a natural generalization of the RT formula to the bulk regions, allowing the optimized correlation measures to be calculated. In the case of pure AdS$_3$ dual to CFT$_2$, geometric duals for all the bipartite correlation measures were obtained in \cite{LSD20}, associating both $E_P$ and $E_R$ with the EWCS, $E_{\rm sq}$ with half the mutual information as in \cite{umemoto2018entanglement}, and $E_Q$ with a novel combination of bulk surfaces; the same bulk dual for $E_Q$ was independently proposed and studied in \cite{umemoto2019quantum}. In all cases, the optimal purification involved closing the boundary regions with surfaces stretching along the boundary of the entanglement wedge. The location of the ``optimization points" dividing the ancilla degrees of freedom into a set associated with one region, and a set associated with the other region, depends on the measure, sometimes lying in the middle of the entanglement wedge boundary, and sometimes being driven to the boundary of spacetime.

One may then wonder about holographic realizations of correlation measures with more than two parties.  Steps in this direction were taken by \cite{balasubramanian2014multiboundary,bao2019entanglement}, where holographic multipartite correlations were studied in the context of multiboundary wormhole geometries. A reason to go beyond bipartite correlations in AdS/CFT is the observation by \cite{akers2020} that if the EWCS is dual to either $E_P$ or the reflected entropy, then holographic CFT states must have significant amounts of tripartite entanglement in the limit of classical geometry.  This is in sharp contrast to the conjecture of \cite{cui2019bit} that entanglement in holographic CFT states should be ``mostly bipartite", which was motivated by the bit threads formalism of \cite{freedman2017bit} as well as the random stabilizer tensor networks of \cite{hayden2016holographic} and \cite{nezami2016multipartite}.  The tension between the ``mostly bipartite" conjecture and the conjectured duality of the EWCS to either $E_P$ or the reflected entropy, forces us to consider the possibility that tripartite correlations in holographic CFT states may be highly quantum.  In order to better understand the nature of multipartite entanglement in AdS/CFT, the first step is to examine tripartite correlations in this setting.

In this paper, we continue the program of \cite{LS20} and \cite{LSD20} to the mutlipartite case, focusing on tripartite correlation measures. The first half of the paper involves no holography. There we discuss a systematic procedure for constructing optimized correlation measures that are symmetric between all parties, involving imposing both monotonicity and a boundedness condition that ensures the correlation measure is well-defined. After identifying generalizations of the bipartite correlation measures to any number of parties, we systematically construct a large class of symmetric tripartite correlation measures. We find fifteen tripartite correlation measures, two of which do not require optimization, while the other thirteen do. These fifteen are ``extremal" in the sense that they cannot be made from positive linear combinations of each other; an infinite set of other correlation measures can in principle be made from their linear combinations. We investigate these measures, illustrating their non-negativity by writing them in terms of manifestly positive quantities, and discuss how they reduce to bipartite measures in appropriate limits. We also discuss the kinds of quantum states on which different tripartite correlation measures vanish; while all correlation measures vanish on product states, those that vanish only on product states can be viewed as measuring all correlations, classical and quantum, while those that additionally vanish on separable states ignore classical correlations, and are thus measures of quantum entanglement.

In the remainder of the paper, we turn to obtaining proposals for holographic duals of these correlation measures in the AdS$_3$ geometry. Building on the bipartite cases, we take the optimal purification to lie along the boundary of the entanglement wedge while allowing the optimization points dividing the ancilla degrees of freedom to vary, and use the generalization of the RT formula suitable to the surface-state correspondence to calculate the relevant entropies and minimize the measure. Focusing on the case of three equal-sized, equally-spaced boundary regions, we show that the resulting holographic duals have a number of interesting properties. They may preserve, partially break, or totally break the geometric symmetries of the regions alone. The optimization points may sit symmetrically, may be driven to the spacetime boundary, or may be constrained at intermediate values; we find these intermediate locations are associated with nontrivial relations between the spacetime surfaces. The tripartite generalization of $E_P$ indeed is realized as the tripartite entanglement wedge cross-section as proposed in \cite{umemoto2018entanglement}, while other measures have jumps in the combination of surfaces they are realized by as the sizes of the boundary regions vary, and some can be quite intricate.

The structure of this paper is as follows.  In Sec.~II we give basic overview of symmetric optimized correlation measures and discuss a procedure to construct them systematically.  In Sec.~III we review the optimized bipartite correlation measures studied in \cite{LS20}.  In Sec.~IV we discuss the generalization of bipartite optimized correlation measures to multipartite optimized correlation measures, and the reduction of higher-party measures to lower-party ones.  In Sec.~V we derive fifteen extremal optimized tripartite correlation measures and discuss some of their properties, and in Sec.~VI we derive conditions under which the correlation measures vanish. In Sec.~VII we review the general method for assigning a holographic dual to an optimized correlation measure and the bipartite holographic results of \cite{LSD20}, which we then utilize in Sec.~VIII to construct duals for thirteen of the optimized tripartite correlation measures derived in Sec.~V.  We wrap things up in Sec.~IX with concluding remarks and some comments on future work.

\section{Optimized correlation measures}

Optimized information measures are functions of a multipartite quantum state calculated by taking some linear combination of entropies and optimizing it over purifications of the state and over partitions of the purifying system. For a bipartite state $\rho_{AB}$, the information measure $E_\alpha (\rho_{AB})$ can be written 
\begin{align}
E_\alpha(\rho_{AB}) = \inf_{\ket\psi_{ABab}}f^\alpha(\ket\psi_{ABab})
\end{align}
where $\ket\psi_{ABab}$ is a purification of $\rho_{AB}$ by adding ancilla degrees of freedom $a$ and $b$, and the optimization function $f^\alpha(\ket\psi_{ABab})$ is a linear combination of entropies involving either or both the parties $A$, $B$ and the ancilla $a$, $b$,
\begin{align}
    f^\alpha(\ket\psi_{ABab}) = \sum_{\emptyset\neq \mathcal{J} \subset \{A,B,a,b\}} \alpha_\mathcal{J} \, S(\mathcal{J}) \,.
\end{align}
Here $\alpha_\mathcal{J}$ are real coefficients and $\mathcal{J}$ runs over combinations of entropies, with the 7 independent possibilities $\mathcal{J} = A, B, AB, Aa, Ab, a, b$; since $\ket\psi_{ABab}$ is pure, other entropies are not independent of these due to relations like $S(ab) = S(AB)$. To have a well-defined optimized information measure, one must ensure the optimization function $f^\alpha(\ket\psi_{ABab})$ is bounded below so the optimization minimizes to a finite value.  The multipartite case on $n$ parties is defined analogously, with a set of parties $A_i$ and purifying ancilla $a_i$, $i = 1 \ldots n$, and the optimized information measure
\begin{align}
E_\alpha(\rho_{A_i}) = \inf_{\ket\psi_{A_i a_i }}f^\alpha(\ket\psi_{A_i, a_i})
\end{align}
given in terms of the optimization function $f^\alpha(\ket\psi_{A_i a_i})$
\begin{align}
    f^\alpha(\ket\psi_{A_i a_i}) = \sum_{\emptyset\neq \mathcal{J} \subset \{A_i a_i\}} \alpha_\mathcal{J} \, S(\mathcal{J}) \,,
\end{align}
where now there are $2^{2n-1} -1$ independent entropies $S(\mathcal{J})$. In what follows, we will abbreviate $E_\alpha(\rho_{A_i})$ as $E_\alpha(A_i)$, and $f^\alpha(\ket\psi_{A_i a_i})$ as $f^\alpha(A_i, a_i)$.

In \cite{LS20}, an optimized {\it correlation measure} of a multipartite state  was defined as an optimized information measure which in addition is monotonically decreasing under local processing, that is
\begin{eqnarray}
\label{Monotonicity}
	E_\alpha(A, B, \ldots) \leq E_\alpha(AA', B, \ldots) \,,
\end{eqnarray}
and similarly for $B$ and any other parties. 

First, we will review bipartite correlation measures and their holographic dual descriptions, and then obtain a set of tripartite correlation measures, most of them novel, and their holographic duals. We will discuss exclusively symmetric correlation measures that treat all the parties equally,
\begin{eqnarray}
	E_\alpha(A, B, \ldots) = E_\alpha(B, A, \ldots) \,,
\end{eqnarray}
and similarly for exchange of any two parties.
Our method is to construct the optimization functions $f^\alpha(A_i, a_i)$ by first assuming symmetry, and then to impose monotonicity and boundedness, as we now discuss.

\subsection{Monotonicity conditions}

We can obtain conditions on the entropies entering into the optimization function $f^\alpha(A_i, a_i)$ that enforce monotonicity as follows. For some correlation measure $E_\alpha(A_i)$, let the minimizing value of $E_\alpha(AA',B,C \ldots)$ be given by purifying ancilla $a, b, c \ldots$, so 
\begin{eqnarray}
	E_\alpha(AA',B,C \ldots) = f^\alpha(AA', B, C \ldots a, b, c\ldots) \,.
\end{eqnarray}
If we purify $\rho_{ABC\ldots}$ with ancilla $(A' a), b, c \ldots$ we get the same pure state. But that might not be the minimizing purification for $E_\alpha(A,B,C \ldots)$, so in general we have
\begin{eqnarray}
	E_\alpha(A, B, C, \ldots) \leq f^\alpha (A, B, C \ldots (A' a), b, c\ldots) \,.
\end{eqnarray}
Combining these two relations, we see that monotonicity (\ref{Monotonicity}) always holds for the correlation measure if its optimization function $f^\alpha$ satisfies
\begin{eqnarray}
\label{MonotonicityCond}
	f^\alpha(AA', B, C,\ldots a, b, c, \ldots) \geq f^\alpha(A, B, C, \ldots (A' a), b, c, \ldots) \,.
\end{eqnarray}
That is, if removing degrees of freedom from a party and giving it to the corresponding ancilla cannot increase the optimization function, then the correlation measure must be monotonic, since the result is a possible purification for the processed state, and thus an upper bound.

Thus an information measure $E_\alpha(A, B, \ldots)$ will be monotonic in $A$ if its objective function $f^\alpha(A_i, a_i)$ is composed of combinations of entropies $S(\{A_i\}, \{a_i\})$ satisfying (\ref{MonotonicityCond}). This occurs if each entropy or pair of entropies in $f^\alpha$ satisfies one of the following conditions:
\begin{enumerate}
	{\item The entropy in the term is either independent of $A$ and $a$ or contains both $Aa$.}
	{\item A pair of terms is of the form $S(AX) - S(aY)$, where $X$ and $Y$ are disjoint and contain neither $A$ nor $a$. }
\end{enumerate}
For condition 1, the inequality (\ref{MonotonicityCond}) is an equality; such terms are invariant under processing, and thus can appear in $f^\alpha$ with either sign. For terms of the form in condition 2, (\ref{MonotonicityCond}) is equivalent to
\begin{eqnarray}
	S(A'|A X) + S(A'|aY)  \geq 0\,.
\end{eqnarray}
Given $X$ and $Y$ are disjoint and do not contain $A$ or $a$, this holds by weak monotonicity. Such terms must appear in $f^\alpha$ with the given sign. We will ensure monotonicity by constructing the optimization function only out of entropy terms satisfying one of the two conditions. 

We will be interested in correlation measures which are symmetric in the parties. One way to implement monotonicity in a symmetric fashion is if $A$ is monotonic due to the appearance of ancilla $a$ in the optimization function $f^\alpha(A_i, a_i)$, there will be another ancilla $b$ playing the analogous role for $B$, and so on. This ``symmetric-ancilla" monotonicity is implemented by imposing a symmetry of $f^\alpha(A_i, a_i)$ under exchange of any party-ancilla pairs $(A_i, a_i)$ and $(A_j, a_j)$, and imposing monotonicity for any one party, for example $A$; in the language of \cite{LS20} this is ``012..." monotonicity. 

Another way to implement monotonicity is for the single ancilla $a$ to implement monotonicity on all parties $A$, $B \ldots$. In this ``universal ancilla" case, corresponding to ``000..." in the language of \cite{LS20}, we impose symmetry only under exchange of the $A_i$ in $f^\alpha$. The $A_i$ and $a$ need not be pure by themselves, so a purification involves a further ancilla $v$. In the holographic sections we will focus on the symmetric-ancilla implementation of monotonicity, but we will find correlation measures using the universal-aniclla case as well.\footnote{In principle one could consider a more general case encompassing both symmetric-ancilla and universal-ancilla as special cases, where the symmetric ansatz $A,B,C,\ldots a, b, c\ldots$ is not pure but is joined by an additional purifying ancilla $v$. We leave this more general case for future work.}

\subsection{Boundedness conditions}

Since our correlation measures involve taking an infimum, we need to guarantee that they cannot become arbitrarily negative as the purifications are varied. Consider the limit as the parties go to zero,
\begin{eqnarray}
	\lim_{A_i \to 0} E_\alpha (A_i) &=& \lim_{A_i \to 0} \inf_{\ket\psi_{A_i a_i}} f^\alpha(A_i, a_i)\\
	&=& \inf_{\ket\psi_{a_i}} f^\alpha(0, a_i) \,,\nonumber
\end{eqnarray}
where now $\ket\psi_{a_i}$ is a pure state involving the ancilla alone. In general $f^\alpha(0, a_i)$ will be some symmetric combination of the entropies of the ancilla,
\begin{eqnarray}
	f^\alpha(0, a_i) = x_1 \sum_i S(a_i) + x_2 \sum_{(ij)} S(a_i a_j) + \ldots \,.
\end{eqnarray}
The coefficients $x_1$, $x_2, \ldots$ then must take values such that the right-hand-side cannot become arbitrarily negative as the optimization samples over arbitrarily large pure states $\ket\psi_{a_i}$ with arbitrarily large entanglement between the $a_i$. 

Considering up to tripartite correlation measures as we do in this paper, the only relevant coefficient is $x_1 \equiv x$,
\begin{eqnarray}
	f^\alpha(0, a_i) = x \sum_i S(a_i)\quad\quad \quad (n \leq 3) \,,
\end{eqnarray}
since for up to $n=3$ parties any nonzero entropies of two or more ancilla equivalent to the entropies of one ancilla, given the purity of the state $\ket\psi_{a_i}$. Thus for us to achieve boundedness we simply require $x \geq 0$.

We can further show that once this is imposed, the correlation measures are non-negative. Terms with $x  =0$ contribute nothing to the correlation measure as $A_i \to 0$, while those with positive $x$ will be minimized by unentangled $a_i$, and we find
\begin{eqnarray}
	\lim_{A_i \to 0} E_\alpha(A_i) = 0\,.
\end{eqnarray}
Since the correlation measures are monotonic we have $ E_\alpha(A_i) \geq \lim_{A_i \to 0} E_\alpha(A_i)$, and thus indeed for properly bounded correlation measures,
\begin{eqnarray}
	E_\alpha(A_i) \geq 0 \,.
	\label{Positivity}
\end{eqnarray}
Thus we have developed a procedure for constructing symmetric optimized correlation measures: construct a trial $f^\alpha$ with an appropriate ansatz symmetric in the parties, require the terms to satisfy one of the two monotonicity constraints, and impose that $x \geq 0$ to ensure boundedness.

\section{Review of bipartite correlation measures}

As a review and warm-up, consider the case of two parties $E_\alpha(A, B)$. Including the purifying degrees of freedom $a$ and $b$, we end up with a seven-dimensional entropy cone,
\begin{eqnarray}
	S(A),\ S(B),\ S(a),\ S(b),\ S(AB) = S(ab),\ S(Aa) = S(Bb),\ S(Ab) = S(aB) \,.
\end{eqnarray}
We will first implement symmetry with the symmetric-ancilla ansatz by using entropy combinations invariant under $(A,a) \leftrightarrow (B,b)$, of which there are five: the first four entropies in the list above symmetrize into the two combinations $S(A) + S(B)$ and $S(a) + S(b)$, while the final three are already symmetric once their purity relationships are taken into account.

Now let us look for combinations of the five symmetrized entropies that are monotonic. $S(Aa)$ is monotonic by itself according to monotonicity condition 1, while the remaining entropies must be combined into combinations of the form $S(AX) - S(aY)$ to satisfy monotonicity condition 2. We find the five ``extremal" monotone combinations
\begin{eqnarray}
\setlength{\extrarowheight}{.05em}
	\begin{tabular}{cc}
	Monotone & $x$ \\ \hline
		$S(Aa)$ & $1$ \\\hline
		$S(A) + S(B) - S(AB) = I(A:B)$ & $0$ \\\hline 
		$S(A) + S(B) - S(Ba)$ & $-1$\\
		$S(Ab) - S(a)- S(b)$ & $-1$ \\ \hline 
		$S(AB) - S(a) - S(b) = - I(a:b)$ & $-2$ \\
	\end{tabular}
\end{eqnarray}
All other symmetric monotones can be built from linear combinations of these, with either sign coefficient for $S(Aa)$ and only conic combinations of the others.

Now let us find correlation measures by taking linear combinations of these quantities that are bounded below, while preserving monotonicity. In the table we  indicate the value of $x$ each monotone would contribute to a quantity $f^\alpha(A=0, B=0, a, b) \equiv x S(a)$, with $S(b) = S(a)$ by purity once we take $A$, $B$ to zero;\footnote{The normalization of $x$ differs from the general remarks in the previous section since we use $S(a)$ instead of $S(a) + S(b) = 2 S(a)$.} the optimization function will be bounded below if the net $x$ for all the terms is non-negative. We see the first two monotones are already bounded below, and thus are already correlation measures. The first optimization function gives $E_P$,
\begin{eqnarray}
	f^P(A,B,a,b) =  S(Aa)\,.
\end{eqnarray}
It is monotonic with either sign, but must have positive sign to be bounded below; in fact since there are no other bipartite monotones with $x > 0$, it will always only appear with positive coefficient in bipartite correlation measures. (As we will see, something more general happens in the tripartite case.) The second optimization function is the mutual information of $A$ and $B$, which is independent of the ancilla. Thus the optimization is trivial, and the correlation measure is also the mutual information,
\begin{eqnarray}
	\inf_{\ket\psi_{ABab}} I(A:B) = I(A:B) \,.
\end{eqnarray}
Thus following our method we recover non-optimized correlation measures as well.

The final three monotones are not bounded below, but they can be made into bounded correlation measures by adding a positive coefficient times $S(Aa)$, preserving monotonicity while shifting $x$ to 0. This gives us the remaining three correlation measures $E_Q$, $E_R$ and $E_{\rm sq}$ from optimizing
\begin{eqnarray}
	f^Q(A,B,a,b) &=&  {1 \over 2} (S(A)+S(B) + S(Aa) - S(Ba))\\
	f^R(A,B,a,b) &=&  {1 \over 2} (2 S(Aa) + S(AB) - S(a) - S(b)) \\
	f^{\rm sq}(A,B,a,b) &=& {1\over 2} (S(Aa) + S(Ba) - S(a) - S(b) ) \,,
\end{eqnarray}
where the normalizations are conventional. These quanities are summarized in table~\ref{Table1}.

We can show $f^Q$, $f^R$, and $f^{\rm sq}$ are manifestly non-negative by expressing them as conic combinations of mutual informations and conditional mutual informations; $f^P = S(Aa)$ is obviously already non-negative, but we can write it in a presentation similar to that for $f^Q$ and $f^R$. The expressions appear in table~\ref{Table1}. 
The quantity $I(Aa:B)$ appearing in the expressions for $f^P$, $f^Q$ and $f^R$ is monotonic and bounded below, but not symmetric between $A$ and $B$, as was observed in \cite{LS20}. Additionally, we can express $f^{\rm sq} = (I(Aa:B) - I(a:B))/2$ and $I(A:B) = I(Aa:B) - I(a:B|A)$, which implies that $E_P$, $E_Q$ and $E_R$ are all bounded below by both $I(A:B)/2$ and $E_{\rm sq}$. One can also see that
\begin{eqnarray}
	f^P(A,B,a,b) = f^R(A,B,a,b) + {1 \over 2} I(a:b) \,,
\end{eqnarray}
and thus we have $E_P \geq E_R$.

\begin{table}
\begin{center}	
\setlength{\extrarowheight}{.25em}
	\begin{tabular}{|c|c|c|}\hline
\begin{tabular}{c}Correlation\cr measure \end{tabular}& As entropies & As MIs and CMIs  \cr \hline\hline
$I(A:B)$ &   $S(A) + S(B) - S(AB)$  &$I(A:B)$  \cr\hline\hline
	$2f^P$ &  $S(Aa) + S(Bb)$ &$I(Aa:B) + I(Aa:b)$  \cr\hline
$2f^Q$	 &  $S(A)+S(B) + S(Aa) - S(Ba)$  & $I(Aa:B) + I(A:b)$\cr\hline
$2f^R$ &  $2 S(Aa) + S(AB) - S(a) - S(b)$ &$I(Aa:B) + I(A:b|a)$  \cr\hline
$2f^{\rm sq}$ &  $S(Aa) + S(Ba) - S(a) - S(b)$  &$I(A:B|a)$  \cr\hline\hline
$f^{v1}_2$ &  $S(ABa)$ &$I(ABa:v)/2$  \cr\hline
$f^{v2}_2$ &  $S(ABa) + S(AB) - S(a)$ &$I(AB:v)$  \cr\hline
$f^{v3}_2$ &  \begin{tabular}{c} $2S(ABa) + S(A) +S(B) $ \cr $ - S(Aa) - S(Ba)$\end{tabular} &$I(A :v) + I(B:v)$\cr\hline
	\end{tabular}	
\end{center}
\caption{Bipartite extremal correlation measures. The mutual information is independent of optimization. The next four come from the symmetric ancilla ansatz. The final three come from the universal ancilla ansatz and optimize to zero. \label{Table1}}
\end{table}

We can also consider imposing symmetry using the universal-ancilla ansatz. Now the ancilla are $a$ (implementing monotonicity for both $A$ and $B$) and $v$, and the symmetric entropies are $S(A)+ S(B)$, $S(AB)$, $S(Aa)+ S(Ba)$, $S(a)$ and $S(ABa)$. We find the monotones
\begin{eqnarray}
\setlength{\extrarowheight}{.05em}
	\begin{tabular}{cc}
	Monotone & $x$ \\ \hline
		$S(ABa)$ & $1$ \\
		$S(Aa) + S(Ba)- S(a)$ & $1$\\ \hline
		$S(A) + S(B) - S(AB) = I(A:B)$ & $0$ \\\hline 
		$S(AB) - S(a)$ & $-1$ \\ \hline 
		$S(A) + S(B) - S(Aa) - S(Ba)$ & $-2$ \\
	\end{tabular}
\end{eqnarray}
and the resulting extremal correlation measures $I(A:B)$, $S(Aa) + S(Ba) - S(a) - S(ABa)$,  $S(ABa)$, $S(ABa) + S(AB) - S(a)$ and $2 S(ABa) + S(A) + S(B) - S(Aa) - S(Ba)$. The first two are the mutual information and squashed entanglement again. One can see that for the other three, any purification where $ABa$ is pure alone gives zero, and since correlation measures are non-negative, these correlation measures are all trivial, optimizing to zero on all states \cite{LS20}. However, they are worth remarking on since conic combinations of extremal correlation measures are also correlation measures, and if one of these is added to a measure that is optimized by a different purification it would create something nontrivial. We call these measures $f^{v1}_2$, $f^{v2}_2$ and $f^{v3}_2$, the three Vanishing bipartite measures, and also include them in table~\ref{Table1}.

\section{Multipartite correlation measures}

Before we turn to a thorough study of the tripartite case, we make a few remarks about general multipartite correlation measures.

\subsection{Generalizing bipartite correlation measures to $n$ parties}

We can determine natural generalizations of the bipartite correlation measures to any number of parties. For the normalization of what follows we have not added overall fractional coefficients to these multipartite measures, unlike in the bipartite case, which will be convenient for our holographic presentations.

The $n$-party entanglement of purification $f_n^P$, as is well-known, is naturally just
\begin{eqnarray}
	f^P_n(A_i , a_i) &=&  \sum_i S(A_ia_i) \,,
\end{eqnarray}
We can also obtain generalizations of the $Q$-correlation and $R$-correlation. For two-party $f^Q$ we can write it more symmetrically between $a$ and $b$ as,
\begin{eqnarray}
	f^Q(A,B,a,b) &=& {1 \over 4} [ 2S(A)  +2S(B) + S(Aa) + S(Bb) -S(Ba) - S(Ab)] \\
	 &=& {1 \over 4} [ S(A) +S(B) + S(Aa) + S(Bb) -S(a|B) - S(b|A)]\,.\nonumber
\end{eqnarray}
and then a higher-party generalization is
\begin{eqnarray}
	f^Q_n(A_i , a_i) &=&  \sum_i [S(A_i) + S(A_ia_i) - S(a_i \Big|\prod_{j\neq i} A_j)]\\
	  &=&\sum_i [S(A_i) + S(A_ia_i) + S( \prod_{j\neq i} A_j) - S(a_i \prod_{j\neq i} A_j)]\,.\nonumber
\end{eqnarray}
Two-party $f^R$ can be written
\begin{eqnarray}
	f^R(A,B,a,b) &=& {1 \over 2} [ S(AB) +S(Aa) + S(Bb) -S(a)-S(b)]\\
	 &=& {1 \over 2} [ S(AB) +S(A|a) + S(B|b) ]\,.\nonumber
\end{eqnarray}
A higher-party generalization is then
\begin{eqnarray}
	f^R_n(A_i , a_i) &=&   S(\prod_i A_i) +\sum_i [S(A_ia_i) - S(a_i)]\\
	 &=& S(\prod_i A_i) +\sum_i S(A_i|a_i) \,.\nonumber
\end{eqnarray}
These generalizations of $E_P$, $E_Q$ and $E_R$ are all in the symmetric-ancilla ansatz.

A generalization of the squashed entanglement has appeared in the literature \cite{avis2008distributed} \cite{yang2009squashed}, which takes the universal-ancilla form
\begin{eqnarray}	
\label{SquashedFamily}
f^{\rm sq}_n(A_i, a) = K(A_1:A_2:\dots :A_n|a) \equiv \sum_{i=1}^n S(A_i| a) - S(\prod_i A_i|a) \,, 	
\end{eqnarray}
where $K(A_1:A_2:\dots :A_n|a)$ is the conditional total correlation, a multipartite generalization of the conditional mutual information. One may in general find multiple inequivalent ways to generalize a measure. For example, \cite{yang2009squashed} pointed out an alternate definition of a multipartite squashed entanglement,
\begin{eqnarray}	
\label{SquashedFamilyTwo}
{f'}^{\rm sq}_n(A_i, a) = J(A_1:A_2:\dots :A_n|a) \equiv \sum_{i=1}^n S(\prod_{j\neq i}A_j| a) - (n-1) S(\prod_i A_i|a) \,, 	
\end{eqnarray}
using the conditional dual total correlation $J(A_1:A_2:\dots :A_n|a)$, but it was shown in \cite{Davis:2018ydj} that when optimized, the definitions (\ref{SquashedFamily}) and (\ref{SquashedFamilyTwo}) produce the same correlation measure, due to the relation $J(A_1:A_2:\dots :A_n|a) = K(A_1:A_2:\dots :A_n|v)$ with $A_1A_2\ldots A_nav$ pure. We can also find a symmetric-ancilla family that also generalizes the squashed entanglement,
\begin{eqnarray}
\label{TildeSquashed}
	 \tilde{f}^{\rm sq}_n(A_i, a_i) &=& \sum_i  I(A_i:\prod_{j\neq i} A_j|a_i) \\
	 &=& \sum_i  [S(A_i a_i) + S(a_i \prod_{j\neq i} A_j) - S(a_i) -S(\prod_{j\neq i} a_j) ] \,.\nonumber
\end{eqnarray}
There is no reason to think that for some general $n$ we will exhaust all correlation measures just by these generalizations of the bipartite cases, and indeed for three parties we will find a host of other correlation measures as well.

\subsection{Decoupling of correlation measures}

It is natural to ask how higher-party correlation measures reduce to lower-party measures. We can show that when a party $A$ has no mutual information with the $(n-1)$ other parties $B_j$, the $n$-party correlation measure $E_\alpha(A, B_j)$ satisfying the symmetric-ancilla ansatz reduces to an $(n-1)$-party correlation measure $E_\alpha(0, B_j)$, even if $A$ is not zero.

The decoupling condition for $A$ is the vanishing mutual information
\begin{eqnarray}
	I(A : \prod_{j=1}^{n-1} B_j) =0 \,.
\end{eqnarray}
Monotonicity of mutual information, equivalent to strong subadditivity, then implies that $A$ has vanishing mutual information with any subset of the $B_j$, and the density matrix factorizes,
\begin{eqnarray}
	\rho_{A B_j} = \rho_A \otimes \rho_{B_j} \,.
\end{eqnarray}
One class of purifications of this density matrix is to
\begin{eqnarray}
	\ket\psi_{AaB_jb_j} = \ket\psi_{Aa} \otimes \ket\psi_{B_jb_j} \,,
\end{eqnarray}
for various $a$ and $b_j$ purifying $A$ and $B_j$ separately.
In these states the entropies decompose
\begin{eqnarray}
	S(A, B_j, a, b_j) \to S(A, 0, a, 0) + S(0,  B_j, 0, b_j) \,,
\end{eqnarray}
and so in this class of states 
\begin{eqnarray}
	f^\alpha(A, B_j, a, b_j) = f^\alpha(A, 0, a, 0) + f^\alpha(0,  B_j, 0, b_j) \,.
\end{eqnarray}
Since these states are acceptable purifications of $\rho_{A B_j}$ they will enter into the optimization for $E_\alpha(A, B_j)$, and thus give an upper bound on the correlation measure. We find
\begin{eqnarray}
\label{DecouplingSum}
	E_\alpha (A, B_j) \leq E_\alpha(A, 0) + E_\alpha(0, B_j) \,,
\end{eqnarray}
where $E_\alpha(A, 0)$ and $E_\alpha(0, B_j)$ are 
\begin{eqnarray}
	E_\alpha(A_i, 0) \equiv \inf_{\ket\psi_{Aa}} f^\alpha(A, 0, a, 0) \,, \quad \quad 
	E_\alpha(0, B_j) \equiv \inf_{\ket\psi_{B_jb_j}} f^\alpha(0, B_j, 0, b_j) \,, 
\end{eqnarray}
which are necessarily correlation measures on $1$ and $n-1$ parties respectively, since they inherit the monotonicity and boundedness properties from the original symmetric $n$-party correlation measure. However, there is no one-party correlation measure; for one party $A$ and its ancilla $a$ in a pure state $\ket\psi_{Aa}$, the only nonzero entropy is $S(A) = S(a)$, and so we would have to have $f^\alpha(A, a) \propto S(A)$, but $S(A)$ is not monotonic. (Monotonicity of $S(A)$ would be equivalent to positivity of conditional entropy, which does not hold in general.) Thus the one-party correlation measure must vanish,
\begin{eqnarray*}
	E_\alpha(A, 0) = 0\,,
\end{eqnarray*}
and hence (\ref{DecouplingSum}) becomes 
\begin{eqnarray}
	E_\alpha (A, B_j) \leq  E_\alpha(0, B_j) \,.
\end{eqnarray}
However monotonicity of $E_\alpha$ in $A$ immediately gives us the opposite,
\begin{eqnarray}
	E_\alpha (A, B_j) \geq  E_\alpha(0, B_j)   \,,
\end{eqnarray}
and thus we find when $I(A:B_1 B_2 \ldots ) =0$, we must indeed have
\begin{eqnarray}
	E_\alpha (A, B_j) =  E_\alpha(0, B_j) \,.
	\label{Reduction}
\end{eqnarray}

\section{Optimized tripartite correlation measures}

We now turn to our main focus, the case with three parties $A$, $B$ and $C$. For the symmetric-ancilla ansatz with purifying ancilla $abc$ we have six regions, and thus 31 independent entropies. However, imposing the complete permutation symmetry under the exchange of any pairs $(A,a)$, $(B,b)$ and $(C,c)$, we are left with nine symmetric combinations:
\begin{eqnarray}
{\cal S}_{Aa}  &=&	S(Aa) + S(Bb)+ S(Cc) \nonumber\\
{\cal S}_{A}  &=&S(A) + S(B) + S(C)\nonumber \\
{\cal S}_{AB}  &=& S(AB) + S(AC) + S(BC)\nonumber \\
{\cal S}_{ABC}  &=& S(ABC)\nonumber\\
\label{SymEntopies}
{\cal S}_{ABc}  &=& S(ABc) + S(ACb) + S(BCa) \\
{\cal S}_{Ab}  &=& S(Ab) + S(Ba) + S(Ac) + S(Ca) + S(Bc) + S(Cb)\nonumber \\
{\cal S}_{ABa}  &=& S(ABa) + S(ABb) + S(ACa) + S(ACc) + S(BCb) + S(BCc)\nonumber \\
{\cal S}_{ab}  &=& S(ab) + S(ac) + S(bc)\nonumber \\
{\cal S}_{a}  &=& S(a) + S(b) + S(c) \,.\nonumber
\end{eqnarray}
The first combination is just the tripartite entanglement of purification, and as in the bipartite case it is monotonic and bounded below by itself. The remaining symmetrized entropies are not monotonic on their own, but can be combined into twelve more extremal monotones, resulting in the list:
\begin{eqnarray}
\setlength{\extrarowheight}{.05em}
\begin{tabular}{cc}
	Monotone  & $x$ \\ \hline
		${\cal S}_{Aa}$   & $1$ \\ 
		${\cal S}_{ABa} - {\cal S}_{ab}$ & $1$\\ \hline
				$2 {\cal S}_A - {\cal S}_{AB} $   & 0 \\ 
		${\cal S}_{AB} - 2 {\cal S}_{ABC}$  & $0$ \\ 
		$2 {\cal S}_A + {\cal S}_{ABa} - {\cal S}_{Ab}$  &  $0$\\ \hline
		${\cal S}_A + {\cal S}_{AB} - {\cal S}_{ABc}$ & $-1$ \\ 
	${\cal S}_{ab} - 2 {\cal S}_a$  & $-1$ \\ 
	$2 {\cal S}_{ABC} - {\cal S}_{ab}$ & $-1$ \\ 
	${\cal S}_{ABc} - {\cal S}_{ab} - {\cal S}_a$  & $-1$ \\ 
	${\cal S}_A + {\cal S}_{Ab} - {\cal S}_{ABc} - 2 {\cal S}_a$ & $-1$ \\ \hline
		$2 {\cal S}_A + {\cal S}_{ABc}- {\cal S}_{Ab}- {\cal S}_a $   & $-2$ \\  
	${\cal S}_{Ab} - {\cal S}_{ABa} - 2{\cal S}_a$ & $-2$ \\ 
	 	${\cal S}_{AB} - {\cal S}_{ABa}$  & $-2$ \\
	\end{tabular}
\end{eqnarray}
Other monotones can be generated by taking linear combinations of these; ${\cal S}_{Aa} \equiv S(Aa) + S(Bb) + S(Cc)$ satisfies monotonicity condition 1 and hence remains monotonic with either sign, while the remaining extremal monotones satisfy monotonicity condition 2 and thus must have positive coefficients to remain monotonic. In the list we also include $x$, which as discussed previously is the coefficient of $S(a) + S(b) +S(c)$ when we take $A, B, C \to 0$ in each monotone. To find monotones that are bounded below, and hence are correlation measures, we must take monotonic linear combinations such that the net $x$ coefficient is nonnegative.

Because there are now two extremal monotones with $x = 1$, and one of them can appear with either sign and remain monotonic, we can create a bounded monotone that includes ${\cal S}_{Aa}$ with negative coefficient, namely ${\cal S}_{ABa} - {\cal S}_{ab} - {\cal S}_{Aa}$; this and ${\cal S}_{Aa}$ are extremal correlation measures. The third through fifth monotones are already bounded below and are also extremal correlation measures. The remaining monotones can be made bounded below by adding a coefficient times ${\cal S}_{Aa}$. (We could also make them bounded below by adding ${\cal S}_{ABa} - {\cal S}_{ab}$, but the result is not extremal, since it can be obtained from the former case by addition of the monotone ${\cal S}_{ABa} - {\cal S}_{ab} - {\cal S}_{Aa}$.) In the end we obtain a list of thirteen extremal tripartite correlation measures in the symmetric-ancilla ansatz, listed in table~\ref{Table2} and table~\ref{Table3}.

\begin{table}
\begin{center}
\setlength{\extrarowheight}{.25em}
\begin{tabular}{|c|c|c|c|c|}\hline
\begin{tabular}{c}Correlation\cr measure \end{tabular}&\begin{tabular}{c}As symmetrized\cr entropies \end{tabular}& As entropies & As MIs and CMIs & \begin{tabular}{c} Bipartite \cr reduction\end{tabular} \cr \hline\hline
	$\sum I(A_i:A_j) $  &$2 {\cal S}_A - {\cal S}_{AB} $ & \begin{tabular}{c} $2S(A) + 2S(B) + 2S(C)$ \cr $-S(AB) - S(AC) - S(BC)$ \end{tabular} &\begin{tabular}{c}$I(A:B) + I(A:C)$ \cr $ + I(B:C)$ \end{tabular}&  $I(A:B)$\cr\hline
	$J(A:B:C)$  & ${\cal S}_{AB} - 2 {\cal S}_{ABC}$& \begin{tabular}{c}$S(AB) + S(AC) + S(BC)$ \cr $- 2 S(ABC)$ \end{tabular}& $ I(AB:C) + I(A:B|C)$ & $I(A:B)$ \cr\hline \hline
   $f_{c_1} = f^P_3$ &${\cal S}_{Aa}$& $S(Aa) + S(Bb)+ S(Cc)$&\begin{tabular}{c}$I(Aa:Bb) + I(Aa:Cc)$ \cr $ + I(Bb:Cc)$\end{tabular} &  $2 f^P$\cr\hline
\begin{tabular}{c}$f_{c_2}$  \cr (part  of $f^R_3$) \end{tabular}	 & \begin{tabular}{c}${\cal S}_{Aa}  +  2 {\cal S}_{ABC}$ \cr $ - {\cal S}_{ab}  $ \end{tabular}&  \begin{tabular}{c} $S(Aa) + S(Bb) + S(Cc)$\cr $ + 2S(ABC)$ \cr $ - S(ab) - S(ac) - S(bc)$ \end{tabular} &\begin{tabular}{c}$I(a:BC|A) + I(b:AC|B)$ \cr $ + I(c:AB|C)$ \cr $+ I(A:B) + I(AB:C)$ \end{tabular}  & $2 f^R$ \cr\hline
\begin{tabular}{c}$f_{c_3}$  \cr (part  of $f^R_3$) \end{tabular} &\begin{tabular}{c}${\cal S}_{Aa}  + {\cal S}_{ab}$\cr $ - 2 {\cal S}_a  $ \end{tabular}&  \begin{tabular}{c} $S(Aa) + S(Bb) + S(Cc)$ \cr $+ S(ab) + S(ac) + S(bc) $ \cr $-2S(a) - 2 S(b) - 2 S(c)$\end{tabular} &\begin{tabular}{c}$I(a:B|b) + I(a:C|c)$ \cr $ + I(b:A|a)+ I(b:C|c)$ \cr $ + I(c:A|a) + I(c:B|b)$\cr $+ I(A:B|ab) + I(A:C|ac)$ \cr $ + I(B:C|bc) $ \end{tabular} & $2 f^R$ \cr\hline \hline
$f_{s_1} = f^Q_3$ &\begin{tabular}{c}${\cal S}_{Aa}  + {\cal S}_A $ \cr $ + {\cal S}_{AB}- {\cal S}_{ABc}$ \end{tabular}& \begin{tabular}{c} $S(Aa) + S(Bb) + S(Cc)$ \cr $+ S(A)  + S(B) + S(C) $ \cr $+ S(AB) + S(AC) + S(BC)$ \cr $-S(ABc) - S(ACb) - S(BCa)$ \end{tabular}  &\begin{tabular}{c}$I(AB:c) + I(AC:b)$ \cr $ + I(BC:a)+ I(A:BCbc)$ \cr $ + I(B:ACac) + I(C:ABab)$ \end{tabular} &  $4 f^Q$ \cr\hline
  $f_{s_2}$	 & \begin{tabular}{c}$2{\cal S}_{Aa}  + {\cal S}_{AB}$ \cr $ - {\cal S}_{ABa}$ \end{tabular}& \begin{tabular}{c} $2S(Aa) + 2S(Bb)+ 2S(Cc)$ \cr $+S(AB) + S(AC) + S(BC)$ \cr $-S(ABa) - S(ABb) - S(ACa)$ \cr $ - S(ACc) - S(BCb) - S(BCc)$ \end{tabular} &\begin{tabular}{c}$I(A:b|B) + I(A:Cc) $ \cr $+ I(B:c|C) + I(B:Aa)$ \cr $ + I(C:a|A) + I(C:Bb)$ \end{tabular} &  $ 2 f^R$ \cr\hline
	\end{tabular}
	\end{center}
	\caption{Tripartite correlation measures in the symmetric-ancilla ansatz, part 1.   \label{Table2}}
	\end{table}

Two of the correlation measures are entirely independent of the ancilla, and thus independent of the optimization, analogous to the mutual information in the bipartite case; one of these is the sum of the bipartite mutual informations,
\begin{eqnarray}
	2 {\cal S}_A - {\cal S}_{AB} = I(A:B) + I(A:C) + I(B:C) \,,
\end{eqnarray}
and the other is a tripartite monotone called the dual total correlation \cite{Han75},
\begin{eqnarray}
	J(A:B:C) \equiv {\cal S}_{AB} - 2 {\cal S}_{ABC} =  S(AB) + S(AC) + S(BC) - 2 S(ABC) \,.
\end{eqnarray}
We note that half the sum of these two is another monotone appearing in the literature, the total correlation \cite{Watanabe60},
\begin{eqnarray}
K(A:B:C) \equiv {\cal S}_{A} - {\cal S}_{ABC} = S(A) + S(B) + S(C) - S(ABC)	\,,
\end{eqnarray}
where all three of these quantities can be thought of as valid ways to generalize the ordinary mutual information to three parties, preserving its monotonicity property. Another generalization of the mutual information, the tripartite information $I_3$, can be expressed as the difference $I_3(A:B:C) = J(A:B:C) - K(A:B:C)$; while $I_3$ is a difference of monotones, it is not monotonic itself.

For the remaining eleven correlation measures, we have adopted a naming convention that anticipates the geometric presentation of the holographic dual descriptions. The names are
\begin{itemize}
	{\item $f_{c_1}$,  $f_{c_2}$, $f_{c_3}$: entanglement wedge Cross-section}
	{\item $f_{s_1}$, $f_{s_2}$: Symmetric presentation}
	{\item $f_{a_1}$,  $f_{a_2}$: Axially symmetric presentation}
	{\item $f_{r_1}$: Rotationally symmetric presentation}
	{\item $f_{b_1}$,  $f_{b_2}$, $f_{b_3}$: optimization points at the Boundary}
\end{itemize}
and we shall discuss the meaning behind these names in section~\ref{HoloTripartiteSec}.

\begin{table}
\begin{center}	
\setlength{\extrarowheight}{.25em}
	\begin{tabular}{|c|c|c|c|c|}\hline
\begin{tabular}{c}Correlation\cr measure \end{tabular}&\begin{tabular}{c}As symmetrized\cr entropies \end{tabular}& As entropies & As MIs and CMIs &  \begin{tabular}{c} Bipartite \cr reduction\end{tabular} \cr \hline\hline
	$f_{a_1}$  & \begin{tabular}{c}$2 {\cal S}_A + {\cal S}_{ABa}$ \cr $ - {\cal S}_{Ab}$\end{tabular} &\begin{tabular}{c} $2S(A) + 2S(B) + 2S(C)$ \cr $+S(ABa) + S(ABb) + S(ACa)$\cr $ + S(ACc) + S(BCb) + S(BCc)$\cr $-S(Ab) - S(Ba) - S(Ac) $\cr $- S(Ca) - S(Bc) - S(Cb)$ \end{tabular} &\begin{tabular}{c}$I(A : Cbc) + I(A: Bbc)$ \cr $ + I(B:Cac)+ I(B:Aac)$\cr $ + I(C:Bab) + I(C : Aab)$ \end{tabular} &   $2f^Q$\cr\hline
$f_{a_2}$	 &\begin{tabular}{c} $2{\cal S}_{Aa}  + 2 {\cal S}_A $\cr $+ {\cal S}_{ABc}$ \cr $- {\cal S}_{Ab}- {\cal S}_a   $\end{tabular} & \begin{tabular}{c} $2S(Aa) + 2S(Bb) + 2S(Cc)$ \cr $+ 2 S(A) + 2 S(B) + 2 S(C)$\cr $+S(ABc) + S(ACb) + S(BCa)$ \cr $-S(Ab) - S(Ba) - S(Ac) $\cr $- S(Ca) - S(Bc) - S(Cb)$\cr $-S(a) - S(b)- S(c)$\end{tabular}  &\begin{tabular}{c}$I(A:Bc|a) + I(A:Cb|a)$ \cr $ + I(B:Ac|b)+ I(B:Ca|b)$ \cr $ + I(C:Ab|c) + I(C:Ba|c)$ \cr $+ I(A:B) + I(A:C)$ \cr $ + I(B:C)+ I(AB:c) $ \cr $+ I(AC:b) + I(BC:a)$ \end{tabular} &  \begin{tabular}{c}$4 f^R+$ \cr $  I(A:B)$ \end{tabular}\cr\hline\hline
$f_{r_1}$	 &\begin{tabular}{c}${\cal S}_{Aa} $ \cr $ + {\cal S}_A + {\cal S}_{Ab}$ \cr $ - {\cal S}_{ABc} - 2 {\cal S}_a  $ \end{tabular}& \begin{tabular}{c} $S(Aa) + S(Bb) + S(Cc)$ \cr $+S(A) + S(B) + S(C)$ \cr $+S(Ab) + S(Ba) + S(Ac)$\cr $ + S(Ca) + S(Bc) + S(Cb)$ \cr $-S(ABc) - S(ACb) - S(BCa)$ \cr $-2S(a) - 2 S(b) - 2 S(c)$\end{tabular}  &\begin{tabular}{c}$I(A:B|c) + I(A:C|b)$ \cr $ + I(B:C|a)+ I(A:BCbc)$ \cr $ + I(B:ACac) + I(C:ABab)$ \end{tabular} & \begin{tabular}{c}$2 f^R +$ \cr $ 2I(A:B)$ \end{tabular} \cr\hline \hline
$f_{b_1} = \tilde{f}^{\rm sq}_3$	 & \begin{tabular}{c}${\cal S}_{Aa}  + {\cal S}_{ABc}$\cr $ - {\cal S}_{ab} - {\cal S}_a  $ \end{tabular}& \begin{tabular}{c} $S(Aa) + S(Bb) + S(Cc)$ \cr $+S(ABc) + S(ACb) + S(BCa)$\cr $- S(ab) - S(ac) - S(bc)$ \cr $- S(a) - S(b) - S(c)$ \end{tabular}  &\begin{tabular}{c}$I(A:BC|a) + I(B:AC|b)$\cr $ + I(C:AB|c)$\end{tabular}  &$4 f^{\rm sq}$\cr\hline
 $f_{b_2}$   & \begin{tabular}{c}${\cal S}_{ABa}$ \cr $ - {\cal S}_{ab}-{\cal S}_{Aa}$\end{tabular} & \begin{tabular}{c}$S(ABa) + S(ABb) + S(ACa)$ \cr $ + S(ACc) + S(BCb) + S(BCc)$\cr $ - S(ab) - S(ac) - S(bc)$ \cr $ - S(Aa) - S(Bb)- S(Cc)$ \end{tabular} &\begin{tabular}{c}$I(B:C|Aa) + I(C:A|Bb)$ \cr $ + I(A:B|Cc)$\end{tabular}&  $ I(A:B)$ \cr\hline
 $f_{b_3}$ &\begin{tabular}{c}$2{\cal S}_{Aa}  + {\cal S}_{Ab}$\cr $ - {\cal S}_{ABa} - 2{\cal S}_a $\end{tabular} & \begin{tabular}{c} $2S(Aa) + 2S(Bb)+ 2S(Cc)$ \cr $+S(Ab) + S(Ba) + S(Ac)$ \cr $+S(Ca) + S(Bc) + S(Cb)$ \cr $-S(ABa) - S(ABb) - S(ACa)$ \cr $ - S(ACc) - S(BCb) - S(BCc)$ \cr $-2 S(a)- 2 S(b) - 2 S(c)$\end{tabular}&\begin{tabular}{c}$I(A:B|b) + I(A:C|c)$ \cr  $+ I(B:A|a)+ I(B:C|c)$ \cr $ + I(C:A|a)+ I(C:B|b)$ \end{tabular}  & $4 f^{\rm sq}$ \cr\hline
	\end{tabular}	
\end{center}
\caption{Tripartite correlation measures in the symmetric-ancilla ansatz, part 2.  \label{Table3}}
\end{table}

The tripartite measures  $f^P_3$, $f^Q_3$ and $f^R_3$, as defined in the previous section, can all be identified in terms of these correlation measures. The tripartite entanglement of purification $f^P_3$ and the tripartite $Q$-correlation $f^Q_3$ are both extremal measures,
\begin{eqnarray}
	f^P_3 &=& f_{c_1} = {\cal S}_{Aa} \,, \\
	f^Q_3 &=& f_{s_1} = {\cal S}_{Aa}  + {\cal S}_A  + {\cal S}_{AB}- {\cal S}_{ABc} \,, 
\end{eqnarray}
while the tripartite $R$-correlation $f^R_3$ is not extremal, but is a linear combination with unit coefficients of two of the extremal measures we find,
\begin{eqnarray}
	f^R_3 = f_{c_2} + f_{c_3} = 2 {\cal S}_{Aa}  +  2 {\cal S}_{ABC} - 2 {\cal S}_{a}    \,.
\end{eqnarray}
The symmetric-ancilla generalization of the squashed entanglement $\tilde{f}^{\rm sq}_3$ we defined in (\ref{TildeSquashed}) is also an extremal correlation measure,
\begin{eqnarray}
	\tilde{f}^{\rm sq}_3 = f_{b_1} = {\cal S}_{Aa}  + {\cal S}_{ABc} - {\cal S}_{ab} - {\cal S}_a  \,.
\end{eqnarray}
The traditional multipartite squashed entanglement $f^{\rm sq}_3$ we will find to be a universal-ancilla correlation measure.

In the tables we present each correlation measure in terms of symmetrized entropies (\ref{SymEntopies}), as well as in terms of the basic entropies themselves. It is also possible to express every correlation measure solely in terms of a conic combination of mutual informations and conditional mutual informations, which makes it manifest that they are non-negative (\ref{Positivity}), and we do this in the tables as well. It is possible that some of the correlation measures have shorter or simpler manifestly positive presentations than the ones given.

Finally, we can also consider the bipartite reduction of each of these measures. As shown previously (\ref{Reduction}), if one region $C$ shares no mutual information with $AB$ each tripartite measure must reduce to a bipartite measure for $A$ and $B$. Which bipartite measure results from the reduction of each tripartite measure is given in the last column of the tables. We see most tripartite measures reduce up to factor to a single bipartite measure, with $f^P_3$, $f^Q_3$, $f^R_3$ and $\tilde{f}^{\rm sq}_3$ all reducing to their bipartite counterparts; there are also two measures that reduce to a linear combination of bipartite $f^R$ and $I(A:B)$.

\begin{table}
\begin{center}	
\setlength{\extrarowheight}{.25em}
	\begin{tabular}{|c|c|c|c|}\hline
\begin{tabular}{c}Correlation\cr measure \end{tabular}& As entropies & As MIs and CMIs &  \begin{tabular}{c} Bipartite \cr reduction\end{tabular} \cr \hline\hline
	$f_{u1} =f^{\rm sq}_3$ &  \begin{tabular}{c} $S(Aa) + S(Ba) + S(Ca)$ \cr $-2S(a) - S(ABCa)$\end{tabular} &$I(A : B|a) + I(AB:C |a)$  &   $2f^{\rm sq}$\cr\hline
$f_{u2}$	 &  \begin{tabular}{c} $2S(ABa) + 2S(ACa) + 2S(BCa)$ \cr $-  S(Aa) - S(Ba) - S(Ca)$\cr $-3S(ABCa)$ \end{tabular}  & \begin{tabular}{c} $I(A:B|Ca) + I(A:C|Ba)$ \cr $+ I(B:C|Aa)$ \end{tabular}&   $2f^{\rm sq}$\cr\hline\hline
$f^{v1}_3$ &  $S(ABCa)$ &$I(ABCa:v)/2$  &   $f^{v1}_2$\cr\hline
$f^{v2}_3$ &  $S(ABCa) + S(ABC) - S(a)$  &$I(ABC:v)$  &   $f^{v2}_2$\cr\hline
$f^{v3}_3$ &  \begin{tabular}{c} $3S(ABCa) + S(A) +S(B) + S(C)$ \cr $-S(ABa) - S(ACa) - S(BCa)$\end{tabular} &\begin{tabular}{c}$I(A :v) + I(B:v)$ \cr $+I(C:v)$ \end{tabular}&   $f^{v3}_2$\cr\hline
	\end{tabular}	
\end{center}
\caption{Tripartite correlation measures in the universal-ancilla ansatz.  \label{Table4}}
\end{table}

Having discussed the symmmetric-ancilla correlation measures, we can also examine the universal-ancilla ansatz. Here we have only two ancilla, $a$ enforcing monotonicity on all parties, and the remainder $v$. The symmetric combinations are ${\cal S}_A$, ${\cal S}_{AB}$, ${\cal S}_{ABC}$ as well as $S(ABCa)$, $S(Aa) + S(Ba) + S(Ca)$, $S(ABa) + S(ACa) + S(BCa)$ and $S(a)$. We find seven extremal monotones, and construct seven correlation measures. Two of these are again $I(A:B) + I(A:C) + I(B:C)$ and the dual total correlation $J(A:B:C)$, and three more optimize to zero; we call these $f^{v1}_3$, $f^{v2}_3$ and $f^{v3}_3$ as they precisely generalize the analogous measures in the bipartite case.

Finally, two measures are new, which we call $f_{u1}$ and $f_{u2}$. The first is the tripartite squashed entanglement $f^{\rm sq}_3$ (\ref{SquashedFamily}) that appears in the literature. One can also see that the alternate form of (\ref{SquashedFamilyTwo}) can be expressed as ${f'}^{\rm sq}_3 = (f_{u1}+f_{u2})/2$. Both can be written in a manifestly positive form as the sum of conditional mutual informations, and both reduce upon removing the region $C$ to the bipartite squashed entanglement. We list the universal-ancilla tripartite correlation measures in table~\ref{Table4}.

\section{Conditions for Vanishing}

A good way to understand the types of correlation quantified by a correlation measure is to determine the set of states on which the measure vanishes. All correlation measures vanish on product states. If a correlation measure vanishes {\it only} on product states, then it being nonzero captures all correlations, classical and quantum.  On the other hand, if a correlation measure vanishes on all classical states, then it being nonzero indicates the presence of entanglement.  This gives us an information theoretic way of classifying our correlation measures, as opposed to the geometric classification developed in the following sections.

 We find that six of our  extremal measures, the tripartite entanglement of purification $E_P^3$, the tripartite $Q$-correlation $E_Q^3$, the non-optimized dual total correlation $J(A:B:C)$, as well as 
$E_{c_2}$, $E_{r_1}$, and $E_{a_1}$ vanish only on product states, and so capture both classical and quantum correlations.  The non-extremal tripartite $R$-correlation $E^R_3$ also vanishes only on product states. Meanwhile, $E_{b_1}$, $E_{b_2}$, $E_{b_3}$, and $E_{c_3}$ vanish on separable states as well, so they are best thought of as tripartite entanglement measures.  There are two measures, $E_{s_2}$ and $E_{a_2}$, for which we do not yet have conditions for vanishing, and the condition for vanishing of the sum of mutual informations $I(A:B) + I(A:C) + I(B:C)$ is simply that each mutual information is zero.

\subsection{Vanishing only on product states}

Here we show some of our correlation measures {\it only} vanish on product states $\rho_{ABC} = \rho_{A}\otimes \rho_B\otimes \rho_C$, and thus encode both classical and quantum correlations.  Since $\rho_{ABC}$ is a product state if and only if
\begin{align}
 \label{prod}
    I(AB:C) = I(AC:B) = I(BC:A) = 0 \,,
\end{align}
it is equivalent to show that a measure vanishes if and only if (\ref{prod}) is satisfied. For some measures this can be seen directly from its expression as mutual informations and conditional mutual informations, as tabulated in Table~\ref{Table2} and Table~\ref{Table3}.

For example, we see from Table~\ref{Table2} that the dual total correlation $J(A:B:C) = I(AB:C) + I(A:B|C)$, and if $J(A:B:C)$ vanishes, both terms must individually vanish  since they are nonnegative, giving $I(AB:C) =0$. But we can equally well decompose the symmetric $J(A:B:C)$ in terms of $I(AC:B)$ and $I(BC:A)$, showing all three must vanish if $J(A:B:C)$ does, and hence the state is a product. An identical argument holds for $E_{c_2}$.

For the tripartite $Q$-correlation  $E_Q^3 = E_{s_1}$ and for $E_{r_1}$ it is $I(A:BCbc)$ and permutations that appear in the expressions as mutual informations and conditional mutual informations, but their vanishing implies that of $I(A:BC)$ and permutations by monotonicity of mutual information, and again the state must be a product. 

Analogous arguments show the the bipartite $E_P$, $E_Q$ and $E_R$ vanish only on product states. These arguments can also be used for any linear combination of measures vanishing only on product states (even if other members of the combination are not in this list), and hence the tripartite $R$-correlation, involving a sum of $E_{c_2}$ and $E_{c_3}$, also vanishes only on product states.

A different argument works for  the tripartite entanglement of purification $E^3_P = E_{c_1}$. If $E^3_P = 0$ then there is a pure state that satisfies
\begin{align}
    I(Aa:Bb) = I(Aa:Cc) = I(Bb:Cc) = 0\,. \label{EP3}
\end{align}
$I(Aa:Bb)=0$ implies that we can decompose $|\psi\rangle_{AaBbCc}$ as
\begin{align}
    |\psi\rangle_{AaBbCc} = |\psi\rangle_{AaC_1} \otimes |\psi\rangle_{BbC_2} \,,
\end{align}
where $C_1C_2 = Cc$. Furthermore (\ref{EP3}) also requires
\begin{align}
|\psi\rangle_{AaC_1} = |\psi\rangle_{Aa}\otimes |\psi\rangle_{C_1}  
\,, \quad \quad 
|\psi\rangle_{BbC_2} = |\psi\rangle_{Bb}\otimes |\psi\rangle_{C_2}\,,
\end{align}
to impose vanishing $I(Aa:Cc)$ and $I(Bb:Cc)$.  As a result, we see that $|\psi\rangle_{AaBbCc} = |\psi\rangle_{Aa}\otimes |\psi\rangle_{Bb} \otimes |\psi\rangle_{Cc}$. Tracing out $a,b,c$ shows that this is a product state, $\rho_{ABC} = \rho_A\otimes \rho_B\otimes\rho_C$.

We can also show that $E_{a_1}$ vanishes only on product states.  If $E_{a_1}=0$ then there is a pure state that satisfies
\begin{align}
    I(A:Cbc) = I(A:Bbc) = I(B:Cac) = I(B:Aac) = I(C:Bab) = I(C:Aab) = 0\,.
\end{align}
Since $I(A:Cbc) = 0$, we can write the global state as $|\psi\rangle_{AaBbCc} = |\psi\rangle_{AE_1}\otimes |\psi\rangle_{CbcE_2}$ where $E_1E_2 = aB$.  This means that $aB$ contains the purification of $A$, but $I(A:Bbc) = 0$ implies that $I(A:B)=0$, so that the purification of $A$ is found entirely in $a$.  That is, $E_1$ is a subset of $a$ alone.  Similarly, since $f_{a_1}$ is symmetric in ABC, we can conclude that the purification of $B$ lies in $b$, and the purification of $C$ lies in $c$.  Thus, we have 
\begin{align}
    |\psi\rangle_{AaBbCc} = |\psi\rangle_{Aa_1}\otimes|\psi\rangle_{Bb_1}\otimes|\psi\rangle_{Cc_1}\otimes |\psi\rangle_{a_2b_2c_2},
\end{align}
where $a = a_1a_2$, $b = b_1b_2$, $c = c_1c_2$, and again tracing out the ancilla leaves us with a product state.

\subsection{Vanishing on separable states}

The correlation measures $E_{b_1}$, $E_{b_2}$, $E_{b_3}$, and $E_{c_3}$ all vanish on separable states, i.e. states of the form
\begin{align}\label{sep}
    \rho_{ABC} = \sum_{i}p_i \rho_A^i \otimes \rho_B^i \otimes \rho_C^i \,,
\end{align}
and hence are nonzero only when quantum entanglement is present.
To show this, we will show that these measures all vanish on classical states, i.e. states of the form
\begin{align}\label{classical}
\rho_{ABC} = \sum_{i,j,k}p_{ijk}\dyad{i}{i}_A \otimes \dyad{j}{j}_B \otimes \dyad{k}{k}_C.
\end{align}
Classical states form a subset of the separable states, which generates all separable states via local processing.  To see this, note that any state of the form (\ref{sep}) can be extended to the classical state
\begin{align}
\sum_ip_i\dyad{i}{i}_A\otimes \dyad{i}{i}_B\otimes \dyad{i}{i}_C
\end{align}
for orthogonal states $\dyad{i}{i}_A$, $\dyad{i}{i}_B$, and $\dyad{i}{i}_C$, from which $\rho^i_A$, $\rho^i_B$, and $\rho^i_C$ can be prepared, respectively.  Therefore, since correlation measures are monotonically decreasing under local processing, vanishing on classical states implies vanishing on all separable states.

Suppose $\rho_{ABC}$ is of the form (\ref{classical}), and consider the purification
\begin{align}\label{Eq:ClassicalExtension}
    |\psi\rangle_{AaBbCc} = \sum_{i,j,k}\sqrt{p_{ijk}} \ket i_A \ket{ijk}_a \ket j_B \ket{ijk}_b\ket k_C\ket{ijk}_c,
\end{align}
For the state (\ref{Eq:ClassicalExtension}), after tracing out one of the ancillae (for example, tracing out $c$), the state is classical,
\begin{align}
    \rho_{ABCab} = \sum_{ijk} p_{ijk} \dyad{i}{i}_{A}\otimes \dyad{ijk}{ijk}_a \otimes \dyad{j}{j}_{B}\otimes \dyad{ijk}{ijk}_b \otimes \dyad{k}{k}_{C},
\end{align}
and knowing the state of either of the ancilla systems $a$, $b$, is enough to know the rest of the state.  This implies that all conditional entropies conditioning on one of the ancilla is zero.  Therefore  any conditional mutual information which contains any of the ancilla in the conditioned system, must vanish.  In particular, this means that $f_{b_1}$, $f_{b_2}$, $f_{b_3}$, and $f_{c_3}$ all vanish for the state (\ref{Eq:ClassicalExtension}).  This implies that $E_{b_1}$, $E_{b_2}$, $E_{b_3}$, and $E_{c_3}$ all vanish on classical states, and therefore on any separable state. A similar argument shows that the bipartite squashed entanglement $E_{\rm sq}$ vanishes on separable states.

Correlation measures $E_{s_2}$ and $E_{a_2}$, as far as we can see, do not fall into either of these two categories.  The only thing we can say with certainty is that if $E_{s_2}$ or $E_{a_2}$ vanishes on a state, then the state must have $I(A:B) = I(A:C) = I(B:C) = 0$.

Another natural question to ask is what can we learn by examining these correlation measures in the context of holographic states.  One thing we know is that holographic states satisfy MMI \cite{Hayden:2011ag}
\begin{align}
I(A:B) \leq I(A:B|C).
\end{align}
In other words, in a holographic state, mutual informations can only increase under conditioning.  Now consider a holographic state where $E_{b_1} = \tilde{E}_3^{\rm sq}$ vanishes.  Then there is a purification in which all three terms of $\tilde{f}_3^{\rm sq} = I(A:BC|a) + I(B:AC|b) + I(C:AB|c)$ are zero.  But then MMI tells us that $I(A:BC) = I(B:AC) = I(C:AB) = 0$, so the state must be product.  Since we just showed that $\tilde{E}_3^{\rm sq}$ vanishes on any separable state, this implies that any separable holographic state is a product state.  This is already an interesting observation, but if we want to know more we must approach the question geometrically.

\section{Holographic Description of Optimized Correlation Measures}

\subsection{General procedure}

We now argue for a holographic dual presentation of these correlation measures. We focus on the simplest case of the AdS$_3$ geometry, at a constant time-slice corresponding to two-dimensional hyperbolic space. Each party now corresponds to an interval on the $S^1$ boundary, which we call a region; we assume the regions do not overlap or touch.

The RT formula \cite{RT} states that the entropy of a region $A$ of the boundary CFT is proportional to the minimum area of a bulk surface that shares its boundary with $A$ and is homologous with (continuously deformable to) $A$; the proportionality factor is $1/4G_N$ and for simplicity we choose Newton's constant $G_N$ so this factor is one.   The minimal area surface is called the RT surface, and the bulk region bounded by $A$ and its RT surface is called the entanglement wedge of $A$. Any linear entropic formula can be calculated in a gravity dual using the RT formula.

In order to implement the holographic dual of an optimized information measure, we seek a geometric picture of the set of all purifications of a given state. Since a state in AdS$_3$ is associated to a collection of boundary regions, one possible purification is always simply the state associated to the complete boundary circle. But what other geometric purifications are possible?

To address this question we draw motivation from the surface-state correspondence \cite{Miyaji:2015yva}, which posits that any closed convex surface corresponds to a pure state, whether it is at the boundary or moves through the bulk. Geometric purifications of the density matrix corresponding to the parties thus correspond to connecting the $n$ boundary regions into a single closed convex surface. The complete boundary $S^1$ is one such surface, but other configurations connect the regions through the bulk, in general up to the boundary of the entanglement wedge, beyond which the surfaces are not convex. The surfaces added to the boundary regions $A$, $B, \ldots$ to make a closed convex surface correspond to the ancilla $a, b, \ldots$, though which part corresponds to which ancilla is yet to be determined. The entropy of a subregion of such a closed convex curve can then be calculated by a generalization of the RT formula,  where one still finds the area of the minimal area surface sharing a boundary with, and homologous to, the subregion in question, even if the subregion is partially or fully in the bulk.

In what follows we will consider correlation measures in the symmetric-ancilla ansatz. The nontrivial case is where the entanglement wedge consists of one connected component; this corresponds to the mutual informations satisfying $I(A_i : \prod_{j\neq i} A_j) > 0$ for all parties $i$.\footnote{If one party $A$ has a disconnected entanglement wedge, the optimal ancilla $a$ to create a closed surface is just the RT surface for $A$. This purifies separately from the remaining regions, and since there is no nonzero one-party correlation measure, evaluates to zero. The correlation measure then reduces to the $(n-1)$-party measure for the other regions, matching our previous expectation from quantum information arguments.} We will make two assumptions about the minimizing purifications which appear in our optimized correlation measures.  First, we take the optimal purification (before dividing it into $n$ subsystems) to be the closed surface formed by the boundary of the $n$-partite entanglement wedge. This is an extremal purification, minimizing the area in the bulk and leading to vanishing mutual information between any two ancilla regions; we will discuss this motivation more in what follows. The second assumption we make is that the optimal partition of the purifying system is one which places $a_i$ adjacent to $A_i$.  This is motivated because our proof  of monotonicity involves moving degrees of freedom from the region $A_i$ to its corresponding ancilla $a_i$ as in equation~(\ref{MonotonicityCond}); geometrically this is implemented by having them next to each other. This assumption is equivalent to saying that the optimal partition of the RT surface into $n$ subsystems is parameterized by the locations of $n$ points, one lying on each of the $n$ segments of the RT surface, and for each segment the portion bounded by this point and a boundary point of $A_i$ belongs to $a_i$. These points dividing the RT surface segments are called {\it  optimization points}.

The procedure for determining the holographic dual for a given correlation measure is then, for each choice of boundary regions, allow the optimization points to vary over their arcs on the boundary of the entanglement wedge. For each location of the optimization points, calculate the entropies entering the formula for the correlation measure, and find the location of the points that minimize it. The combination of lengths of surfaces calculating these entropies is then the desired holographic dual. 
Thus for a given correlation measure and a given set of boundary regions, the holographic dual is characterized by the data:

\begin{itemize}
	{\item The locations of the optimization points (corresponding to the optimized purification)}
	{\item An undirected multigraph whose vertices are the boundary points and the optimization points, with negative edges allowed (corresponding to the value of the correlation measure). That is, each pair of vertices has an integer associated to the edge between them, which can be positive, negative or zero.}
\end{itemize}
For simplicity, we will call each such collection a ``picture" for the correlation measure evaluated on those regions.

 \subsection{Review of bipartite cases}

\begin{figure}
\begin{center}
\includegraphics[width=0.32\textwidth]{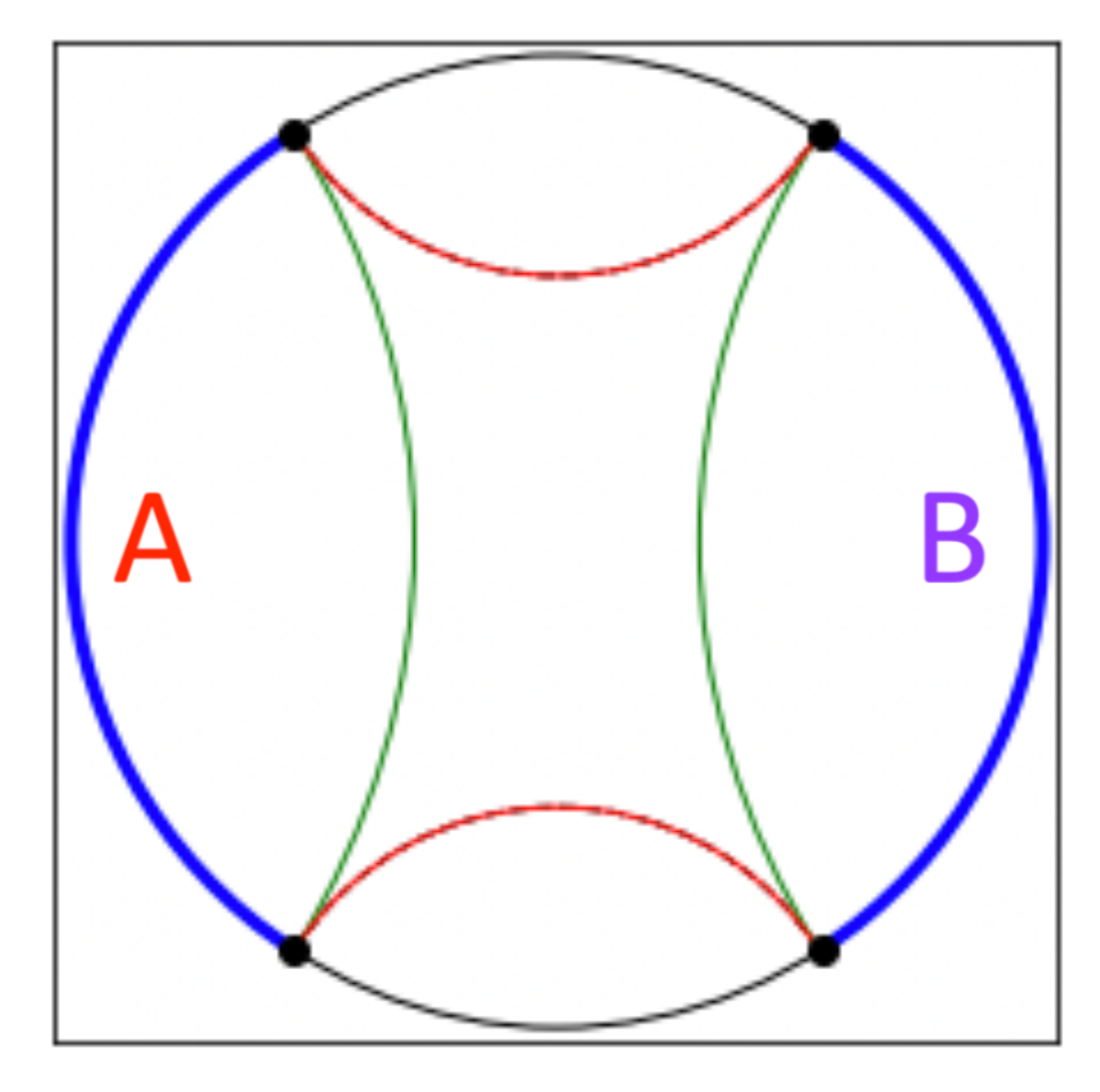}
\includegraphics[width=0.32\textwidth]{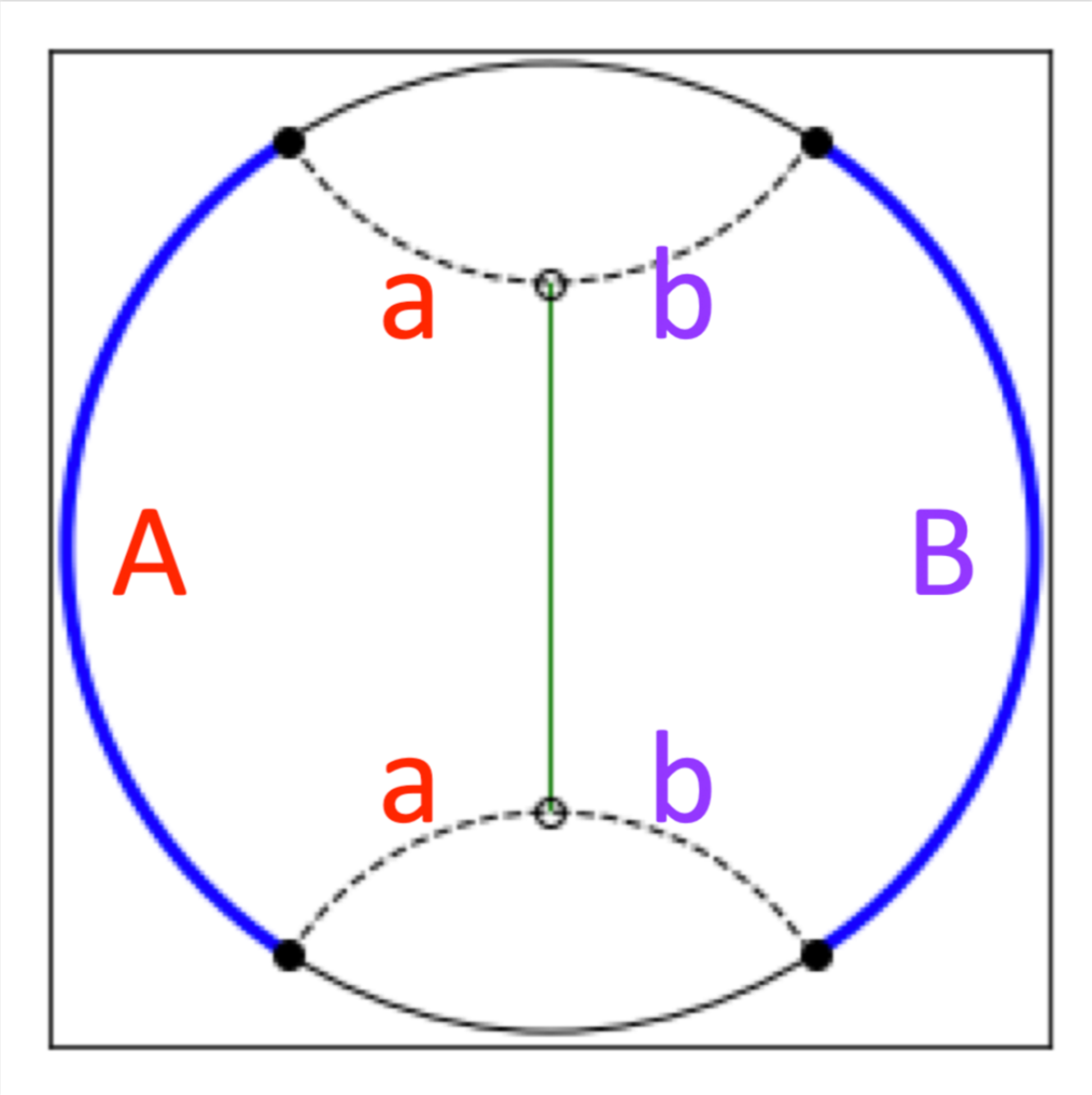}
\caption{Left, the holographic dual of the mutual information $I(A:B)$, where green arcs are added, red arcs subtracted to obtain the value of the measure; this measure is independent of the optimization and so the optimization points are not indicated. Right, the holographic dual of the bipartite entanglement of purification $E_P$, and of the $R$-correlation $E_R$,  calculated by the entanglement wedge cross-section (green arc); the optimization points (open circles) divide the ancilla into $a$ associated with $A$, and $b$ assoociated with $B$.   Both are shown for boundary region size $\theta = 2$.}
\label{Bipartite_EP_labeled}
\end{center}
\end{figure}

Here we review what is already known about the holographic duals of bipartite correlation measures. To begin with, consider the non-optimized correlation measure  the mutual information, $I(A:B) = S(A) + S(B) - S(AB)$, illustrated on the left side of figure~\ref{Bipartite_EP_labeled}. We take the regions $A$ and $B$, indicated by blue  curves with boundary points shown as solid dots, to be diametrically opposite with size $\theta = 2$. Since no optimization is necessary, this measure is calculated simply by the RT formula. The entropies $S(A)$ and $S(B)$ are the lengths of the green RT surfaces, spanning geodesic curves that share the boundaries of $A$ and $B$. Meanwhile, the entropy $S(AB)$ is obtained by the red geodesic curves connecting the boundaries of $A$ and $B$, and forming the boundary of the entanglement wedge of $AB$. Since this measure is independent of the optimization, the ancilla and optimization points are not shown. We will maintain this visual convention that green curves are added, and red curves subtracted, to obtain the value of a correlation measure.

The right-hand picture in figure~\ref{Bipartite_EP_labeled} shows the bipartite entanglement of purification $E_P(A,B)$ for the same two parties. The boundary of the entanglement wedge stretched between them is shown as two dashed arcs, on which live the optimization points, shown as open circles; these divide the entanglement wedge boundary into the ancilla $a$ adjacent to $A$, and the ancilla $b$ adjacent to $B$. For this correlation measure the optimization points are in the middle of their arcs, and the correlation measure itself is calculated by the area (length) of the geodesic arc between the optimization points, the so-called entanglement wedge cross-section, shown here in green. This curve can be interpreted in the surface-state correspondence as the RT surface calculating $S(Aa)$, where $Aa$ is an open convex region, part on the boundary and part in the bulk; the entanglement wedge cross-section is the minimal area curve sharing the boundary of $Aa$ (the optimization points) and homologous to it.

It is not hard to see geometrically that these measures are monotonic. Monotonicity implies that when the regions get smaller, the correlation measure decreases as well. For the mutual information, making the regions smaller decreases the size of the positive green curves spanning the RT surfaces of A and of B, while increasing the size of the negative red curves spanning the entanglement wedge boundary,  decreasing the value of the measure. For the entanglement wedge cross-section, making the regions smaller makes the entanglement wedge narrower, and hence the cross section gets shorter, and the measure decreases. Note that if we simply give the edges of the boundary region $A$ to the ancilla $a$, the optimization points do not move and the value is unchanged, corresponding to the invariance of $S(Aa)$ under the operation (\ref{MonotonicityCond}). However, this is now no longer the optimal purification, which instead uses the boundary of the new entanglement wedge, moving the optimization points closer and decreasing the measure.

The $R$-correlation $E_R$ is also calculated by the EWCS; this can be understood since $f^R(A,B,a,b) = f^P(A,B,a,b) - I(a:b)/2$, and as we have mentioned $I(a:b) = 0$ for optimization points along the entanglement wedge boundary arcs, which are geodesics.

\begin{figure}
\begin{center}
\includegraphics[width=0.24\textwidth]{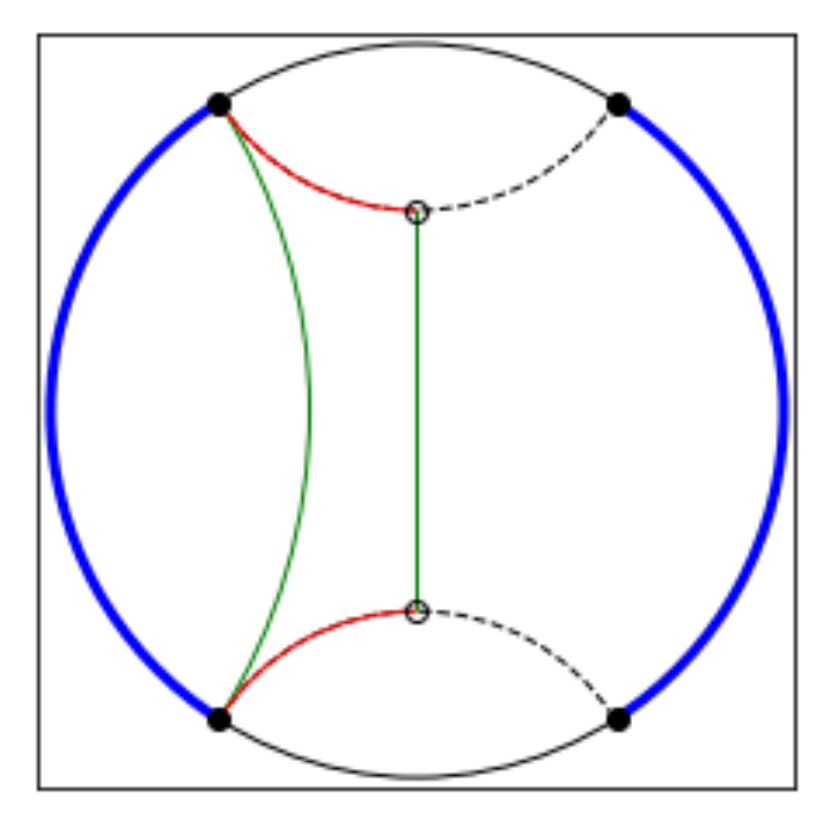}
\includegraphics[width=0.24\textwidth]{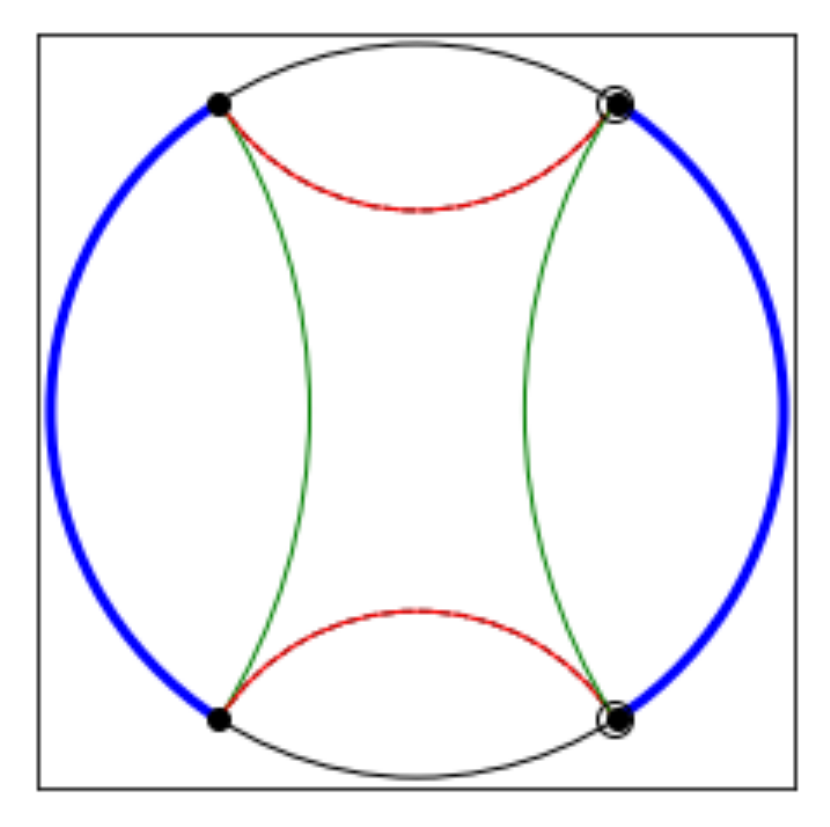}
\caption{The holographic dual of the bipartite $Q$-correlation $E_Q$ (left) and that of the bipartite squashed entanglement $E_{\rm sq}$ (right) at $\theta = 2$. The optimization points in the right-hand figure coincide with some of the boundary points and are visible as open circles surrounding the solid dots.}
\label{Bipartite_EQ_Esq}
\end{center}
\end{figure}

The remaining bipartite correlation measures are shown in figure~\ref{Bipartite_EQ_Esq}. On the left is the $Q$-correlation $E_Q$. As with $E_P$ and $E_R$, the minimizing purification has the optimization points halfway along their arcs, but in this case the curves calculating the measure are the sum of the length of the EWCS and the RT surface for $A$, minus the lengths of the arcs corresponding to the ancilla $a$. The measure is clearly monotonic, similarly to how $I(A:B)$ is; the fact that as the region size decreases, the measure wants to continue purifying along the entanglement wedge cross-section boundary, with the optimization points moving inward, follows from the contribution of the (positive, green) EWCS winning over that of the (negative, red) curve along the boundary.

On the right of figure~\ref{Bipartite_EQ_Esq} is the squashed entanglement $E_{\rm sq}$. In this case, the optimization points are not centered, but move all the way out to the boundary, setting the ancilla $b=0$ and purifying $AB$ entirely with $a$. The result has the same configuration of surfaces as the mutual information, and is hence numerically identical to $I(A:B)$, as discussed in \cite{umemoto2018entanglement}. 

We have displayed these measures for diametrically opposite regions of size $\theta = 2$. For all the bipartite measures,  the qualitative picture of the holographic dual does not change as the boundary region sizes are varied.  In principle each pair of regions is determined by four numbers, the locations of the two pairs of boundary points. However, we can use the conformal group action on the boundary $S^1$ to arbitrarily choose the locations of three of these quantities, leaving only one independent parameter (equivalent to the cross-ratio). We may always choose to locate the centers of the regions diametrically opposite at $0$ and $\pi$, fixing two numbers, and finally to fix the sizes of the two regions to be identical, $\theta_A = \theta_B \equiv \theta$, leaving this size as the only independent parameter; we then consider $\theta \geq \pi/2$ to have a connected entanglement wedge, otherwise all the correlation measures are zero. Any other configuration is equivalent by a conformal transformation to one in this class, and will calculate the same numerical value of the correlation measures. This use of conformal symmetry requires that the correlation measures are cutoff independent, as we demonstrate in the next subsection.

Symmetry is a helpful concept for classifying the pictures of correlation measures, and we introduce this here as a warm-up for the tripartite case, where it is more intricate. Since the two bipartite boundary regions can be taken to have the same size, and the correlation measures are symmetric, we may consider an $S_2$ permutation symmetry exchanging the regions. We see the pictures for the $E_P$ and $E_R$, and for $I(A:B)$, in figure~\ref{Bipartite_EP_labeled} preserve this symmetry. Those for $E_Q$ and $E_{\rm sq}$ in figure~\ref{Bipartite_EQ_Esq}, however, do not; for $E_Q$ the optimization points are symmetric but the surfaces calculating the measure are not, while for $E_{\rm sq}$ the surfaces are symmetric (and match $I(A:B)$), but the optimization points break the symmetry. Whenever a picture ``spontaneously" breaks the permutation symmetry of the regions like this, we can equally well calculate the measure by exchanging the roles of the regions in the picture, that is, we can equally well reflect either of the pictures in figure~\ref{Bipartite_EQ_Esq} across the vertical axis and we will calculate the same number.

\subsection{Cutoff-independence and monotonicity}

The RT formula for the entropy of a region $A$ in general relates two infinite things: the field theory calculation of the entropy is ultraviolet divergent, while the area (length in our case) of the RT curve diverges as it crosses the infinite distance to the AdS boundary. In general one regulates both quantities, the field theory entropy with a lattice cutoff or other ultraviolet regulator, the gravity side with some cutoff on the geometry very close to the boundary.

Certain combinations of entropies, however, may be independent of the cutoff, if the cutoff-dependence cancels. From the geometric point of view, this happens when for any boundary point with RT curves approaching it in the picture of the quantity, an equal number of curves with positive and negative coefficient head to that point. This trivially holds for the entanglement of purification (and the $R$-correlation as well) in figure~\ref{Bipartite_EP_labeled}, since no curves approach the boundary. For the remaining measures, cutoff-independence is evident in the fact that each boundary point has an equal number of positive and negative arcs ending at it.

We now argue that all our correlation measures, and indeed all quantities that are monotonic in the way we have described, are also cutoff-independent. All curves that approach the boundary do so at the endpoints of the regions $A_i$. Consider $A$ for definiteness: it has two endpoints, which are also the near endpoints of the two components of the ancilla $a$; the far endpoints of $a$ are the optimization points. If an entropy contains the combination $Aa$, this entire region is bounded only by these optimization points, which are in the bulk, and thus no curves approach the boundary, as in the entanglement wedge cross-section, dual to $S(Aa)$. If instead we have an entropy combination $S(AX) - S(aY)$ with $X, Y$ not containing $A$ or $a$, then the RT curve going to an endpoint of $A$ with positive coefficient is canceled by an RT curve going to the same point, now thought of as an endpoint of $a$, with negative coefficient. But these are precisely the criteria we use for monotonicity; thus all the monotones, and hence all the correlation measures, are cutoff-independent.

Cutoff-independence is useful for us since it means the correlation measures do not see any breaking of conformal invariance due to the cutoff, and therefore we can use the conformal group acting on the boundary to make canonical choices for the boundary regions, knowing that other boundary choices equivalent under conformal transformations give the same value for the correlation measure. We note that while all monotones are cutoff-independent, not all cutoff-independent quantities are monotones; for example, a difference of monotones is generically not monotonic, since  monotonicity condition 2 does not allow monotones to flip sign, but will remain cutoff-independent. The tripartite mutual information $I_3(A:B:C) = J(A:B:C) - K(A:B:C)$ is an example of a cutoff-independent but non-monotonic quantity.

\section{Holographic duals of tripartite correlation measures}
\label{HoloTripartiteSec}

Now we consider the various tripartite correlation measures in AdS$_3$. As reviewed in the last section, a correlation measure with particular boundary regions is characterized by a picture, consisting of the locations of the optimization points, which specify the optimizing purification, as well as the configuration of curves (surfaces) with associated integers specifying the weights by which their lengths are combined into the value of the measure. Since our correlation measures are cutoff-independent we can again use conformal symmetry on the boundary $S^1$ to fix three of the six parameters; we choose to place the centers of each region at $2\pi/3$-spaced intervals around the circle. The three remaining parameters are the sizes of the three boundary regions.

\subsection{Overview: curve optimization relations and symmetry}

There are a number of different pictures that can arise for our thirteen (symmetric-ancilla) tripartite correlation measures with different sizes of boundary regions, so we will look for a few characteristics to impose some order on the menagerie. One is the location of the optimization points. As in the bipartite case, the optimization points are either pushed to the boundary, or sit at finite locations along the arcs bounding the entanglement wedge. Interestingly, in the tripartite case when the optimization points are at finite distance they need not necessarily sit in the middle of their arcs, even for symmetric region sizes, but sometimes sit at off-center locations. We find that such locations can always be characterized by a nontrivial relation holding between lengths of various curves, some starting or ending at the location of the optimization point. This {\it curve optimization relation} characterizes the location of the optimization point, and hence the optimal purification for that measure. If the boundary regions change size slightly, the optimization point will adjust its location to maintain the relation (unless the change in boundary regions happens to push the measure into favoring a qualitatively new picture). As we will see, there are a number of distinct curve optimization relations, each determining a different location of the optimization point.

To characterize these novel conditions, it is useful to consider the family of cases where all three boundary regions are of the same size, in addition to being equally spaced around the boundary. In this case there is an $S_3$ permutation symmetry exchanging the three boundary regions, acting geometrically like the symmetries of an equilateral triangle. Once the picture for a correlation measure is obtained, the location of the optimization points may respect, partially break, or fully break this geometric $S_3$. The possibilities are:
\begin{itemize}
	{\item $S_3$: full symmetry of $2\pi/3$ rotations and reflections preserved}  
	{\item $A_3$: alternating symmetry of $2\pi/3$ rotations preserved} 
	{\item $Z_2$: reflection symmetry along one axis preserved} 
	{\item None: no symmetry preserved}
\end{itemize}
We find optimization point locations corresponding to each; the behavior under these symmetries is the origin of our labels $f_c$, $f_s$, $f_a$, $f_r$ and $f_b$, as we shall discuss.  Moreover, whenever the optimization points preserve a symmetry, we can find a configuration of curves calculating the correlation measure that preserves the symmetry as well. As we will show, due to the nontrivial curve optimization relations that hold for symmetry-breaking cases, there may also be less symmetric configurations of curves that equally well calculate value for the measure.

When the boundary regions have different sizes, the $S_3$ symmetry no longer obtains. The location of the optimization points, however, can still be characterized by their curve optimization relations. In what follows, we focus on a thorough investigation of the case with all boundary regions the same size, and then give a few examples with discussion for different-sized boundary regions.

As in the bipartite correlation measure figures, we color the boundary regions blue,  the rest of the boundary is a solid black line, and the boundary of the entanglement wedge is three dashed lines. Black dots indicate the boundaries of the regions, and open circles are the optimization points that are varied over in finding the minimum for each measure. The geometric realization of the correlation measure is then the sum of the curves drawn in green, minus the sum of the curves drawn in red. Sometimes a curve is counted more than once, in which case the green or red line becomes thicker. As in the bipartite case, geometric monotonicity is reflected in the fact that positive curves tend to get shorter, and negative curves tend to get longer, as the boundary regions shrink.

\subsection{Correlation measures with same-size boundary regions}

Tripartite correlation measures have a three-parameter family of conformally-inequivalent boundary region configurations. We will focus our attention on the one-parameter family where all three boundary regions have the same size $\theta$. For $\theta < \pi/3$ the entanglement wedge splits into three disconnected parts, and all the measures are zero, while at $\theta = 2\pi/3$ the regions fill the boundary circle, and thus we consider the values $\pi/3 \leq \theta \leq 2 \pi/3$.

\begin{figure}
\begin{center}
\includegraphics[width=0.24\textwidth]{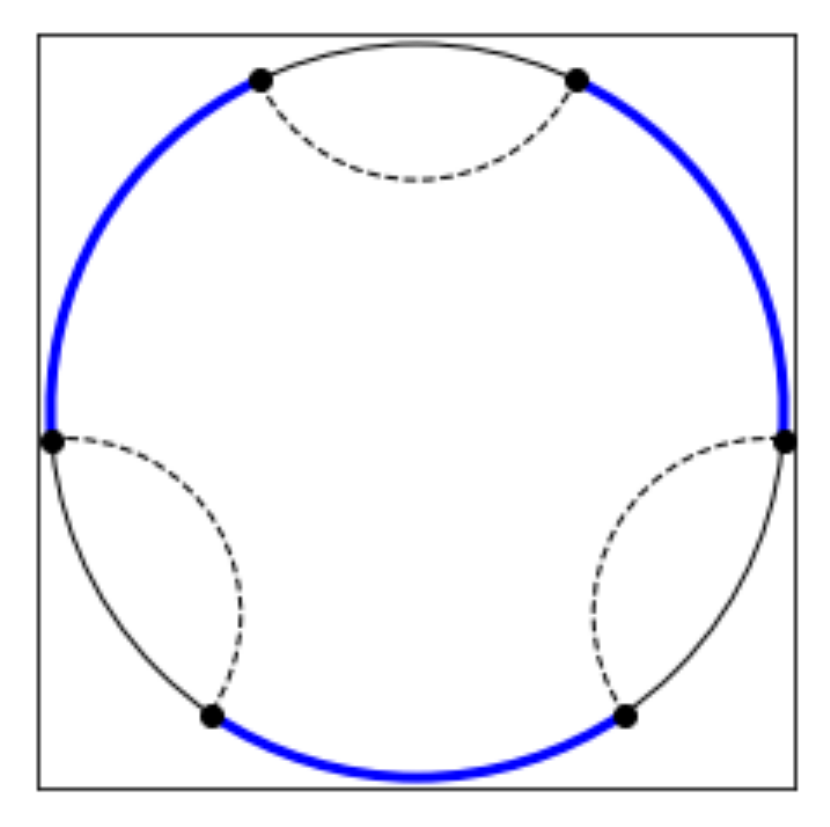}
\includegraphics[width=0.24\textwidth]{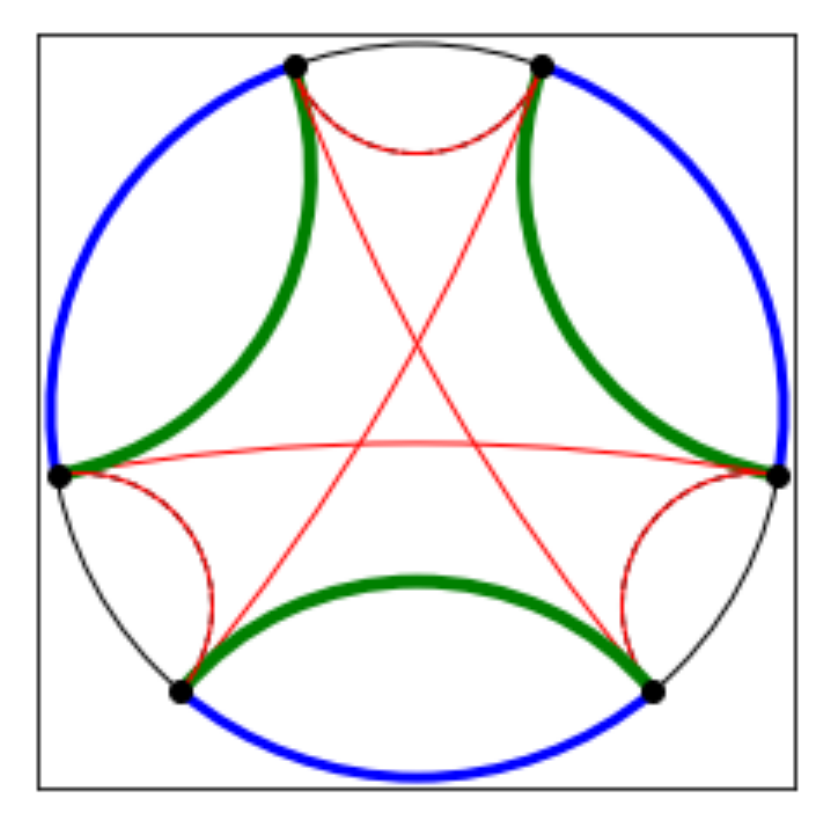}
\includegraphics[width=0.24\textwidth]{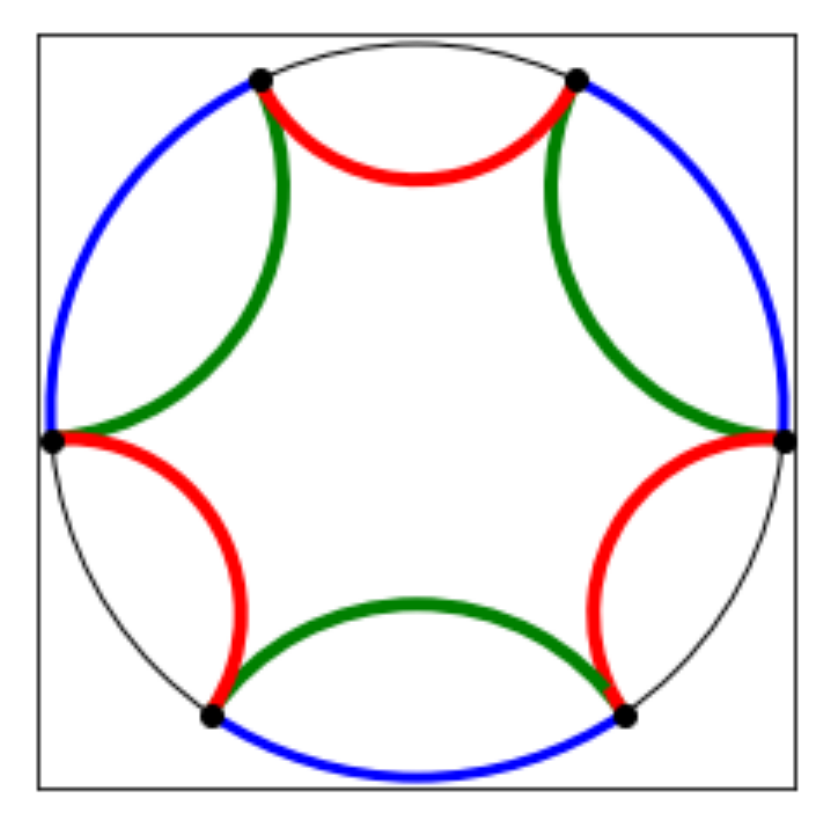}
\includegraphics[width=0.24\textwidth]{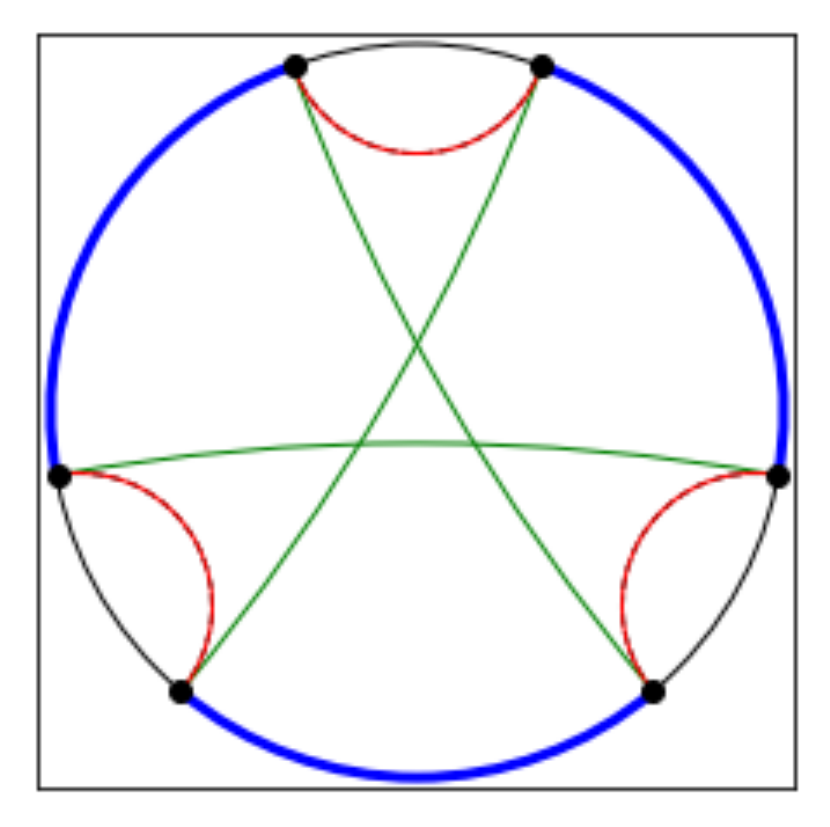}
\caption{Correlation measures $I(A:B) + I(A:C) + I(B:C)$ at $\theta = 1.2$ and $\theta = 1.4$, and the dual total correlation $J(A:B:C) \equiv S(AB) + S(AC) + S(BC) - 2S(ABC)$ at $\theta = 1.2$ and $\theta = 1.4$. They are fully symmetric and independent of the optimization points, which are not shown.}
\label{OptIndepFig}
\end{center}
\end{figure}

\subsubsection{Optimization-independent measures}

First we discuss the two correlation measures independent of the optimization; for these measures the optimization points are not shown, since they are irrelevant.  One is the sum of the pairwise mutual informations $I(A:B) + I(A:C) + I(B:C)$. The fact that the entanglement wedge is connected means that $I(A:BC)$, $I(B:AC)$ and $I(C:AB)$ are nonzero throughout. However, $I(A:B)$, $I(A:C)$ and $I(B:C)$ are zero for $\theta < \theta_I \approx 1.3182$, and nonzero for $\theta > \theta_I$, and this is reflected in the pictures (the left two images in figure~\ref{OptIndepFig}), which are zero for $\theta < \theta_I$, and nonzero for $\theta > \theta_I$. The green curves are the individual entropies $2S(A)$, $2 S(B)$ and $2S(C)$, while each   $- S(AB)$, $- S(AC)$, and $- S(BC)$ is a pair of red curves. The green curves are thicker because they are counted twice. Both pictures are fully $S_3$-symmetric (the former trivially so).

The other optimization-independent correlation measure  is the dual total correlation $J(A:B:C) \equiv S(AB) + S(AC) + S(BC) - 2S(ABC)$. This measure also undergoes a change in picture at $\theta = \theta_I$; both pictures are fully $S_3$-symmetric, as shown in the last two images in figure~\ref{OptIndepFig}. It is easy to see the pictures on either side of the transition are what is predicted by RT, with zero and nonzero pairwise mutual informations, respectively.

\subsubsection{Entanglement wedge cross-section measures}

For these and all remaining measures, the locations of the optimization points, denoted by open circles, are nontrivial.

\begin{figure} 
\begin{center}
\includegraphics[width=0.24\textwidth]{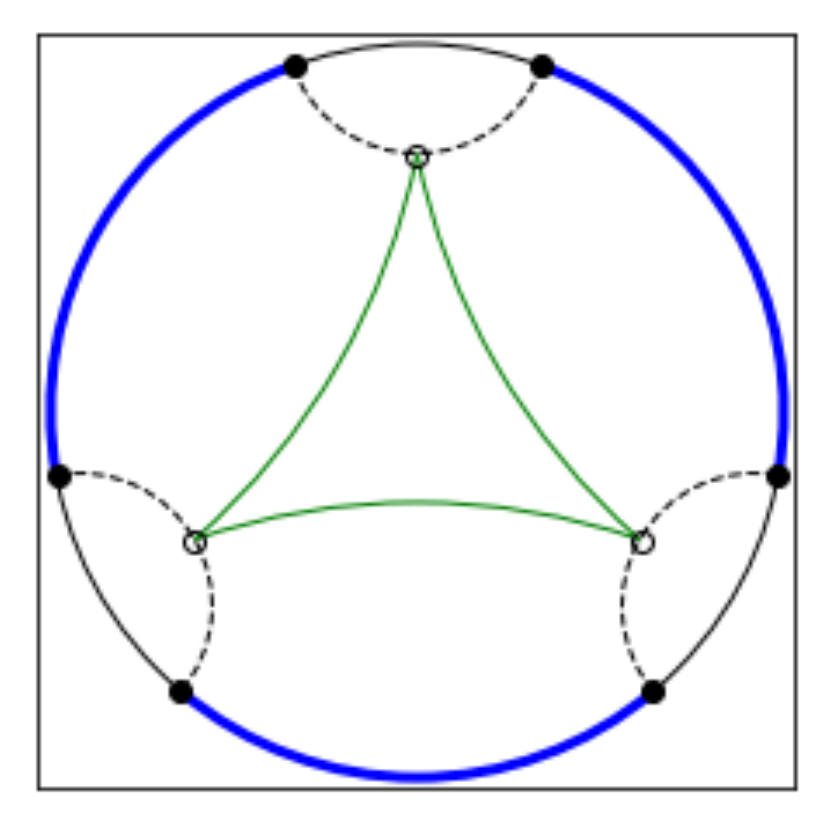}
\caption{Correlation measures $f_{c_1}$, $f_{c_2}$ and  $f_{c_3}$ at 
$\theta = 1.4$; for all $\theta$ the shape remains that of the tripartite entanglement wedge cross section. This class includes  the tripartite entanglement of purification $f^P_3 = f_{c_1}$ and tripartite R-correlation $f^R_3 = f_{c_2} + f_{c_3}$.}
\label{EWCSFig}
\end{center}
\end{figure}

The entanglement-wedge cross section is characterized geometrically by locating the optimization points such that the sum of the three curves between the pairs of points is minimized, as in figure~\ref{EWCSFig} \cite{umemoto2018entanglement}.  For these cases with equal-sized boundary regions, the optimization points are all in the middle of their arcs bounding the entanglement wedge. The configuration is fully $S_3$-symmetric, and undergoes no transition in shape at $\theta = \theta_I$ or anywhere else.

 Three different correlation measures, which we labeled $f_{c_1}$, $f_{c_2}$ and $f_{c_3}$ are all realized geometrically by the entanglement-wedge cross-section. The first is just the tripartite entanglement of purification, $f_{c_1} = f^P_3\equiv S(Aa) + S(Bb) + S(Cc)$. The other two differ from this by
\begin{eqnarray}
	f_{c_2}  =f^P_3  - J(a:b:c) \,,
\end{eqnarray}
and
\begin{eqnarray}
	f_{c_3} =f^P_3 -I(a:b) - I(a:c) - I(b:c) \,.
\end{eqnarray}
Moreover, the tripartite $R$-correlation is the sum of these two,
\begin{eqnarray}
	f^R_3 = f_{c_2} + f_{c_3}= 2 f^P_3   -I(a:b) - I(a:c) - I(b:c) - J(a:b:c) \,.
\end{eqnarray}
The fact that all these measures coincide indicates that the optimization points are chosen so that the ancilla $a$, $b$, and $c$ obey
\begin{eqnarray}
	I(a:b) + I(a:c) + I(b:c)  = J(a:b:c) = 0\,.
\end{eqnarray}
These equations also imply $I(ab:c) = I(ac:b) = I(bc:a) = 0$. In fact, these relations hold for any optimization points located anywhere on the boundary arcs of the entanglement wedge, since these curves are geodesics. This is the generalization to three parties of the fact for two parties that the optimization points there are chosen so that $I(a:b) = 0$, and as a result the two-party $f^R$ coincides with $f^P$ holographically.

\subsubsection{More symmetric measures}

\begin{figure} 
\begin{center}
\includegraphics[width=0.24\textwidth]{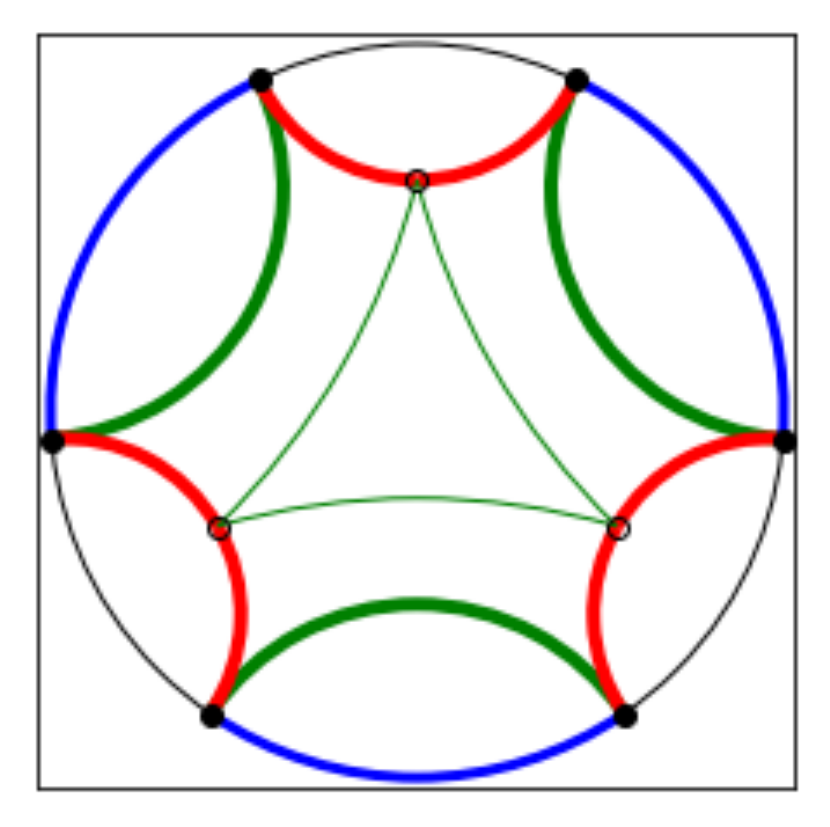}
\includegraphics[width=0.24\textwidth]{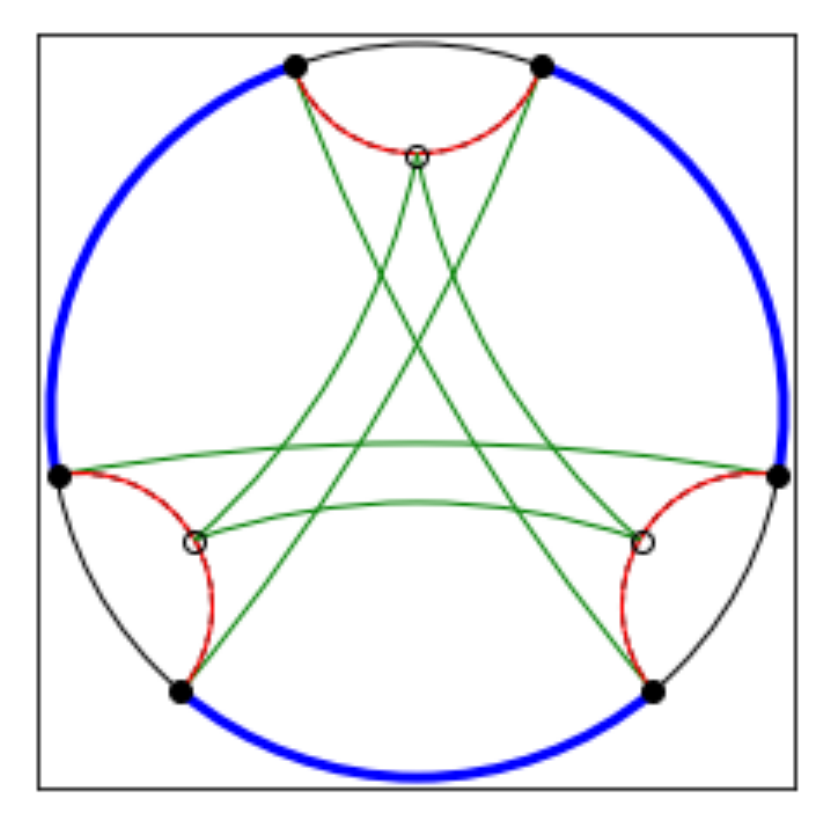}
\includegraphics[width=0.24\textwidth]{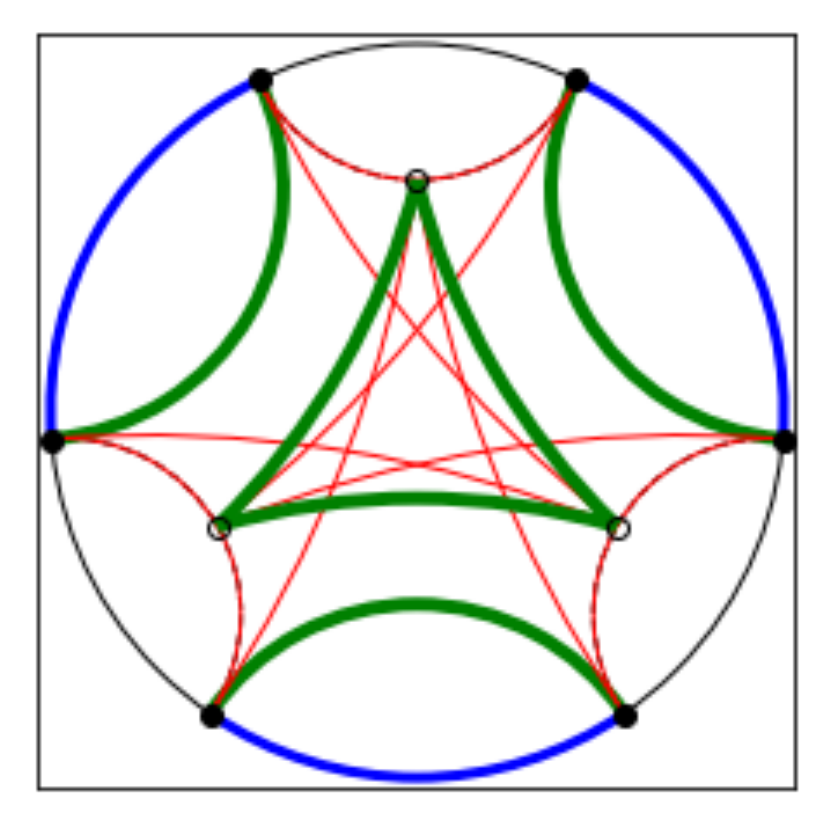}
\includegraphics[width=0.24\textwidth]{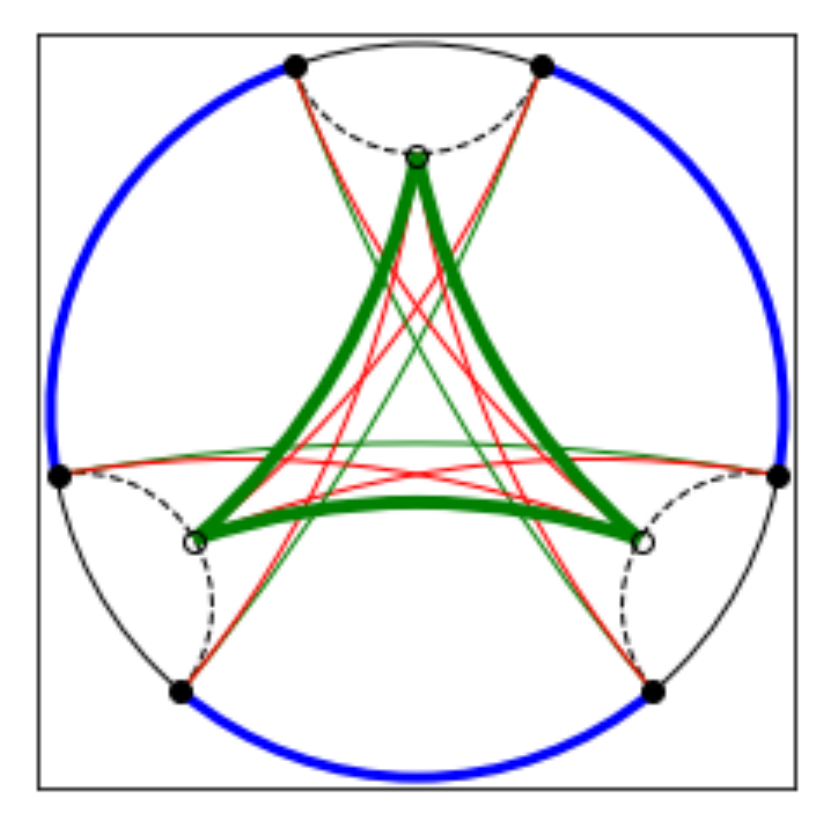}
\caption{Correlation measures $f_{s_1} = f^Q_3$ (the tripartite $Q$-correlation, left pictures) at $\theta = 1.2$ and $\theta = 1.4$, and $f_{s_2}$ at $\theta = 1.2$ and $\theta = 1.4$ (right pictures).}
\label{SymmetricFig}
\end{center}
\end{figure}

There are two more correlation measures where the optimization points are in the center of their arcs and the measures have fully $S_3$-symmetric presentations for all $\theta$. One is the tripartite $Q$-correlation, $f_{s_1} = f^Q_3$. For all these cases with the boundary regions having the same size, this measure is exactly the sum of the pictures of $J(A:B:C)$ and the entanglement wedge cross-section, and thus has a transition at $\theta = \theta_I$ inherited from the transition of $J(A:B:C)$, as shown in the left two pictures in figure~\ref{SymmetricFig}.  Thus for these cases, we have $E_Q = E_P + J(A:B:C)$; however as we will see this equation does not hold for all unequal-size regions, and so is not a general prediction of holographic backgrounds.

The other $S_3$-symmetric measure is $f_{s_2}$. At all values of $\theta$ it has the form $f_{s_2} = 2 f_{s_1} - f_{a_1} - J(A:B:C)$, where it is always the symmetric form of $f_{a_1}$ (discussed in the next subsection) that appears. This also has a transition at $\theta = \theta_I$, as shown in the last two images in figure~\ref{SymmetricFig}.

\subsubsection{Correlation measures with reflection or rotation symmetry; curve optimization relations}

\begin{figure} 
\begin{center}
\includegraphics[width=0.24\textwidth]{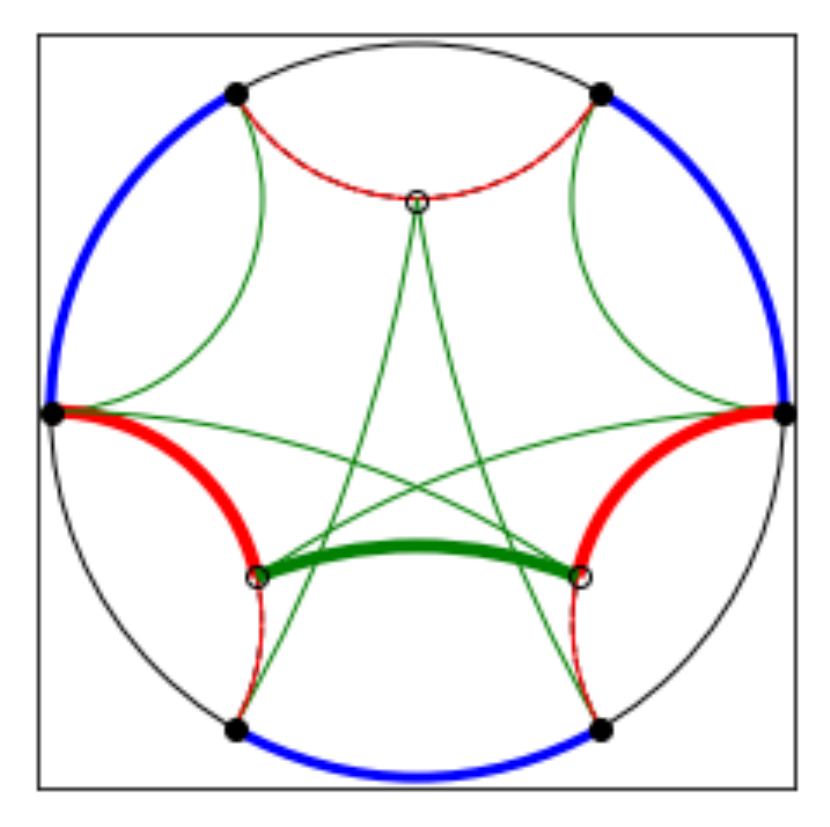}
\includegraphics[width=0.24\textwidth]{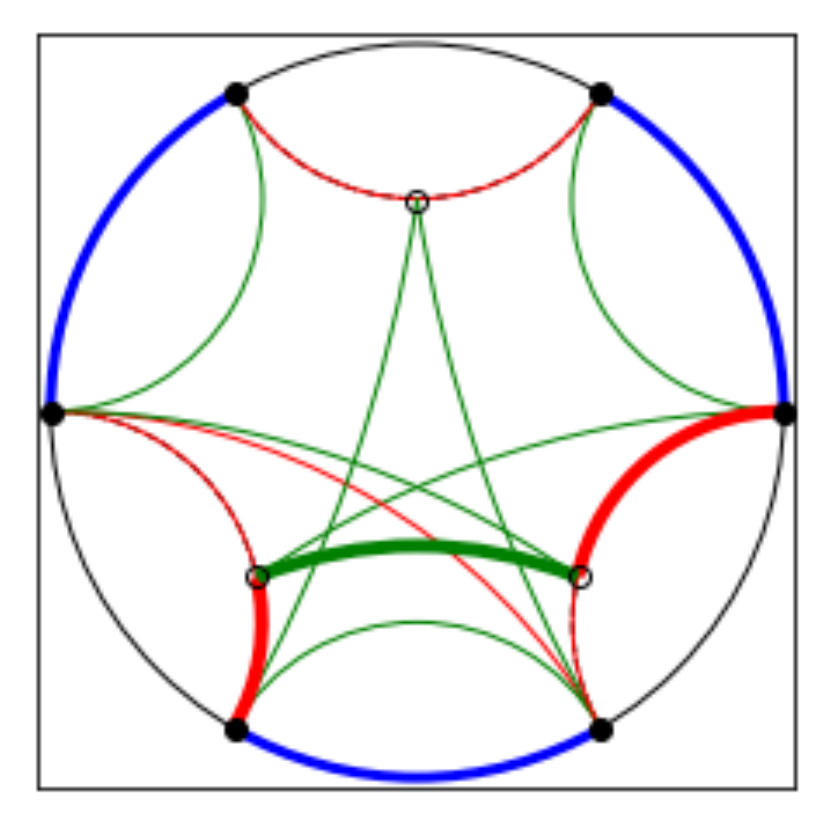}
\includegraphics[width=0.24\textwidth]{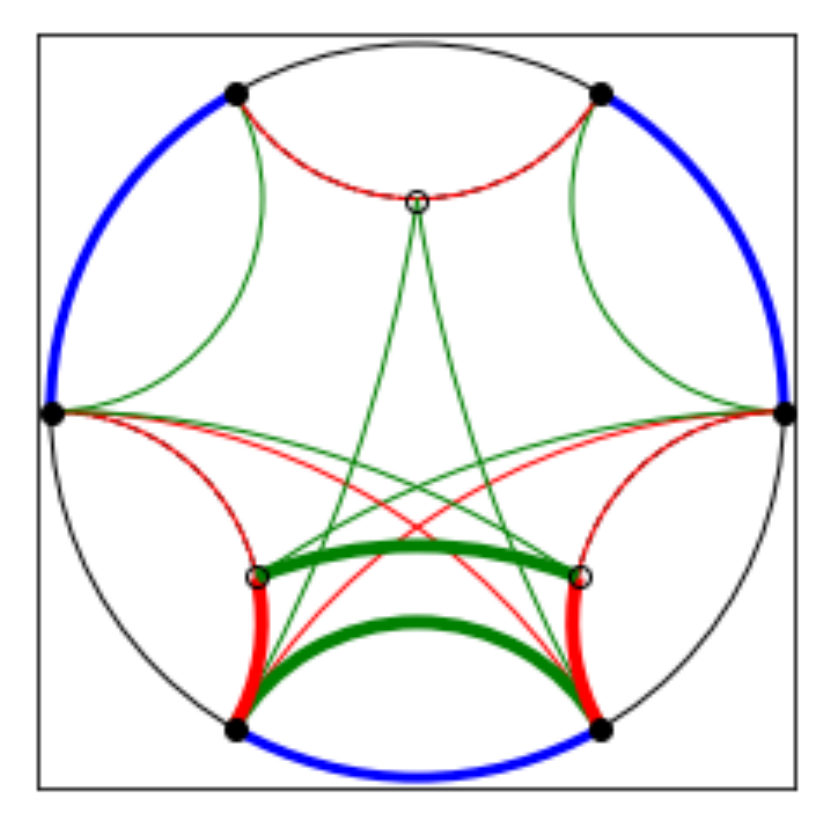}
\includegraphics[width=0.24\textwidth]{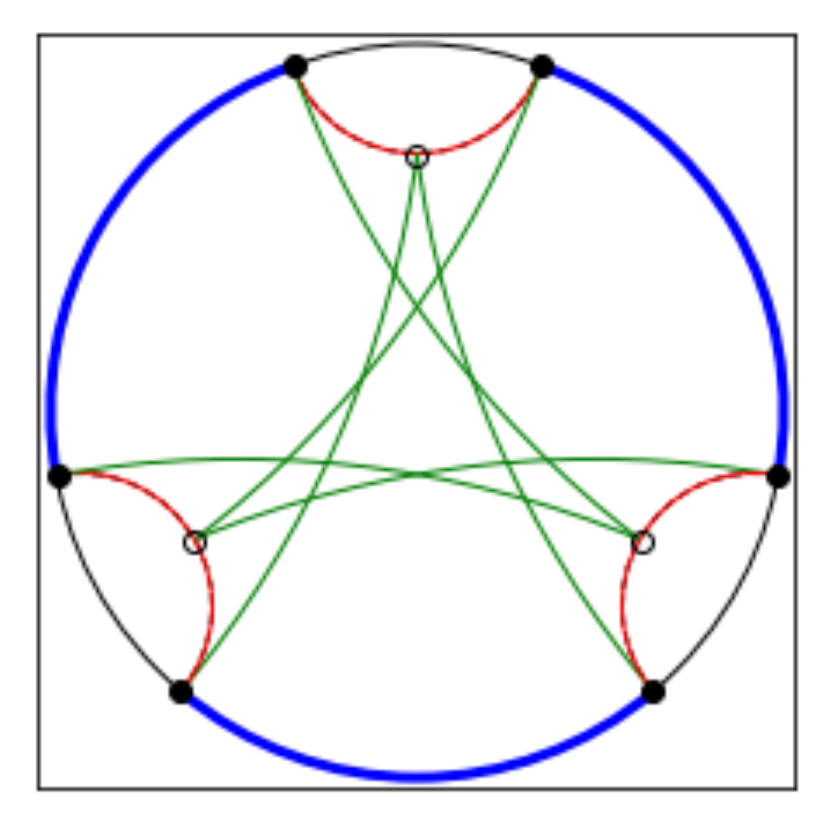}
\caption{Three equivalent pictures for correlation measure $f_{a_1}$ at $\theta = 1.05$ (left) and the unique picture for $\theta = 1.4$ (far right).}
\label{AxialSymFig1}
\end{center}
\end{figure}

Here we consider a set of correlation measures that, at least for some size boundary regions, present a subset of the full $S_3$ symmetry. In the cases of $f_{a_1}$ and $f_{a_2}$ we see a $Z_2$ symmetry under exchange of two boundary regions, where two of the optimization points move away from the centers of the arcs they are sitting on, though they move together in a symmetric fashion, preserving a reflection (``axial symmetry") along the axis through the center of the third boundary region. In the case of $f_{r_1}$, we see all three optimization points move along their arcs in the same rotational sense, preserving a ``rotational symmetry" that is the $A_3 \subset S_3$ alternating group of cyclic permutations of all three boundary regions. These measures also have regions of parameter space where they show the full $S_3$ symmetry.

\begin{figure} 
\begin{center}
\includegraphics[width=0.24\textwidth]{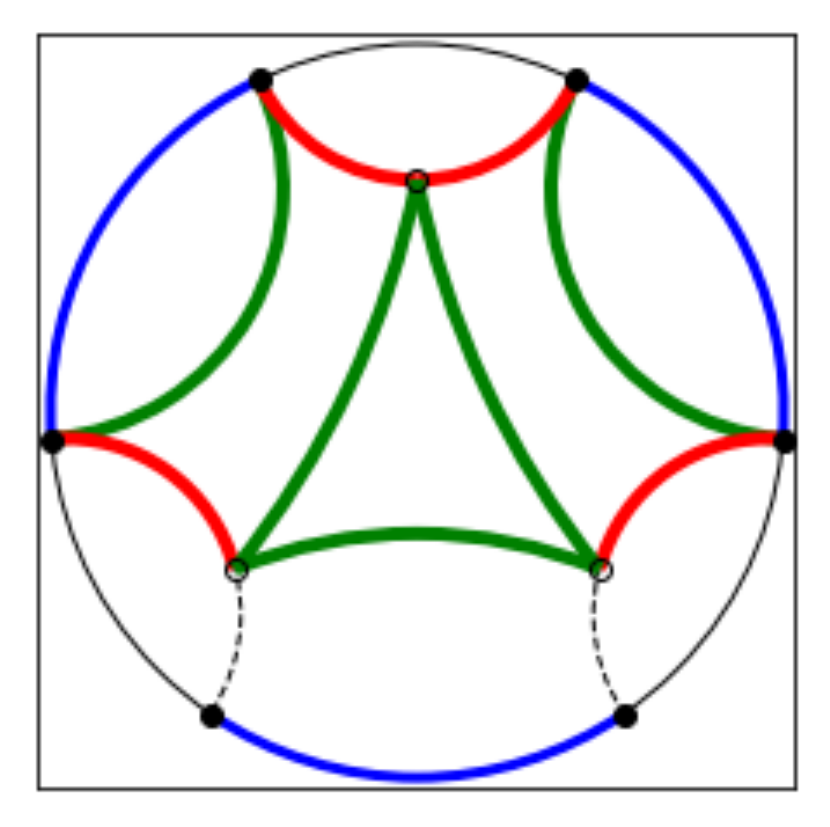}
\includegraphics[width=0.24\textwidth]{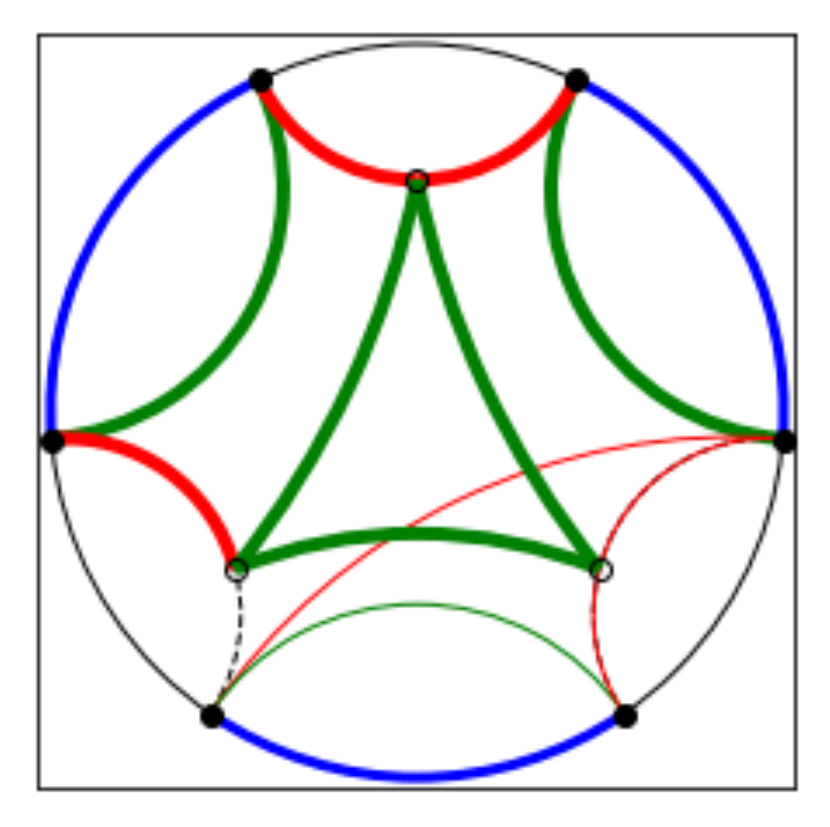}
\includegraphics[width=0.24\textwidth]{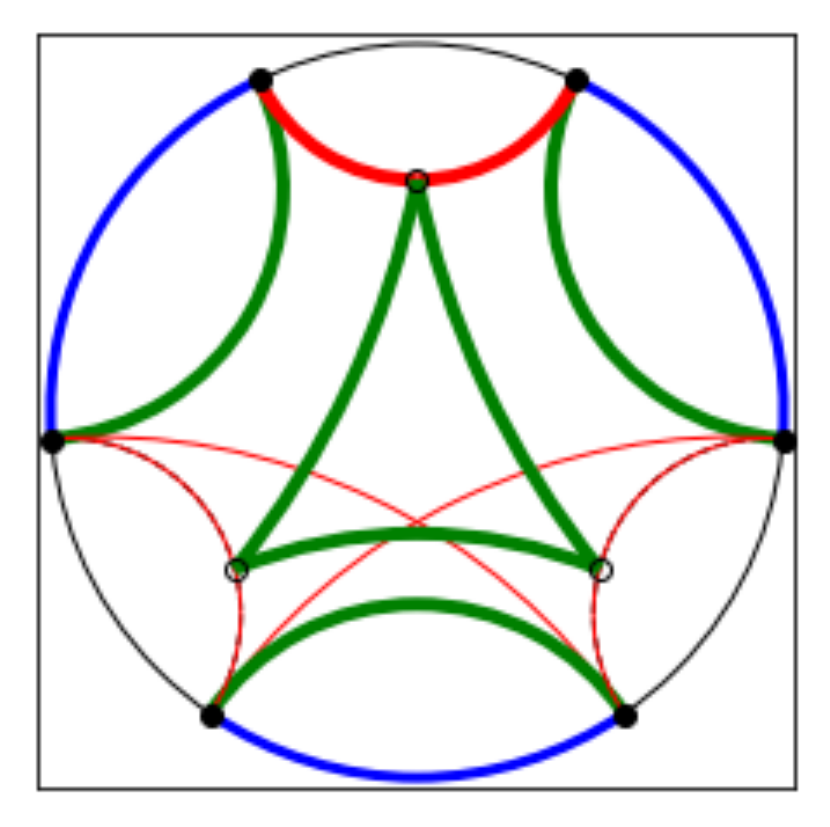}
\includegraphics[width=0.24\textwidth]{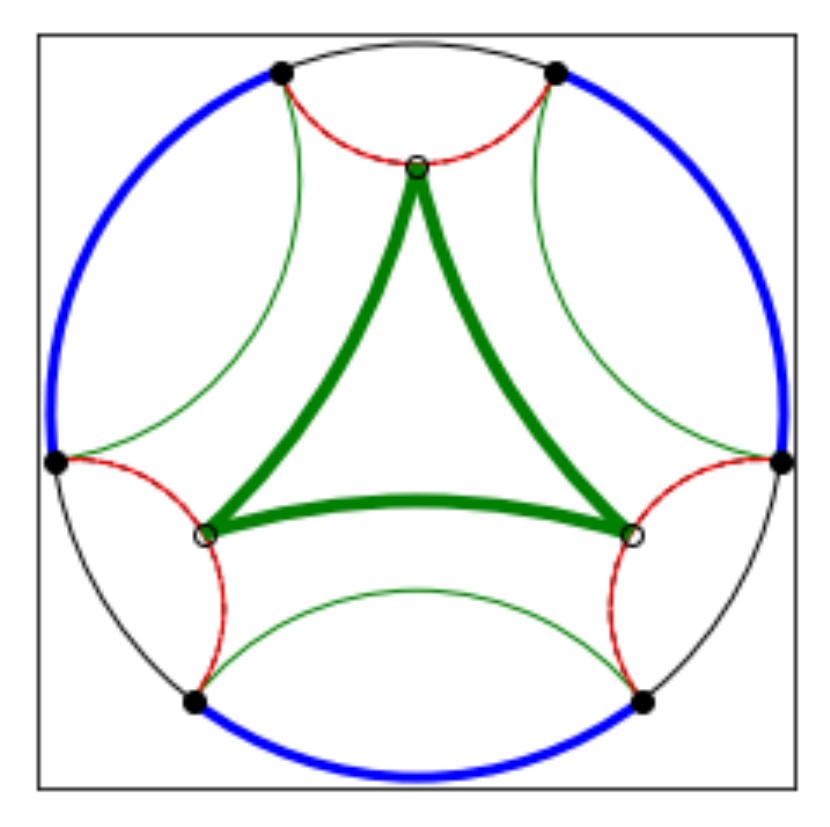}
\includegraphics[width=0.24\textwidth]{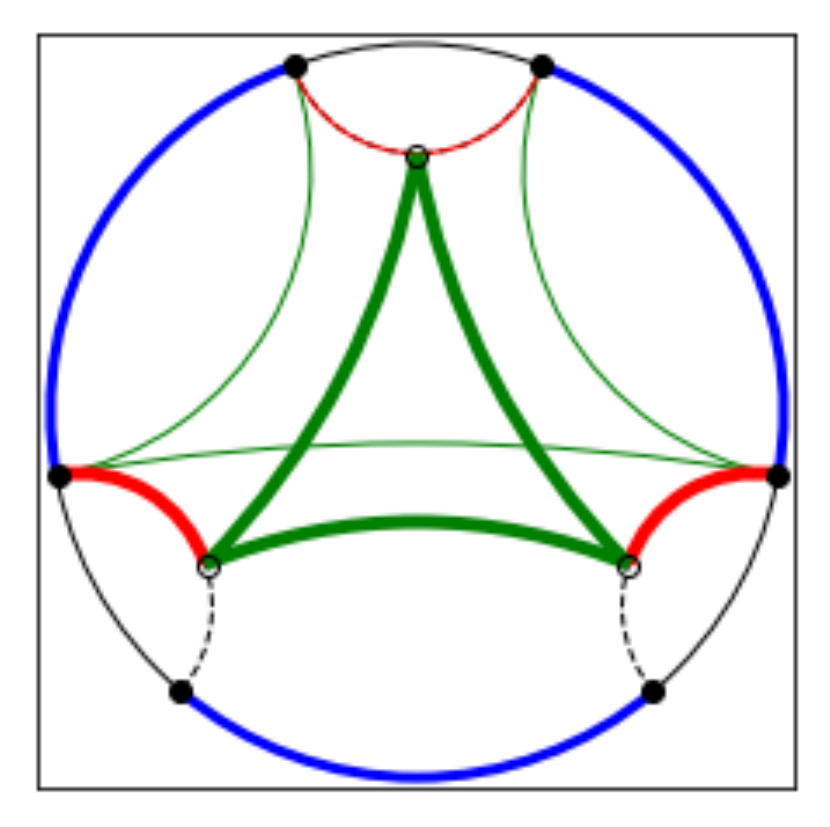}
\includegraphics[width=0.24\textwidth]{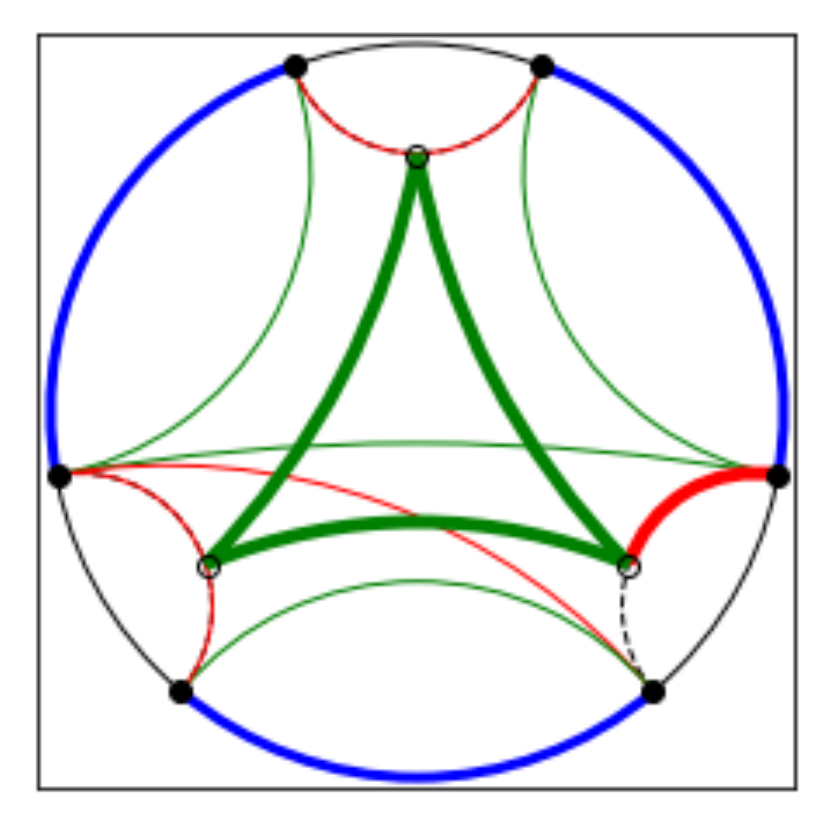}
\includegraphics[width=0.24\textwidth]{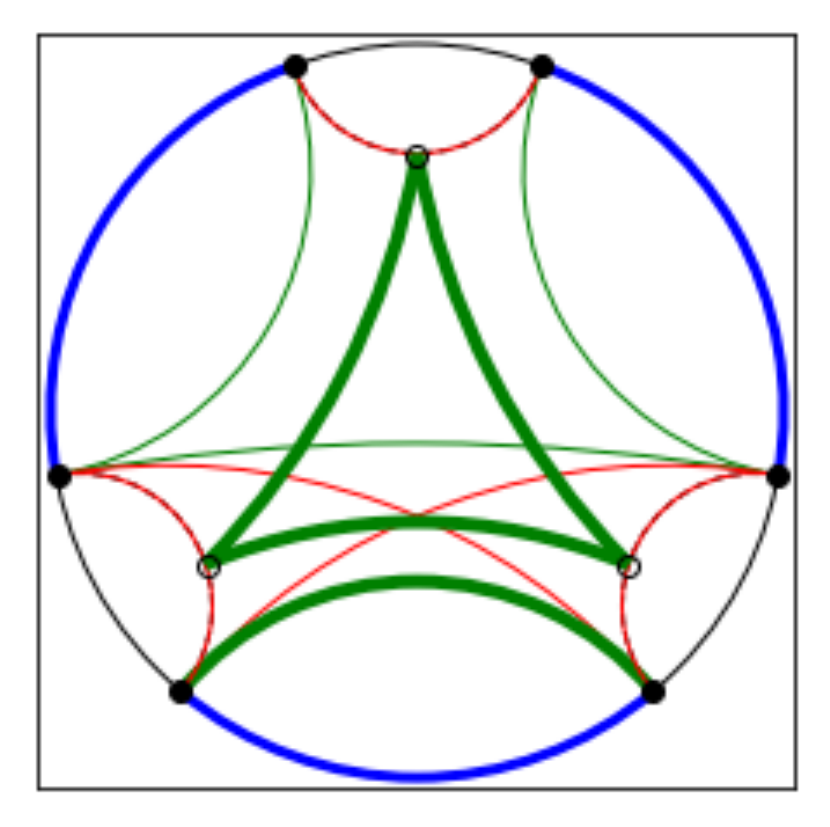}
\caption{Correlation measure $f_{a_2}$. First are three equivalent pictures for $\theta = 1.2$, followed by the unique picture for $\theta = 1.32$, and on the second line three equivalent pictures for $\theta = 1.4$.}
\label{AxialSymFig2}
\end{center}
\end{figure} 

\begin{figure} 
\begin{center}
\includegraphics[width=0.3\textwidth]{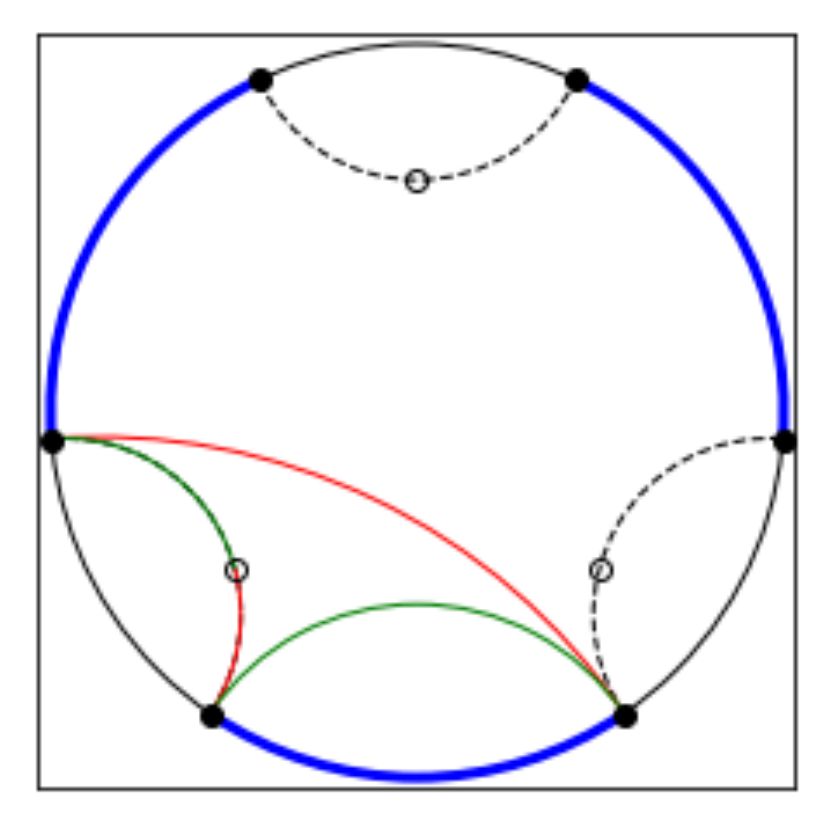}
\caption{The vanishing of the sum of these curves gives the curve optimization relation determining the axial optimization point.}
\label{AxialOptPointFig}
\end{center}
\end{figure}

For most values of $\theta$ except very small ones,  correlation measure $f_{a_1}$ has a fully $S_3$-symmetric presentation where the optimization points are at the centers of the entanglement wedge boundary segments, as in the last image in figure~\ref{AxialSymFig1}. This measure has  no transition at $\theta = \theta_I$; however, below $\theta < 1.0706$ two of the optimization points move off-center on the entanglement wedge boundary arcs, and the curves on either side of these points are weighted differently. For a fixed size of boundary regions three different pictures are possible, the first three images in figure~\ref{AxialSymFig1}, and the value of the correlation measure agrees for all three to many significant digits; this is a signature of the curve optimization relation defining the optimization point, as will see. The optimization points preserve the $Z_2$ reflection symmetry in all three cases. Which of the three symmetry axes is preserved is arbitrary.

The other correlation measure that shows reflection (axial) symmetry is $f_{a_2}$. It has three distinct regions of parameter space as $\theta$ is varied. For $\theta < 1.3173$, it displays a case with the optimization points in a reflection-symmetric position; again three different pictures appear, all calculating the same value for the correlation measure. For the narrow intermediate range of $1.3173< \theta< 1.3327$, the correlation measure jumps to a fully $S_3$-symmetric phase, which resembles the low-$\theta$ picture of $f_{s_1} = f^Q_3$, with different weights for the curves. Finally for $\theta > 1.3327$, the correlation measure returns to a reflection-symmetric configuration with two of the optimization points off-center, again with three numerically identical pictures, looking almost identical to the small-$\theta$ cases, but a little different. Note that the symmetric  case in the middle does not connect smoothly to the picture on either side of it; the optimization points jump from asymmetric to symmetric to asymmetric positions.

\begin{figure} 
\begin{center}
\includegraphics[width=0.24\textwidth]{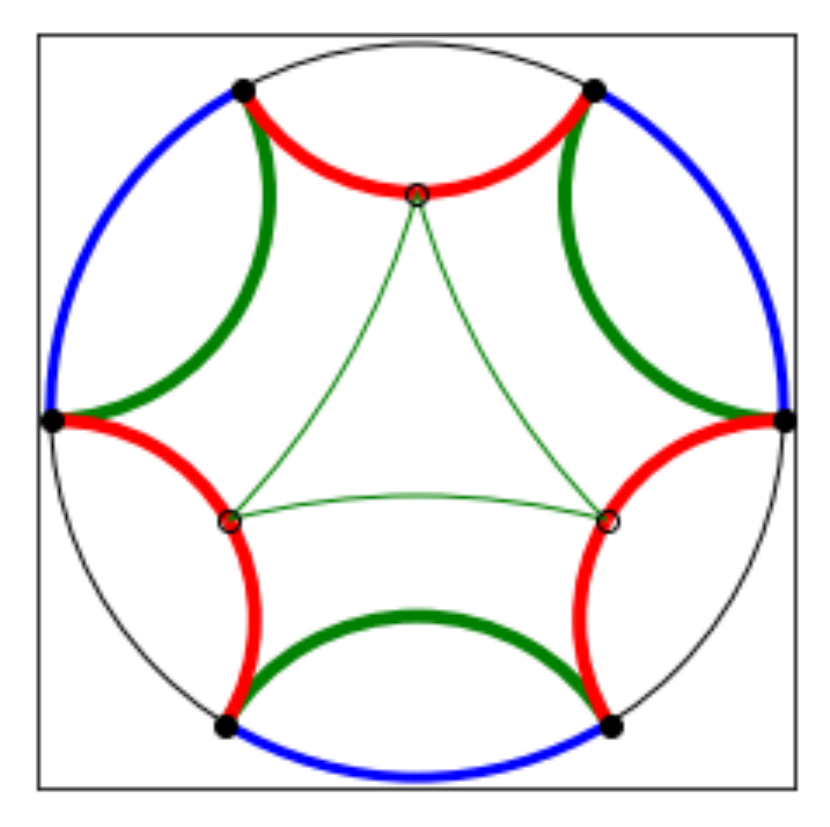}\break
\includegraphics[width=0.24\textwidth]{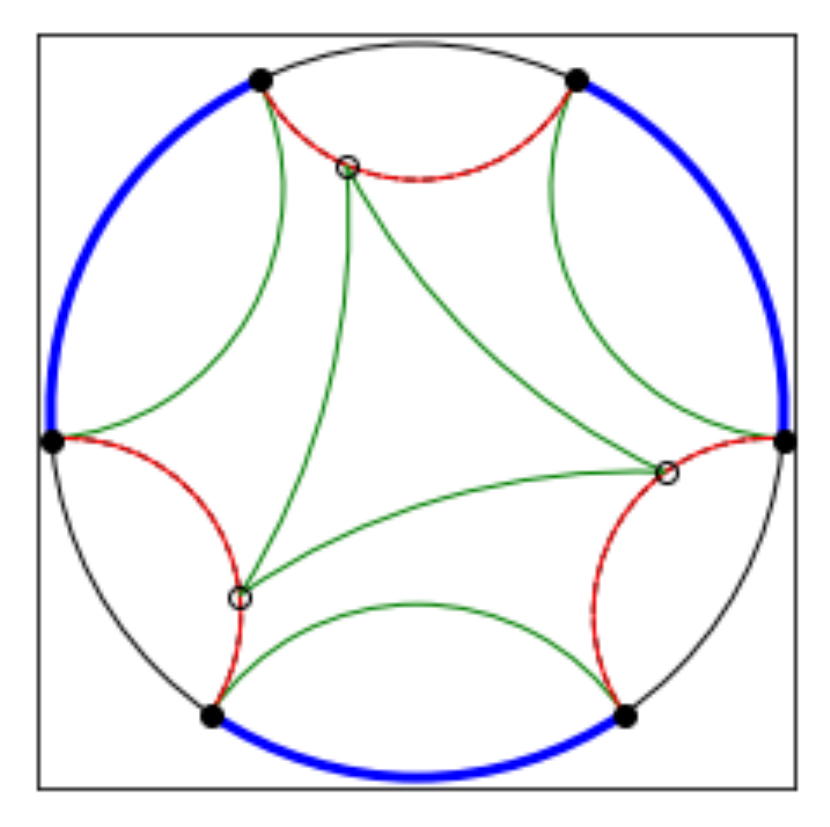}
\includegraphics[width=0.24\textwidth]{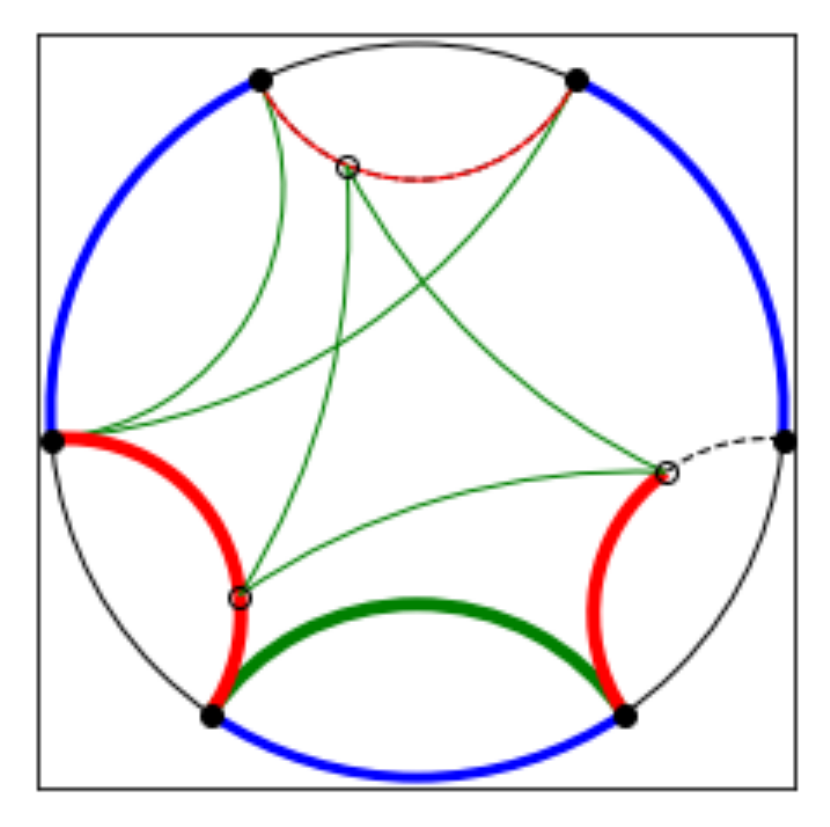}
\includegraphics[width=0.24\textwidth]{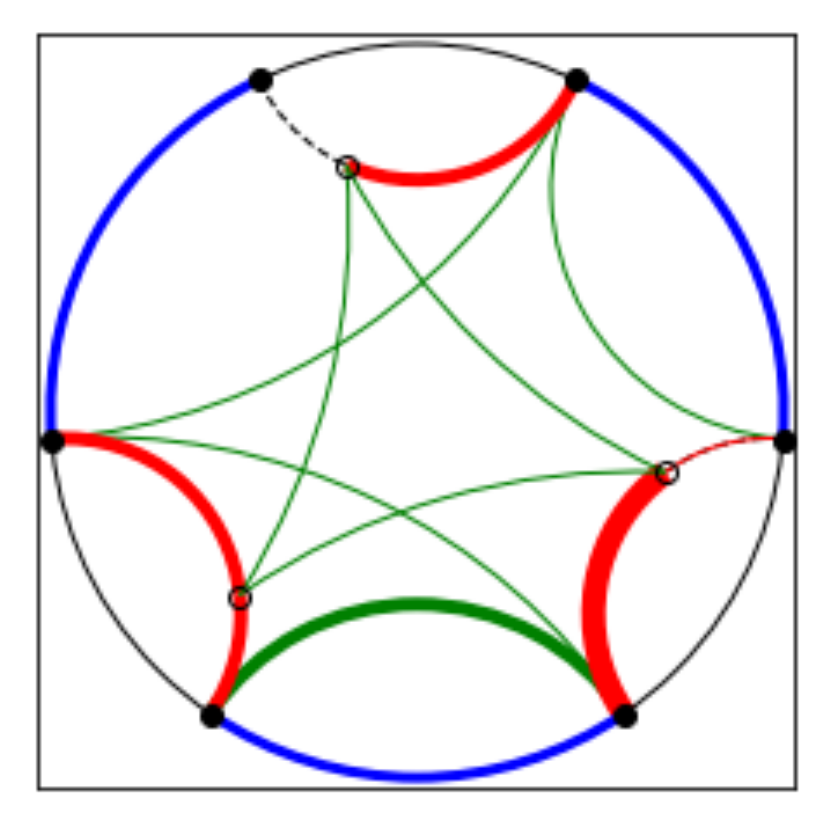}
\includegraphics[width=0.24\textwidth]{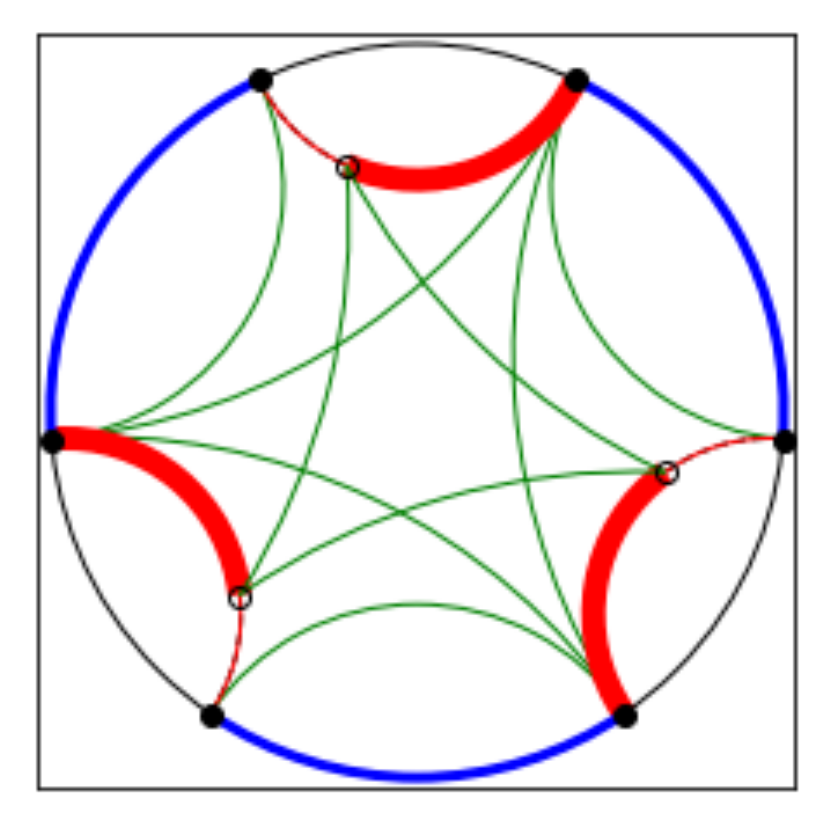}
\caption{Correlation measure $f_{r_1}$. On top, the unique picture at $\theta = 1.1$, identical to that for $f_{s_1}= f^Q_3$. On bottom, four equivalent pictures for $\theta = 1.2$.}
\label{RotationalSymFig}
\end{center}
\end{figure}

\begin{figure} 
\begin{center}
\includegraphics[width=0.3\textwidth]{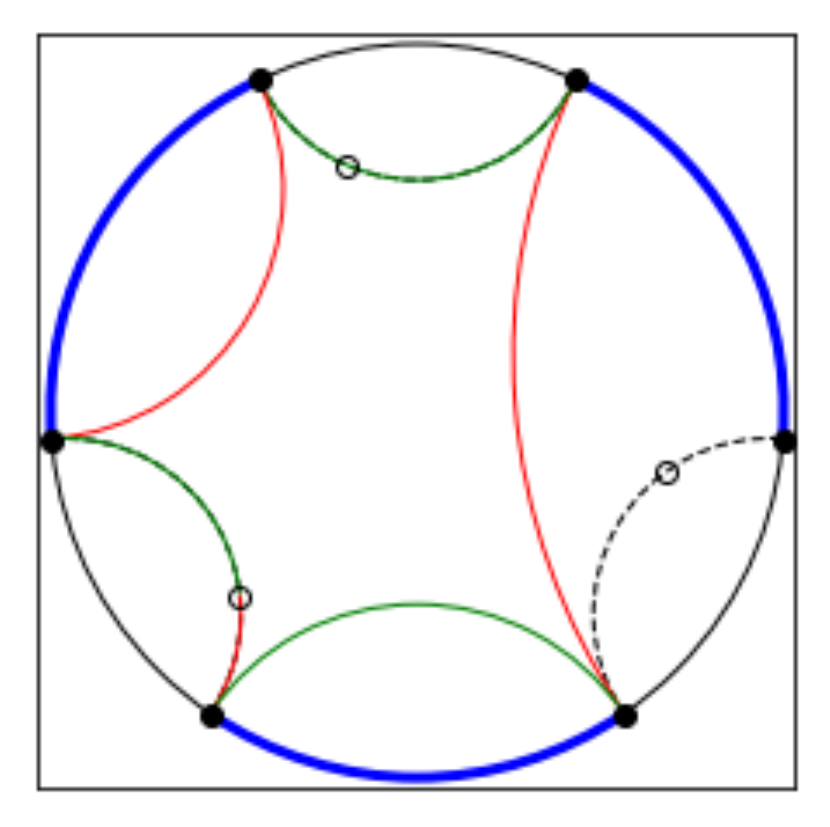}
\caption{The vanishing of the sum of these curves gives the curve optimization relation determining the rotational optimization point.}
\label{RotOptPointFig}
\end{center}
\end{figure}

The fact that for a number of examples, multiple different pictures give the same numerical value of the correlation measure is surprising at first. It is not an artifact of a special value of $\theta$, but persists over the range of $\theta$ for which the axial symmetry is present. For all such cases in measures $f_{a_1}$ and $f_{a_2}$ described above, the existence of the equivalent pictures is the same as the statement that the curve configuration in figure~\ref{AxialOptPointFig} is zero. Because the lengths of the curves in figure~\ref{AxialOptPointFig} depend on the location of the optimization point, the vanishing of their sum is equivalent to a constraint on the location of the optimization point; this is a  curve optimization relation.

One can see that were the lower-left optimization point to move up to the middle of its arc, the net sign of the curves would be negative (since the curves on either size of the optimization point then cancel and the remaining red curve is longer than the green), while if it moved down to the boundary, the net sign would be positive (since the red curve is a shorter path between its two endpoints than the sum of the green curves). Thus there is a point in the middle where the sum of the curves is zero, and the optimization of these measures requires the optimization point to find it. We call this the {\it axial optimization point}.

Next we have correlation measure $f_{r_1}$, showing rotational symmetry. 
This measure is $S_3$-symmetric and identical to $f_{s_1} = f^Q_3$ for $\theta < 1.1746$, as shown on top in figure~\ref{RotationalSymFig}. However, for $\theta > 1.1746$ the optimization points  move cyclically off-center in a rotationally symmetric way. We again find multiple pictures that calculate the correlation measure equally well. This fact indicates that each optimization must be sitting at a location, the {\it rotational optimization point}, where the sum of the curves in figure~\ref{RotOptPointFig} vanishes. One can see this is a different curve optimization relation than the one for the axial point, since many of the curves are the same, but there is one additional curve of each sign that are not of the same length; thus the rotational optimization point needs to move further off-center relative to the axial one to balance this out.

\begin{figure} 
\begin{center}
\includegraphics[width=0.24\textwidth]{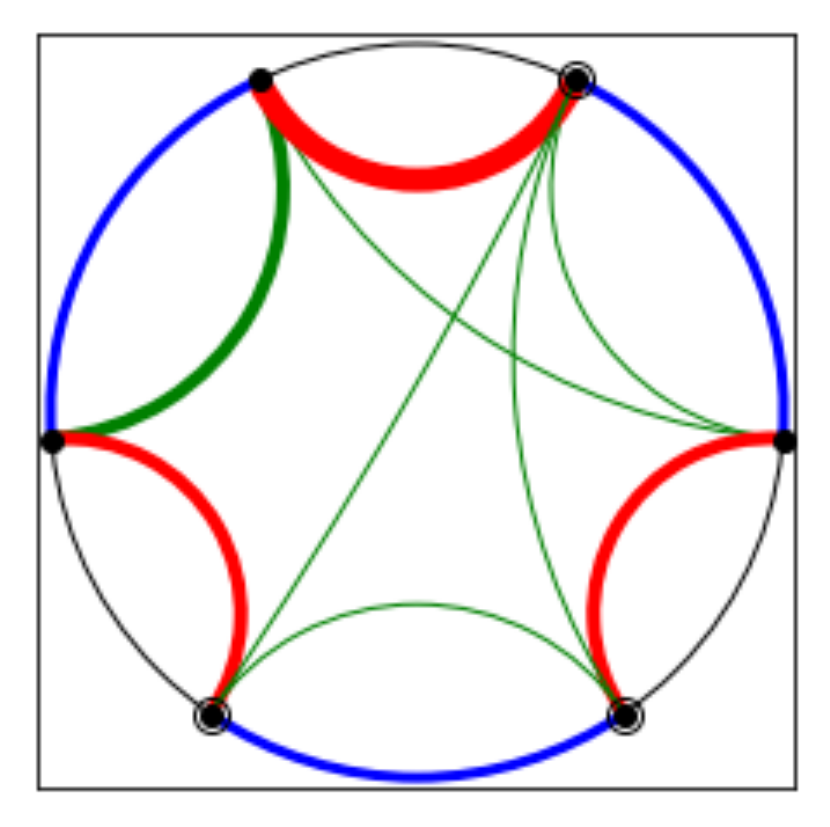}
\includegraphics[width=0.24\textwidth]{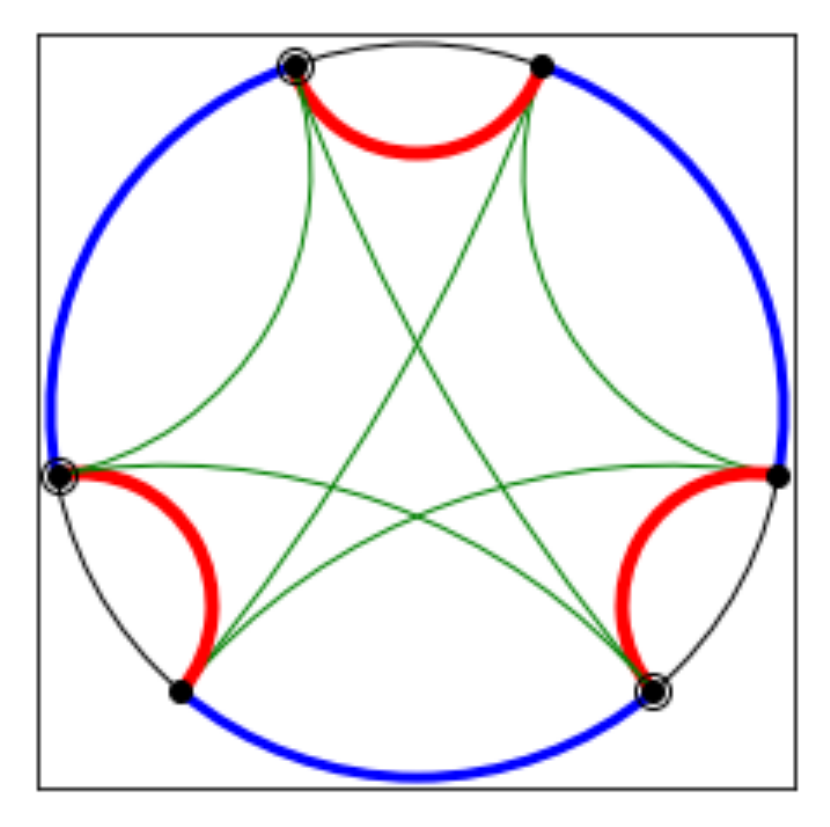}
\includegraphics[width=0.24\textwidth]{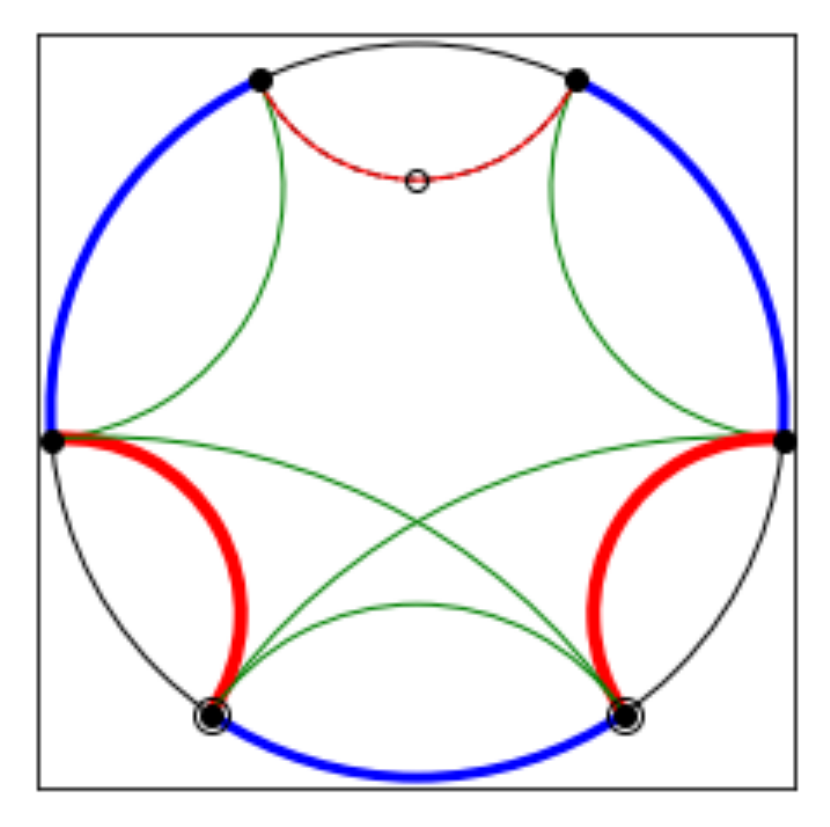}
\includegraphics[width=0.24\textwidth]{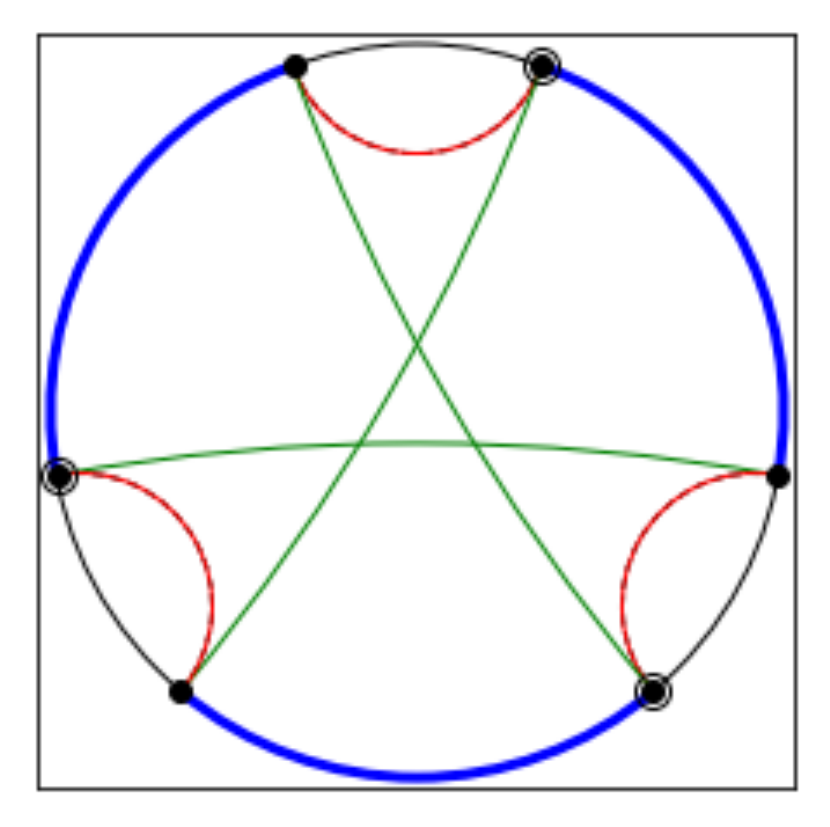}
\caption{Correlation measures $f_{b_1} = \tilde{f}^{\rm sq}_3$ for $\theta = 1.2$ and $\theta = 1.4$ (left pictures) and $f_{b_2}$ for $\theta = 1.2$ and $\theta = 1.4$ (right pictures).}
\label{BoundaryPointsFig}
\end{center}
\end{figure}

\subsubsection{Correlation measures with optimization points at the boundary}

Our last trio of correlation measures shows the phenomenon of some or all of the optimization points move all the way to the boundary. Some of these correlation measures have configurations that manifest no symmetry for the optimization points. When an optimization point coincides with a boundary point, we draw it as an open circle surrounding the black dot of the boundary point.

The first of these measures is $f_{b_1}$, equal to the tripartite version $\tilde{f}^{\rm sq}_3$ of the symmetric-ancilla family generalizing the squashed entanglement we defined in (\ref{TildeSquashed}). Like the two-party squashed entanglement, the minimization sets one of the three sets of ancilla, $a$, $b$ or $c$, to zero size. One of the remaining ancilla is then twice as big as the other; this results in a configuration of optimization points that does not preserve any of the $S_3$ symmetry of the boundary regions. This measure has a minor transition at $\theta = \theta_I$, where the optimization points keep the same pattern but the curves change slightly; the two pictures are shown on the left of figure~\ref{BoundaryPointsFig}.

\begin{figure} 
\begin{center}
\includegraphics[width=0.24\textwidth]{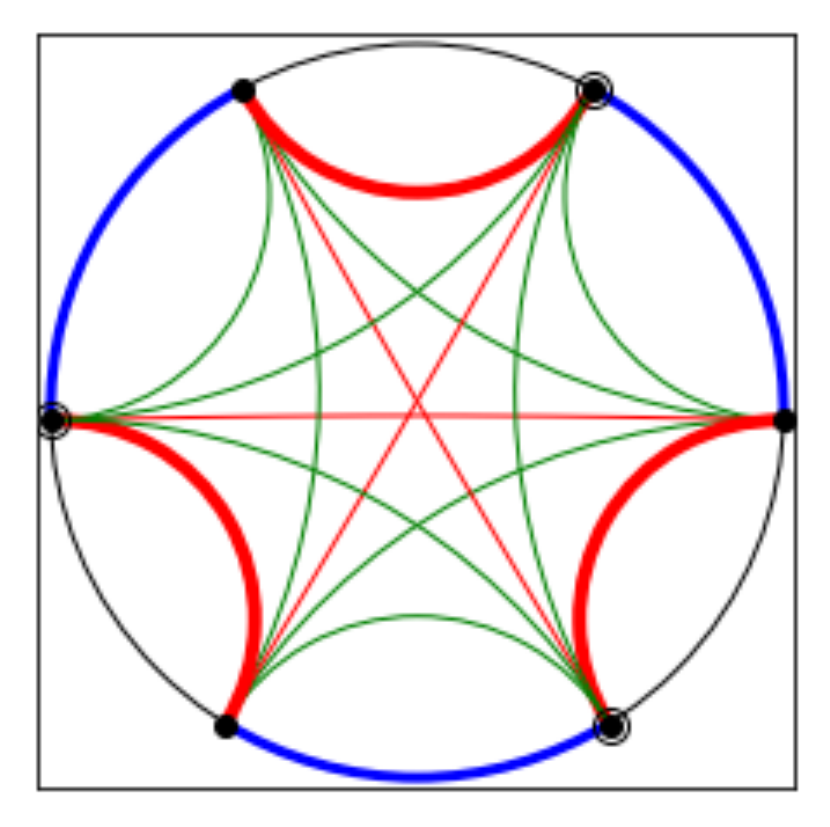}
\includegraphics[width=0.24\textwidth]{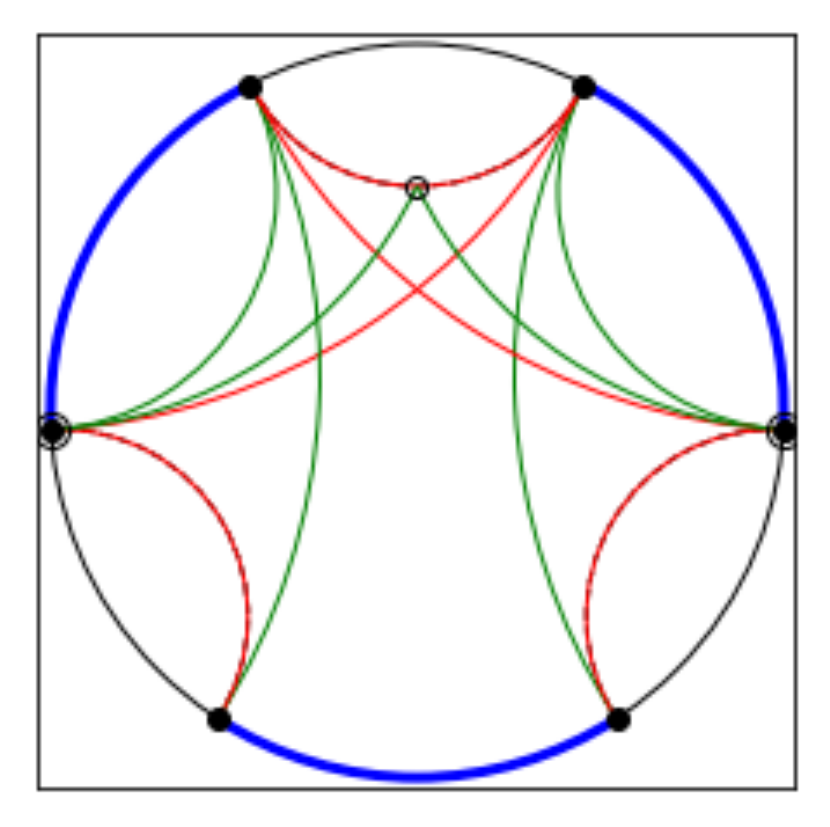}
\includegraphics[width=0.24\textwidth]{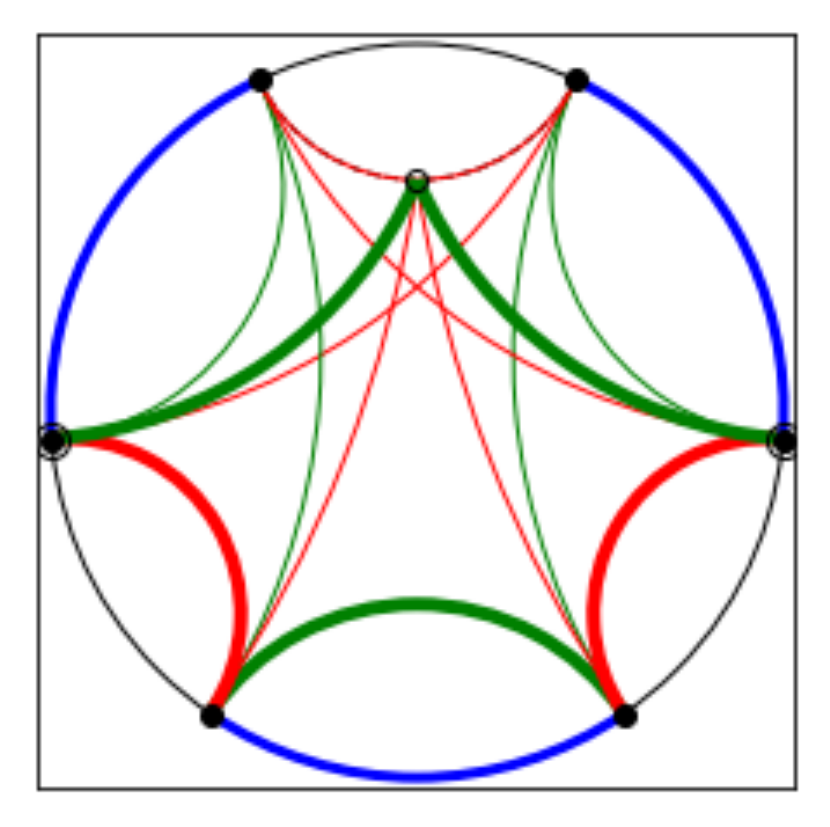}\break
\includegraphics[width=0.24\textwidth]{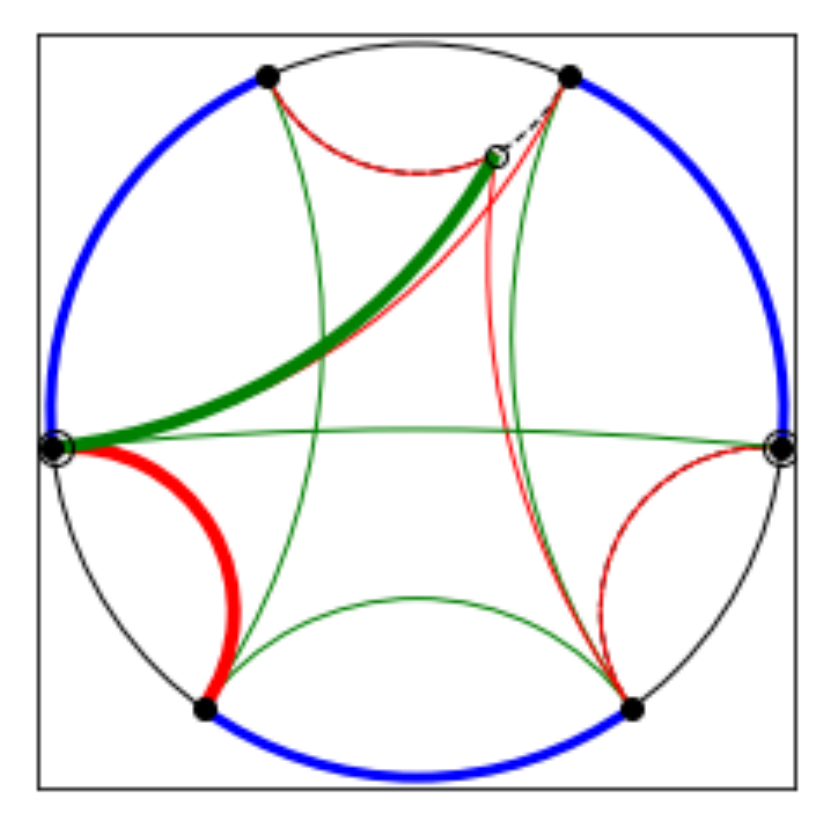}
\includegraphics[width=0.24\textwidth]{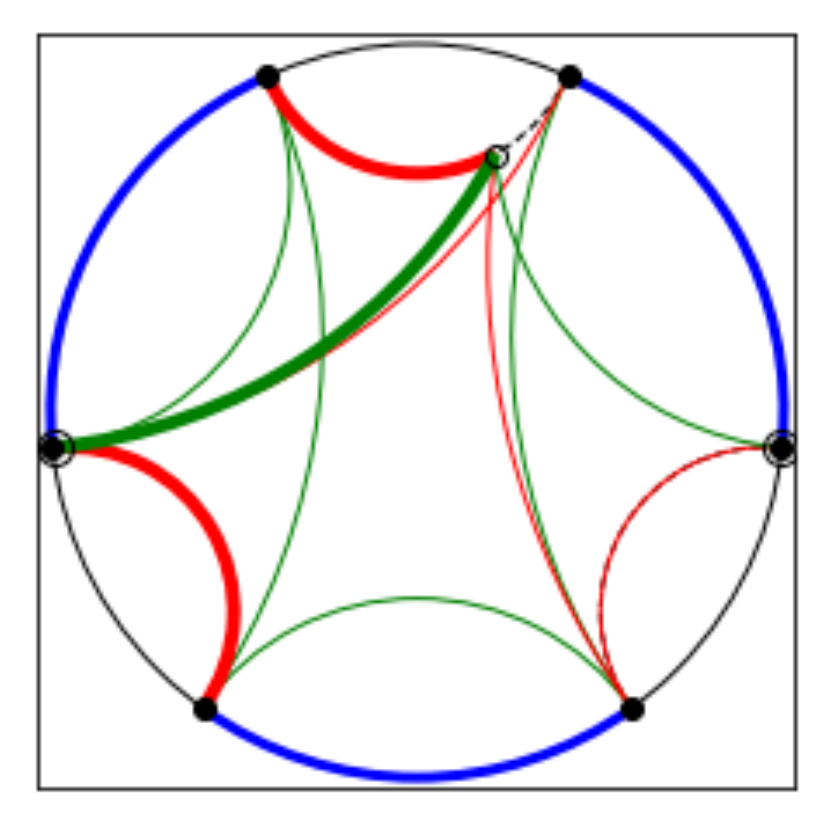}
\includegraphics[width=0.24\textwidth]{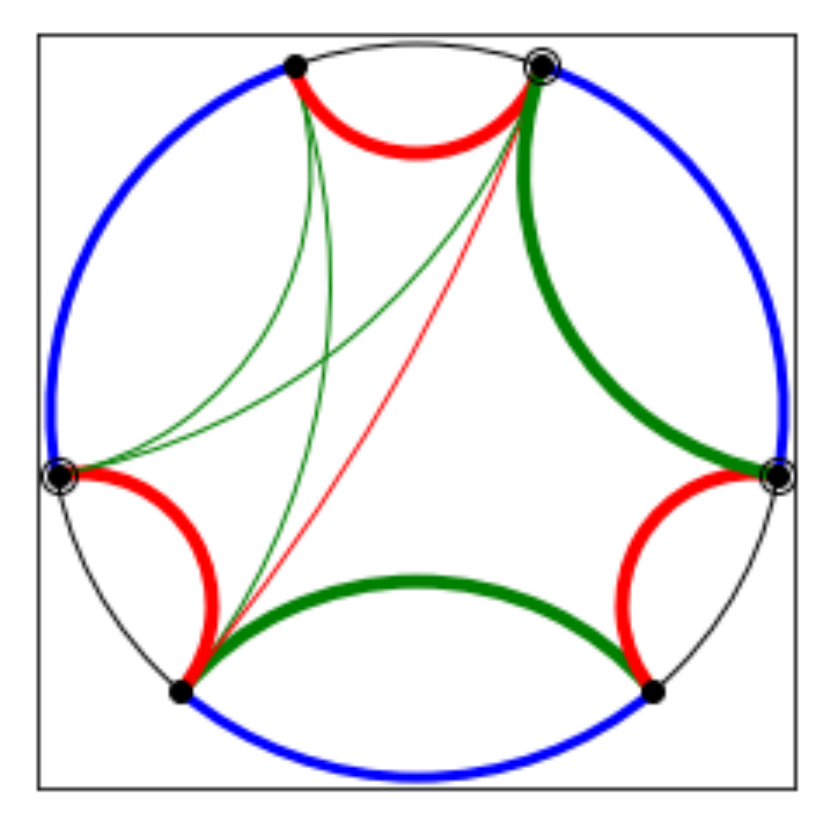}
\caption{Correlation measure $f_{b_3}$ for $\theta = 1.1$, $\theta = 1.15$ and $\theta = 1.2$ (left to right, top line), two equivalent pictures for  $\theta = 1.25$, and finally the unique picture for $\theta = 1.4$ (left to right, bottom line) . }
\label{BoundaryPointsFig2}
\end{center}
\end{figure}

\begin{figure} 
\begin{center}
\includegraphics[width=0.3\textwidth]{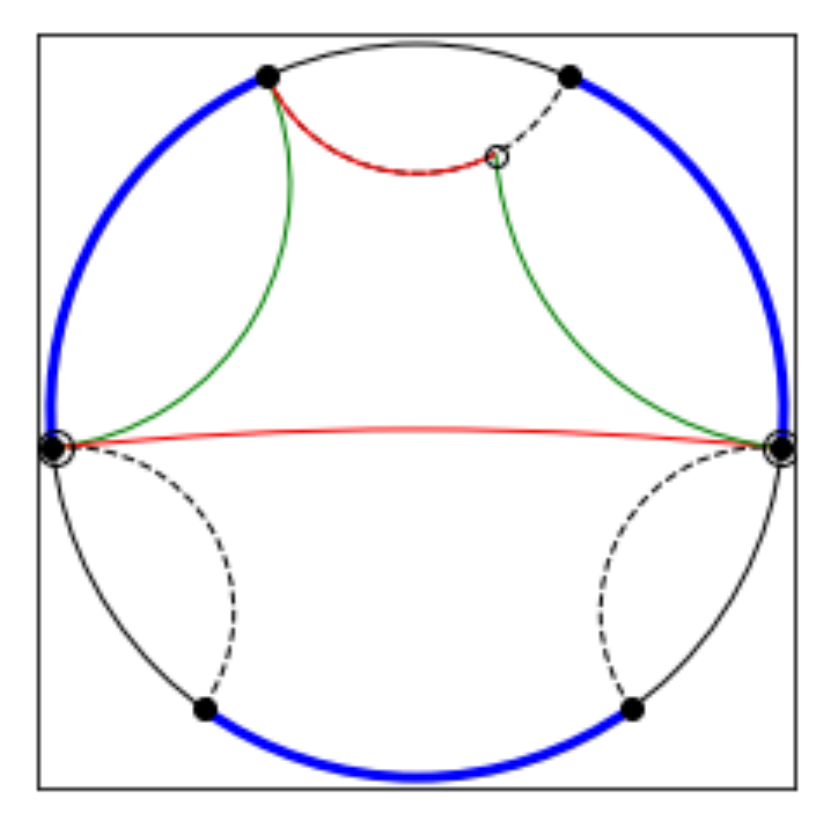}
\caption{The vanishing of the sum of these curves gives the curve optimization relation determining the no symmetry optimization point.}
\label{NoSymOptPointFig}
\end{center}
\end{figure}

Correlation measure $f_{b_2}$ also has configurations where optimization points go to the boundary, but unlike the previous case it does so in a way that preserves some symmetry (see the right two pictures in figure~\ref{BoundaryPointsFig}). For $\theta < 1.2715$ it preserves an axial symmetry, with one optimization point staying in the center of its arc while the other two go to the boundary in a symmetric fashion; thus here one of the ancilla $a$, $b$ or $c$ vanishes, while the other two are equal size.

For $\theta > 1.2715$, $f_{b_2}$ switches to a configuration where all three optimization points go to the boundary in a symmetric fashion, preserving $A_3$ symmetry, and all three ancilla are the same size, each taking one complete arc of the boundary of the entanglement wedge. The curves end up being identical to those for $J(A:B:C)$, as in the last picture in figure~\ref{OptIndepFig}, and hence the two measures are numerically identical for $\theta > \theta_I$, while the difference is that for $f_{b_2}$ the optimization points are constrained to the boundary, while $J(A:B:C)$ is independent of the optimization. Thus the relationship between $f_{b_2}$ and $J(A:B:C)$ for $\theta > \theta_I$ is analogous to the relationship of the bipartite squashed entanglement to $I(A:B)$, where again they have the same set of curves and they numerically coincide, but in the former case the optimization points are fixed at the boundary to get there, while the latter case is optimization-independent.

Finally, correlation measure $f_{b_3}$ has the most varied progression of any of the measures, moving through a number of different configurations with optimization points that can preserve axial symmetry, rotational symmetry or no symmetry, with some or all optimization points at the boundary; these are shown in figure~\ref{BoundaryPointsFig2}. For $\theta < 1.1233$, the optimization points are at the boundary in a sense preserving rotational symmetry, so all three ancilla $a$, $b$ and $c$ are of the same size (each being one arc of the entanglement wedge boundary), as with the large-$\theta$ behavior of $f_{b_2}$.  Then for  $1.1233 < \theta < 1.1892$, the correlation measure changes to a presentation where one optimization point is in the center of its arc, while the other two are at the boundary, in a way that preserves axial symmetry; for the region $1.1892< \theta< 1.2249$, this symmetry pattern is preserved but the presentation of the curves changes slightly. Next, at $\theta = 1.2249$ the optimization point in the middle of its arc jumps a small ways to one side, and as $\theta$ is increased, it migrates closer to the boundary; all $S_3$ symmetry is now broken. As $\theta$ approaches $\theta = 1.3182$, the optimization point gets closer to the boundary, and then for $\theta > 1.3182$ it appears on the boundary, and the shape of the curves change to the final form what persists for $\theta > 1.3182$, which again has no symmetry.

Once again, when the optimization point moves to a finite location off-center, we find multiple equivalent presentations, here the left and middle pictures on the lower row of figure~\ref{BoundaryPointsFig2}. Their numerical equivalence implies the vanishing of the sum of the curves in figure~\ref{NoSymOptPointFig}, a curve optimization relation determining this {\it no symmetry optimization point}.

\subsubsection{Summary of optimization points}

\begin{figure} 
\begin{center}
\includegraphics[width=0.4\textwidth]{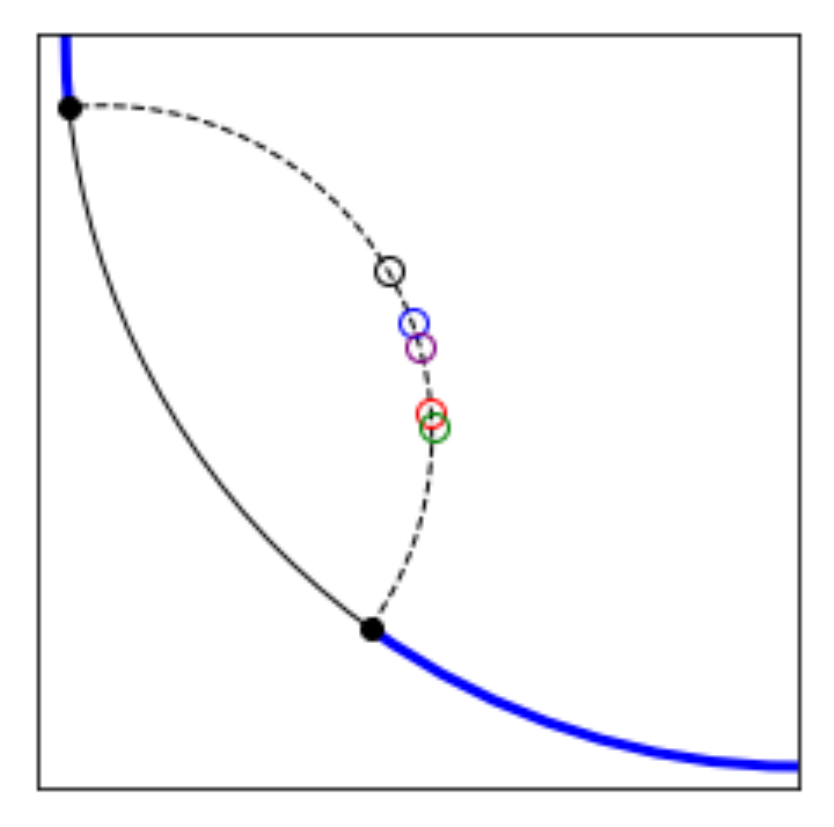}\break
\includegraphics[width=0.19\textwidth]{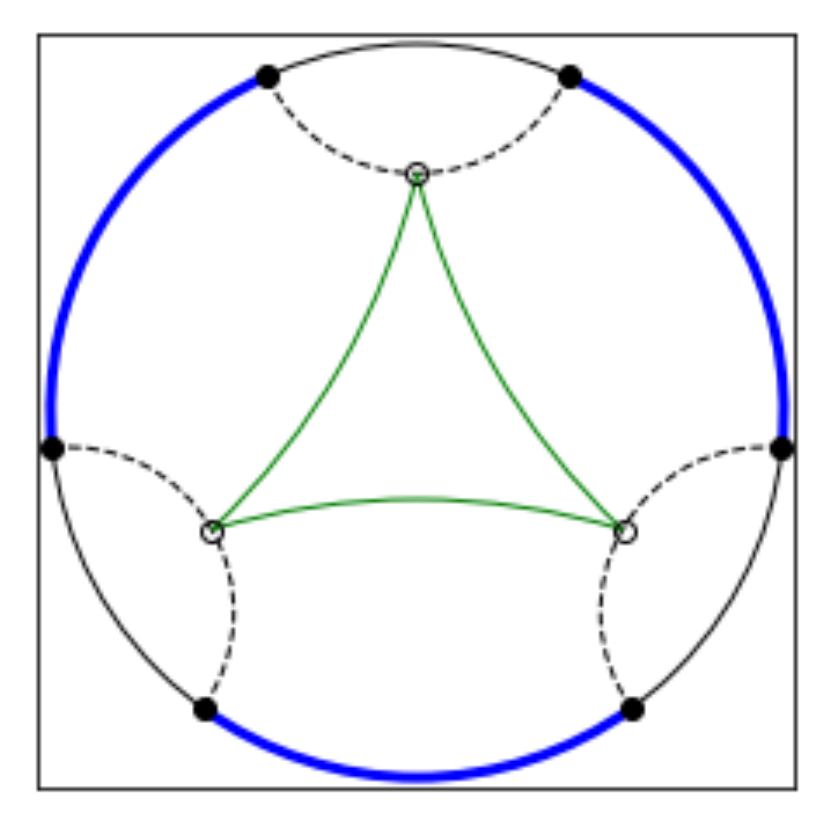}
\includegraphics[width=0.19\textwidth]{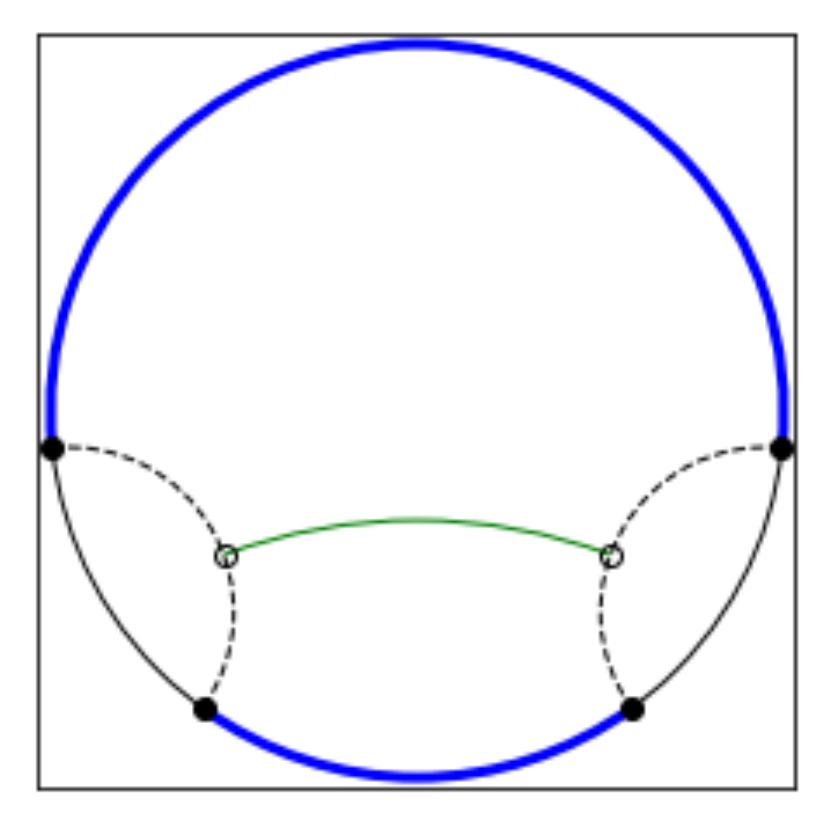}
\includegraphics[width=0.19\textwidth]{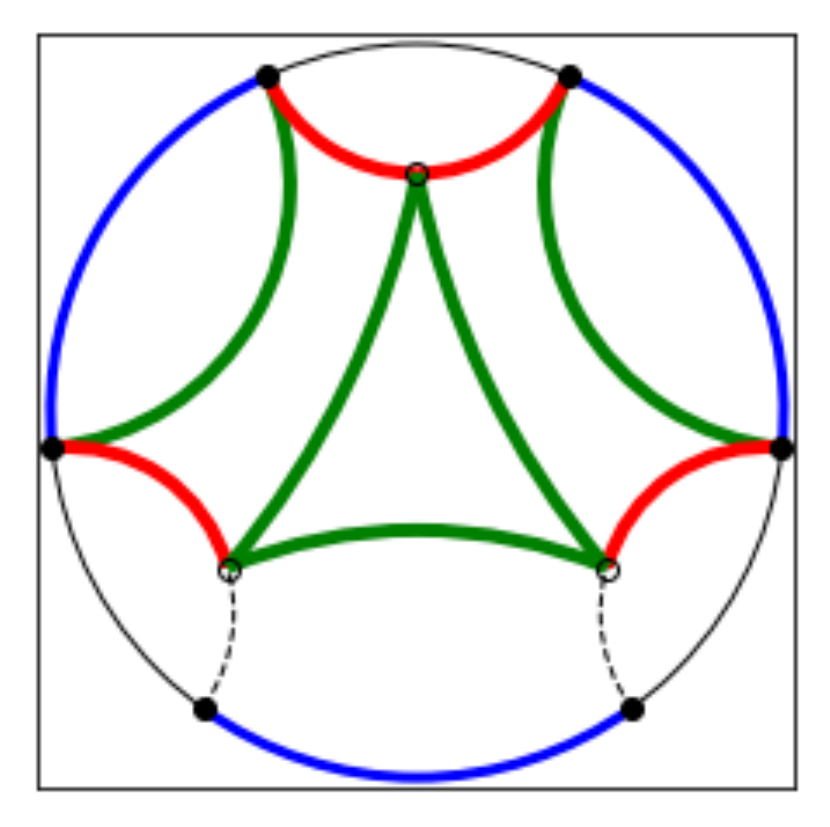}
\includegraphics[width=0.19\textwidth]{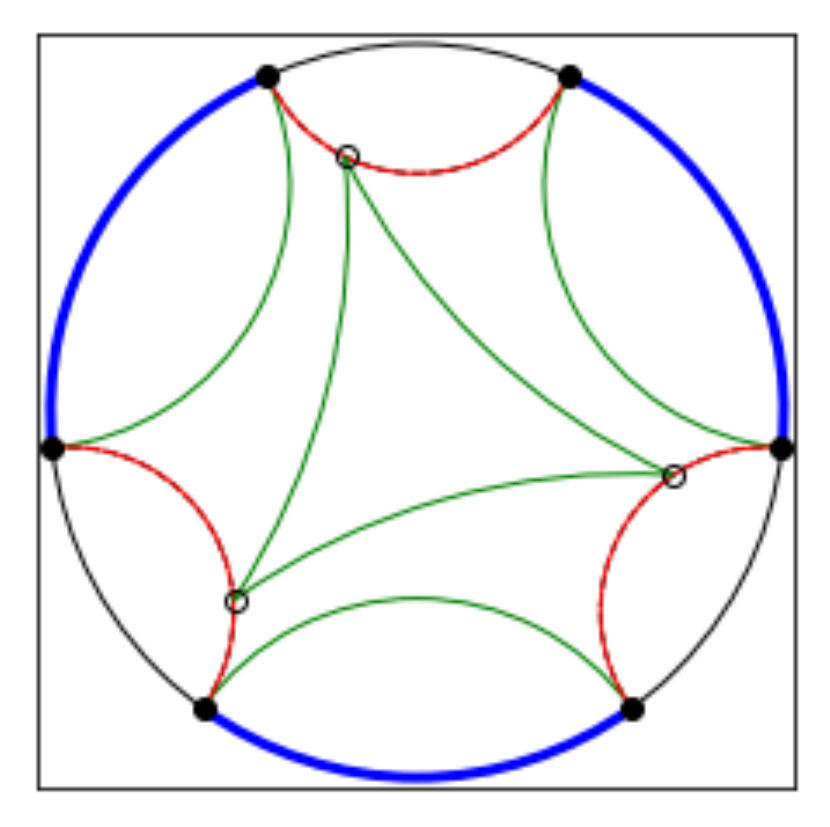}
\includegraphics[width=0.19\textwidth]{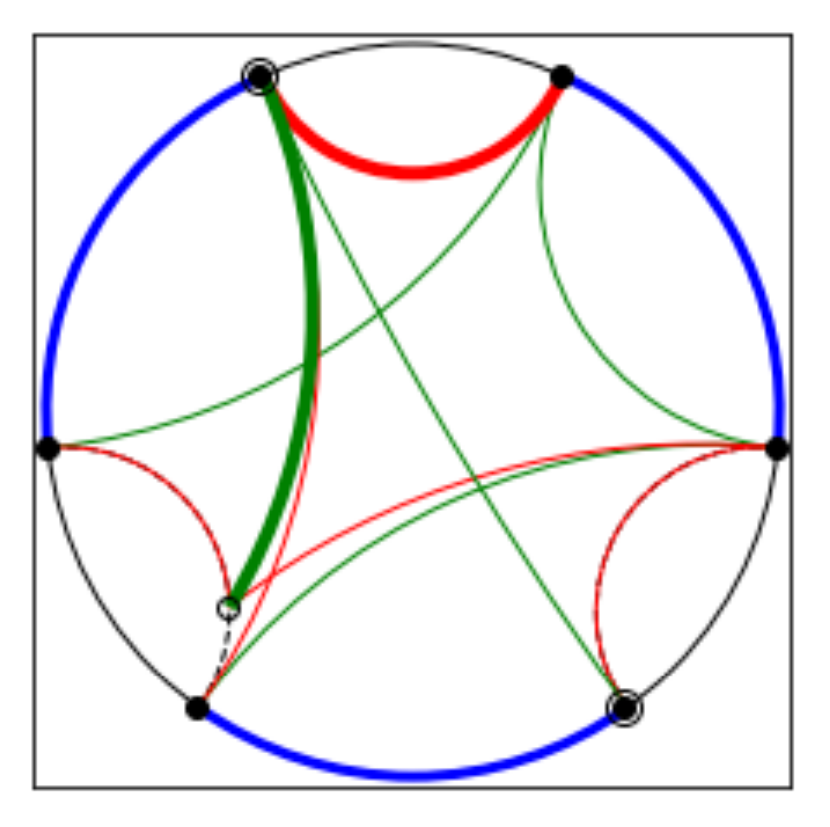}
\caption{Top, five optimization points at $\theta = 1.25$, top-to-bottom the tripartite entanglement wedge cross-section point (black), the bipartite entanglement wedge cross-section point (blue), the axial optimization point (purple), the rotational optimization point (red), and the no symmetry optimization point (green). Bottom, the five pictures these points come from.}
\label{AllOptPointsFig}
\end{center}
\end{figure}

We have seen optimization points not pushed all the way to the boundary at four different locations: the entanglement wedge cross section point, the axial optimization point, the rotational optimization point, and the no symmetry optimization point. Each obeys a different curve optimization relation. The actual location of each depends on the size of the boundary regions. It is interesting to plot them all for the same size boundary regions, which we do in figure~\ref{AllOptPointsFig}. We also plot the bipartite entanglement wedge cross-section point that results when we put regions $A$ and $B$ and the space in between them together as a single region. None of these points are the same.

\subsection{Correlation measures with different-size boundary regions}

We will not thoroughly investigate the whole three-dimensional parameter space of varying boundary region sizes, but we will give a few examples and some discussion.

\begin{figure} 
\begin{center}
\includegraphics[width=0.2\textwidth]{1_1.2.pdf}
\includegraphics[width=0.2\textwidth]{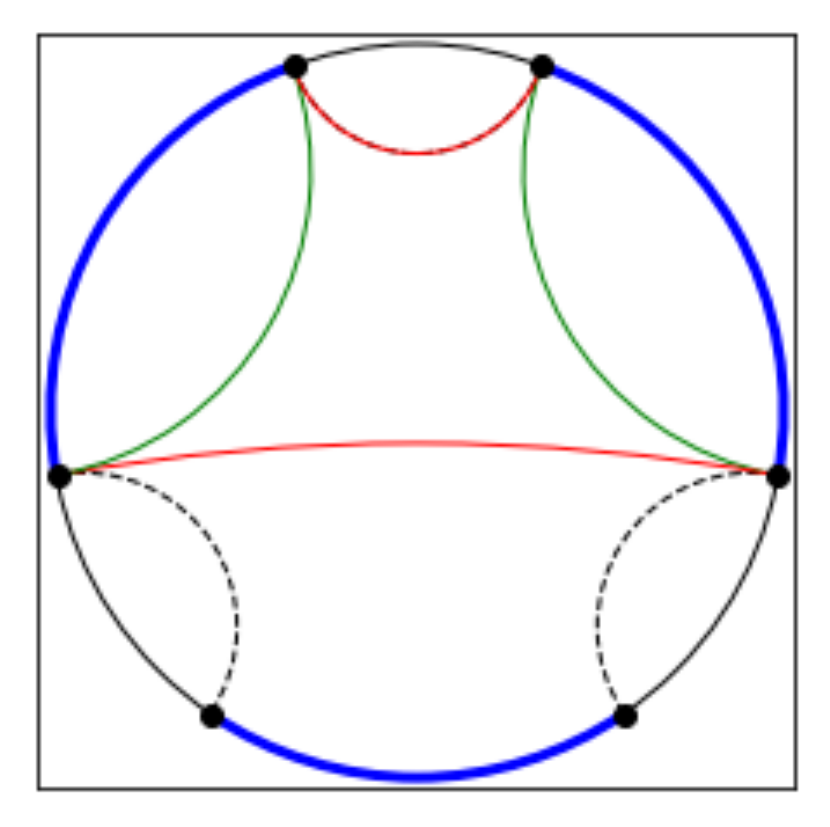}
\includegraphics[width=0.2\textwidth]{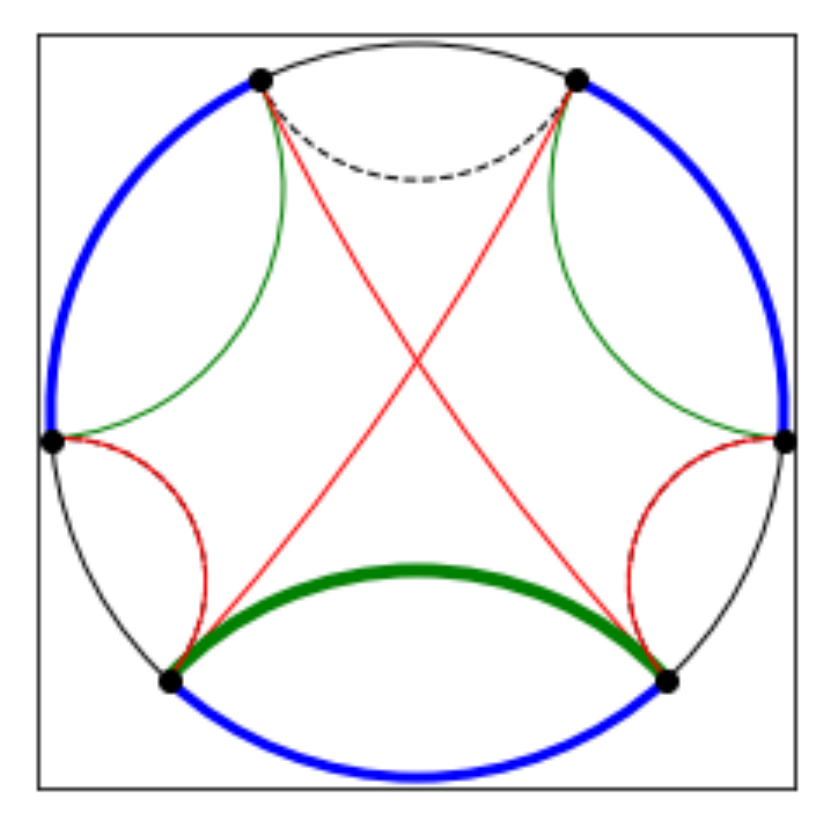}
\includegraphics[width=0.2\textwidth]{1_1.4.pdf}
\includegraphics[width=0.2\textwidth]{3_1.2.pdf}
\includegraphics[width=0.2\textwidth]{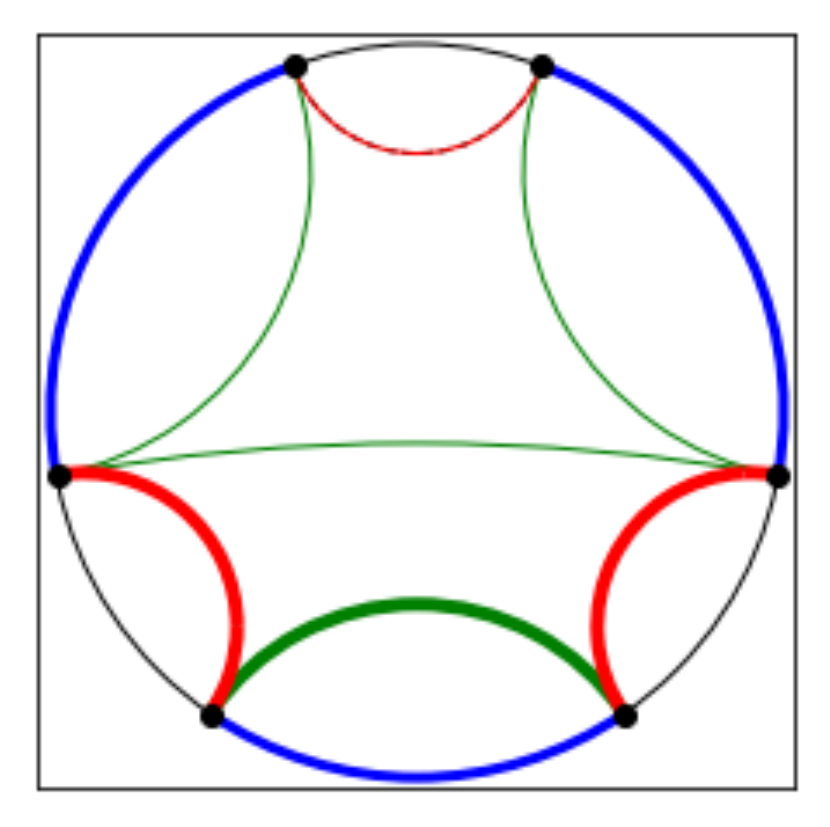}
\includegraphics[width=0.2\textwidth]{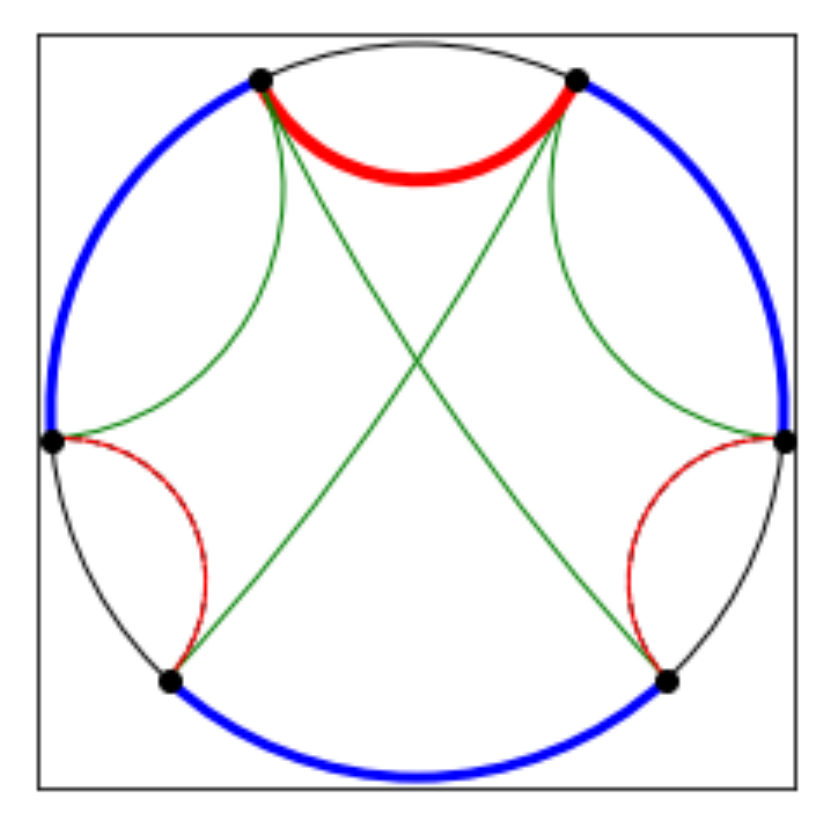}
\includegraphics[width=0.2\textwidth]{3_1.4.pdf}
\caption{The measures $I(A:B) + I(A:C) + I(B:C)$ and $J(A:B:C)$ (top and bottom rows respectively) for $\theta_A = \theta_B = \theta_C = 1.2$ (no nonzero pairwise mutual informations), $\theta_A = 1.4$, $\theta_B =1.4$,  $\theta_C = 1.2$ (one nonzero pairwise mutual information), $\theta_A =1.2$,  $\theta_B =1.2$,  $\theta_C = 1.5$ (two nonzero pairwise mutual informations), and $\theta_A = \theta_B = \theta_C = 1.4$ (three nonzero pairwise mutual informations).}
\label{AsymNonOptimizedFig}
\end{center}
\end{figure}

When all three boundary regions were the same size, we either had all three pairwise mutual informations $I(A:B)$, $I(A:C)$ and $I(B:C)$ zero, or all three nonzero. In general it is possible to have zero, one, two or three of them nonvanishing. The sum $I(A:B) + I(A:C) + I(B:C)$ and its fellow non-optimized measure the dual total correlation $J(A:B:C)$ both go through different presentations for each case, as shown in figure~\ref{AsymNonOptimizedFig}.

\begin{figure} 
\begin{center}
\includegraphics[width=0.24\textwidth]{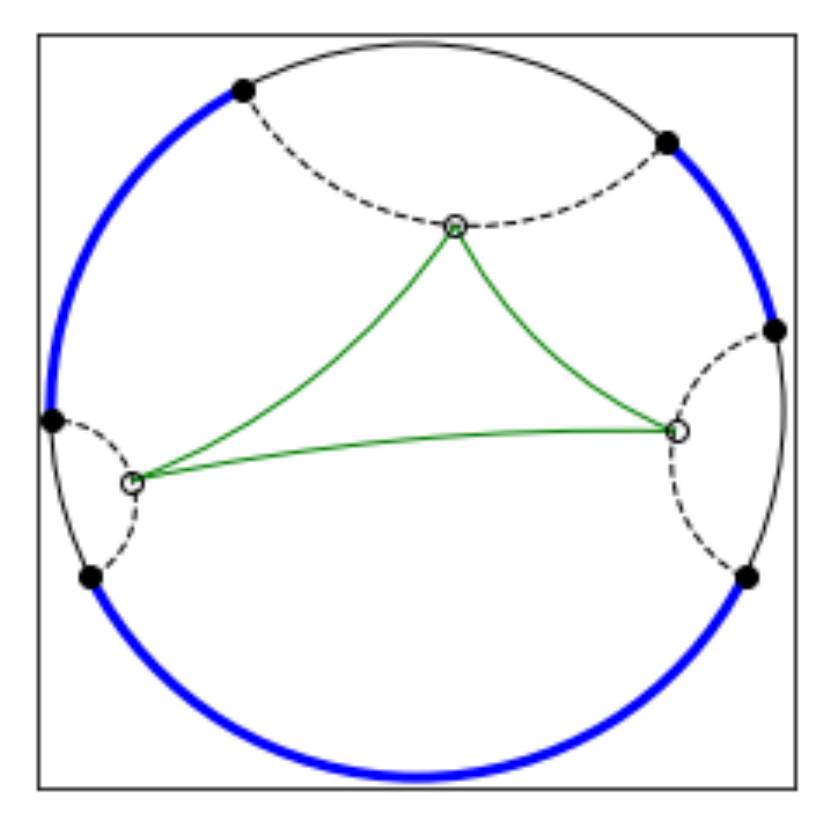}
\includegraphics[width=0.24\textwidth]{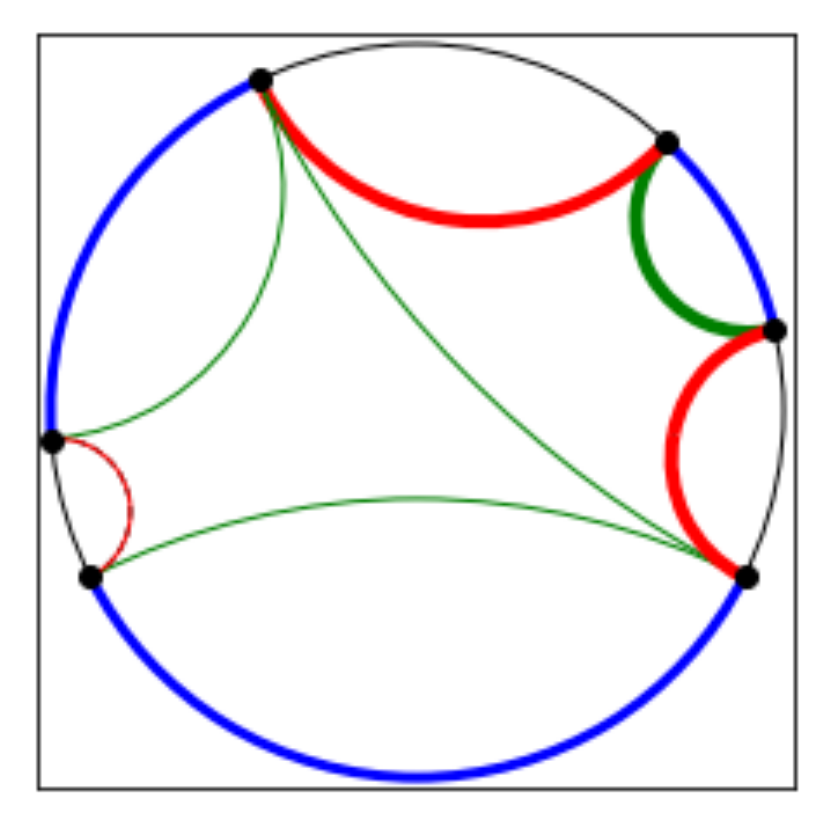}\break
\includegraphics[width=0.24\textwidth]{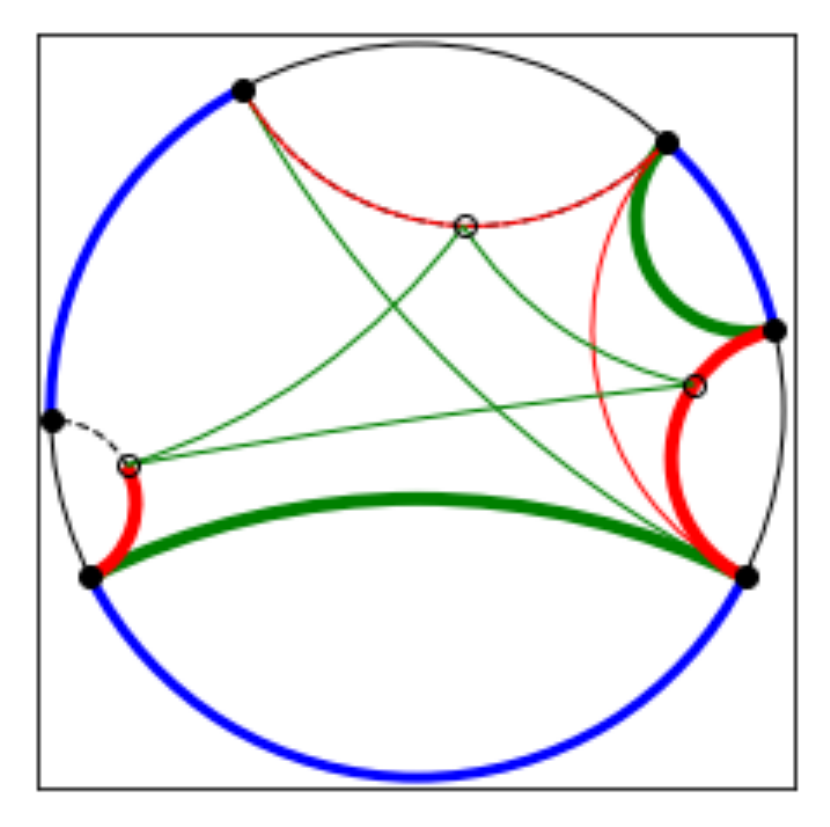}
\includegraphics[width=0.24\textwidth]{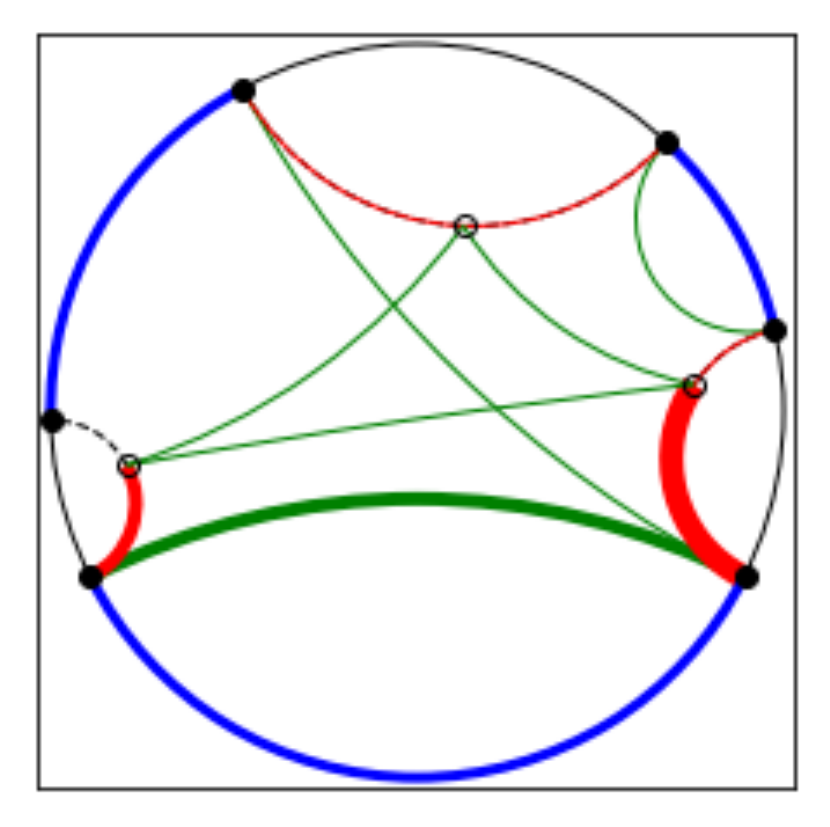}
\includegraphics[width=0.24\textwidth]{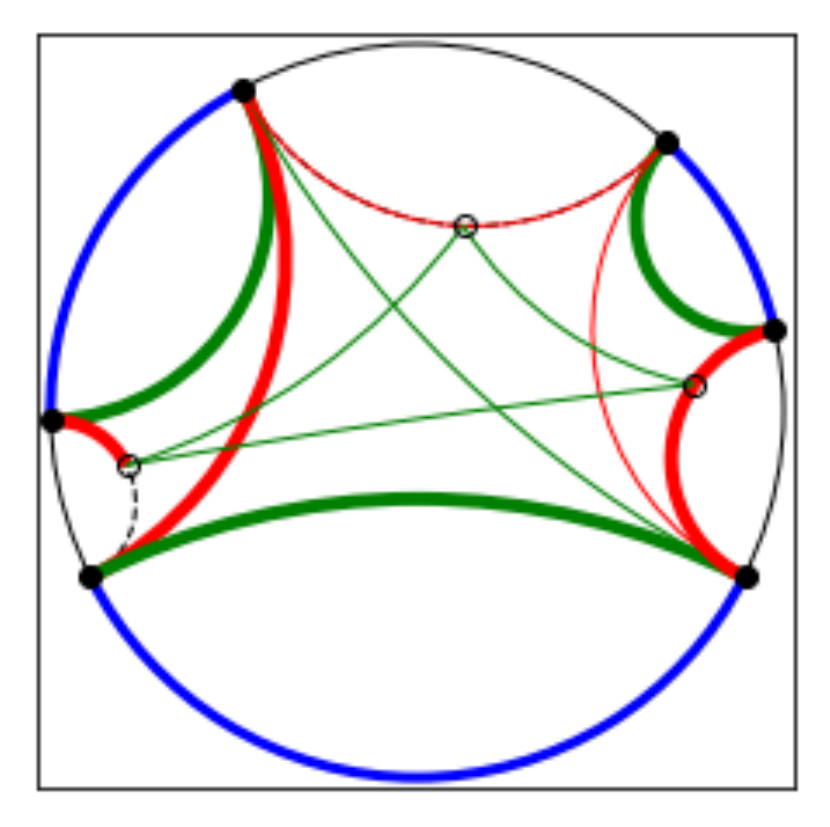}
\caption{Top line, the entanglement wedge cross section (left) and dual total correlation $J(A:B:C)$ (right), bottom line three equivalent pictures for the $Q$-correlation $f_{s_1} = f^Q_3$, none of which are the sum of the top two pictures, all for $\theta_A =0.6$,  $\theta_B =1.1$,  $\theta_C = 2.2$.}
\label{ASymEPandJ}
\end{center}
\end{figure}

As the regions are made to have different sizes, the entanglement wedge cross section stretches but never undergoes any qualitative change in its presentation. For example, it is plotted on the top left of figure~\ref{ASymEPandJ} for the case of regions with size $\theta_A =0.6$,  $\theta_B =1.1$,  and $\theta_C = 2.2$. All three of $f_{c_1} = f^P_3$, $f_{c_2}$ and $f_{c_3}$ (and thus also the tripartite $R$-correlation $f^R_3$) coincide with the entanglement wedge cross-section over all of parameter space.

Let us turn now to the $Q$-correlation. For all cases with the three boundary regions of the same size, we found that $E_Q = E_P + J(A:B:C)$. One might wonder whether this holds over all of parameter space. Moreover, since $E_P$ comes with optimization points at the entanglement wedge positions and $J(A:B:C)$ doesn't depend on optimization points, one might hypothesize that the optimization points remain those of the entanglement wedge over all of parameter space.

However, neither of these turns out to hold. In the bottom line of figure~\ref{ASymEPandJ} we show three equivalent presentations for $f_{s_1} = f^Q_3$ for regions $\theta_A =0.6$,  $\theta_B =1.1$, and $\theta_C = 2.2$. We see that although there is some resemblance to the sum of the pictures of $E_P$ and $J(A:B:C)$ shown on the top row, none matches exactly. Moreover, the fact that there are three equivalent pictures indicates the presence of a curve optimization relation at play. In fact, the condition following from all three being equivalent is nothing else than the axial curve optimization condition, shown in figure~\ref{AxialOptPointFig}. This indicates that the left and right optimization points in this measure are axial optimization points; this is a meaningful statement even though there is no symmetry. Thus the $Q$-correlation, whose optimization points match the entanglement wedge cross section when the boundary regions are all the same size, acquires axial optimization points in these cases of different sized boundary regions.

Many of the other pictures with non-equal boundary regions appear similar to, or to interpolate between, their cousins with equal-sized regions; however none appear so qualitatively different as to require display here, so we will end our tour here.

\section{Discussion}

In this paper, we derived a slew of novel optimized tripartite correlation measures, all of which can be seen as generalizations of known optimized bipartite correlation measures.  By studying the set of states on which measures vanish, we were able to determine that some of the correlation measures quantify only quantum correlations, while others quantify both classical and quantum correlations.  We then used the surface-state correspondence to construct holographic duals for each correlation measure, and saw that the holographic duals can manifest the symmetry of the correlation measures in different ways, either as one unique presentation in terms of bulk surface that respects the complete symmetry of the correlation measure, or as a family of equivalent bulk surface presentations which, as a set, respects the full symmetry of the measure, while each individual presentation does not.  We also saw how the optimal purifications for the correlation measures are encoded in a set of different points in the bulk, each obeying a different constraint on bulk surfaces. This reveals new examples of how measures of correlations in quantum systems are represented as geometry, providing another example of how quantum information theory and high energy physics can have a symbiotic relationship. 

There are a number of questions we would like to address in future work.  It was shown in \cite{LSD20} that for a pure state on $ABE$, the entropy of $A$ can be decomposed as 
\begin{align}
S(A) = E_Q(A:B) + I^{ss}(E\rangle A),
\end{align}
where $I^{ss}(E\rangle A)$ is the symmetric side-channel-assisted distillable entanglement \cite{SSW06} from $E$ to $A$.  This gives operational meaning to the bipartite $Q$-correlation.  Similar operational interpretations of the tripartite correlation measures presented here would be of great interest. Secondly, we would like to find bit thread \cite{freedman2017bit} interpretations of the tripartite correlation measures, as was done for  bipartite $E_P$ in \cite{harper2019bit, Bao19, Du19}. Individual bit threads appear to manifest a bipartite structure, but multipartite correlations may be present by considering families of thread configurations, and the examples developed here may be of considerable use in elucidating how multiparty correlations manifest holographically. Thirdly, in both this work and in \cite{LSD20}, we focused solely on the case of a pure AdS bulk geometry, corresponding to the ground state of the boundary theory.  It will be interesting to consider the evaluation of optimized correlation measures in more general geometries, including spacetimes with horizons such as a black hole geometry, corresponding to a boundary thermal state. Finally, again in both this work and \cite{LSD20}, we have identified convex polyhedral cones of optimized correlation measures but examined only the extreme rays of these cones.  Since the infimum of the sum is not necessarily the sum of the infima, there may be more interesting correlation measures living in the cones' interiors.  

\section*{Acknowledgements}

OD is supported by the Department of Energy under grant DE-SC0010005. OD, JL, and GS are supported by the Department of Energy under grant DE-SC0020386.

\bibliography{Opt_Tri_Corr}

\end{document}